\documentclass[12pt]{report}
\usepackage[french]{babel}
\usepackage{a4}
\usepackage[dvips]{graphics}
\usepackage{times}

\usepackage[T1]{fontenc} 
\def\flqq{\ouvreguillemet\everypar={}}
\def\ouvreguillemet{\leavevmode\hbox{%
     \guillemotleft\kern+0.20em}\nobreak}
\def\frqq{\fermeguillemet\everypar={}}
\def\fermeguillemet{\nobreak\leavevmode\hbox{%
     \kern+0.20em\guillemotright}}

\makeatletter

\def\lesssim{\mathrel{\mathpalette\vereq<}}
\def\vereq#1#2{\lower3pt\vbox{\baselineskip1.5pt \lineskip1.5pt
\ialign{$\m@th#1\hfill##\hfil$\crcr#2\crcr\sim\crcr}}}

\def\lambdabar{%
\relax
\bgroup
\def\@tempa{\kern-.8em\hbox{\raise1ex
\hbox{\kern.25em\vrule width.5em height.3pt depth.1pt}}}%
\mathchoice{\hbox{$\displaystyle\lambda$}\@tempa}%
{\hbox{$\textstyle\lambda$}\@tempa}%
{\hbox{$\scriptstyle\lambda$}\@tempa}%
{\hbox{$\scriptscriptstyle\lambda$}\@tempa}%
\egroup
}

\def\dbar{%
\relax
\bgroup
\def\@tempa{\kern-.6em\hbox{\raise1.2ex
\hbox{\kern.25em\vrule width.4em height.3pt depth.1pt}}}%
\mathchoice{\hbox{\normalsize\mbox{d}}\@tempa}%
{\hbox{\normalsize\mbox{d}}\@tempa}%
{\hbox{\small\mbox{d}}\@tempa}%
{\hbox{\tiny\mbox{d}}\@tempa}%
\egroup
}

\makeatother

\newcommand{\demi}{{\textstyle\frac12}}

\let\oldbf\bf
\newcommand{\mbf}[1]{\ifmmode%
\mathchoice{\mbox{\boldmath{$\displaystyle#1$}}}%
{\mbox{\boldmath{$\textstyle#1$}}}%
{\mbox{\boldmath{$\scriptstyle#1$}}}%
{\mbox{\boldmath{$\scriptscriptstyle#1$}}}%
\else\oldbf#1\fi}

\begin{document}

\thispagestyle{empty}

\vspace*{5cm}
\begin{center}
\large Ecole Centrale Paris\\
Laboratoire EM$^2$C\\[2\baselineskip]
\LARGE\bf 
Radiative transfer and atom transport\\[2\baselineskip]
Transfert radiatif et transport d'atomes\\[2\baselineskip]
\large\rm Postdoc Report of\\[1\baselineskip]
\Large Carsten Henkel\\[2\baselineskip]
\large Mars--Juillet 1997 (preprint date: May 2005)
\end{center}
\vspace*{5cm}

\clearpage

\thispagestyle{empty}
\vspace*{1cm}

\clearpage

\setcounter{page}{1}



\addcontentsline{toc}{section}{Foreword}

\vspace*{1cm}

\begin{footnotesize}

\begin{center}
\bf Foreword to the preprint version
\end{center}

\noindent
This report summarizes a post-doctoral training period of four
months that I spent, eight years ago, in the group working on Propagation
and Scattering of Electromagnetic Waves that is led by Jean-Jacques
Greffet (Ecole Centrale Paris). The report deals with classical and quantum
descriptions of particles that interact with smooth random potentials,
for example ultracold atoms in the dipole potential of an optical speckle 
pattern.
In addition, a discussion of the link between Radiative Transfer
theory and the underlying coherence theory of wave optics is presented.
The Radiative Transfer Equation is shown to emerge as a limiting case
of the Bethe-Salpeter Equation, and some next-order corrections are
discussed. 

I decided to put this material on the arxiv
because of a recent explosion of interest in the transport of 
ultracold atoms in the multiple scattering regime. Also a number
of colleagues have benefited from the parts dealing with 
Radiative Transfer. I hope
that these pages may be useful for other interested scientists
as well.

The report is written in French which may cause difficulties,
although only basic vocabulary is used, and the technical terms
are mainly identical to the English ones (some English equivalents
are given in footnotes, if needed). Just keep in mind that the French
\emph{diffusion} means `scattering', while \emph{diffusion spatiale}
is used for `diffusion'. 
In case you really get stuck, please do not hesitate to contact me at

carsten.henkel@physik.uni-potsdam.de

\bigskip

\noindent
C. Henkel, Potsdam, 03 May 2005.

\addcontentsline{toc}{section}{Remerciements}

\vspace*{1cm}

\begin{center}
\bf Remerciements
\end{center}

Je tiens d'abord \`a remercier Jean-Jacques Greffet de
m'avoir accueilli au Laboratoire \flqq EM2C\frqq\ de
l'Ecole Centrale Paris, ainsi d'avoir pu jouir d'excellentes
conditions de travail. 

Merci \'egalement aux autres
membres de son groupe d'\'electro-magn\'etisme : 
aux th\'esards Olivier Calvo, Laurent Roux, 
Jean-Baptiste Thibaud, \`a R\'emi Bussac, 
Pedro Valle et Anne Sentenac. 
Merci aussi aux \'etudiants du
groupe des \flqq combustionnistes\frqq, et plus
particuli\`erement \`a Metta et  Katja, 
S\'ebastien Pax(ion), Olivier Delabroy, Manuel et Ulrich.

Merci \`a Gilbert pour l'assistance avec le r\'eseau d'ordinateurs.
Merci \`a Caroline et St\'ephanie pour votre aide avec les
d\'emarches administratives.

C'est un temps agr\'eable que j'ai pass\'e avec vous.

\vspace*{1cm}

\end{footnotesize}

\tableofcontents

\chapter*{Introduction}
\addcontentsline{toc}{chapter}{Introduction}

\section*{Le probl\`eme}
\addcontentsline{toc}{section}{Le probl\`eme}

Le pr\'esent rapport r\'esume les quelques id\'ees que j'ai pu 
explorer au sujet du transport, que ce soit de lumi\`ere ou
d'atomes. Ce sujet est n\'e d'une question apparue vers
la fin de ma th\`ese : 
\begin{quote}
lorsque des atomes de basse temp\'erature 
traversent un champ lumineux de tavelures, qu'arrive-t-il \`a
leur distribution de vitesse ? Et qu'en est-il de la coh\'erence
spatiale des ondes de {\sc de Broglie} associ\'ee aux atomes ?
\end{quote}
Une telle question a un int\'er\^et imm\'ediat pour ce que l'on
appelle \flqq l'optique atomique\frqq, qu'il s'agisse de
l'imagerie d'atomes ou de l'interf\'erom\'etrie.
Le groupe d'optique atomique \`a l'Institut d'Optique \`a Orsay 
a pu observer dans l'exp\'erience du miroir \`a atomes 
\cite{Landragin96b} 
que la travers\'ee du champ de tavelures \'elargit la distribution
des vitesses transverses, perpendiculaires \`a la vitesse
moyenne des atomes. Dans une telle exp\'erience,
des champs de tavelures apparaissent \`a cause de faisceaux
lumineux parasites ou bien par diffusion sur une surface
rugueuse. L'\'elargissement de la vitesse transverse
d\'epend en particulier de la quantit\'e de lumi\`ere parasite.  
Dans le cas le moins favorable, 
il emp\^echait d'observer la diffraction d'atomes
par un r\'eseau en r\'eflexion parce que la
largeur angulaire du faisceau atomique devenait trop
importante.

D'un point de vue th\'eorique, le mouvement d'une onde de
mati\`ere dans un champ de tavelures se r\'esume \`a
un probl\`eme de propagation d'une onde dans un milieu
al\'eatoire o\`u l'onde subit des multiples collisions 
\'elastiques. Sous cet \'elairage, l'on con\c coit 
ais\'ement qu'il y a un grand nombre de situations physiques
o\`u appara\^{\i}t un probl\`eme analogue : 
\begin{itemize}
\item
la diffusion de la lumi\`ere dans un milieu fortement
diffusant (atmosph\`eres stellaire ou terrestre, liquides,
peintures, tissus biologiques) ;
\item
le transport des \'electrons dans un solide \flqq r\'eel\frqq\ 
(avec des d\'efauts et des vibrations thermiques), 
avec des applications pour la conductivit\'e ;
\item
la diffusion des neutrons dans les r\'eacteurs \`a fission ;
\item
la propagation des ondes sismiques \`a travers le globe
terrestre ; 
\item
et d'autres exemples encore (arr\^et d'un faisceaux de particules
radioactives dans une matrice solide, mouvement de boules
dans un terrain rugueux, marche au hasard d'une particule,
mouvement d'atomes dans un champ laser r\'esonnant ...)
\end{itemize}
Il est donc naturel d'explorer les analogies avec ces probl\`emes
pour trouver des m\'ethodes qui permettent de r\'epondre aux
questions pos\'ees.

\section*{Approches th\'eoriques}
\addcontentsline{toc}{section}{Approches th\'eoriques}

Une caract\'eristique commune de ces m\'ethodes th\'eoriques 
est qu'elles sont des {\em approches statistiques\/} : au lieu
de d\'ecrire le mouvement d'une particule (ou d'une onde)
donn\'ee dans un
milieu al\'eatoire donn\'ee, l'on se r\'esigne \`a calculer
un {\em mouvement moyen\/}, que l'on esp\`ere \flqq typique\frqq\ 
pour une mesure portant sur un grand nombre de particules (ondes). 
L'on construit ainsi une {\em th\'eorie de transport\/}
pour les particules (ondes).
La quantit\'e dont on calcule la valeur moyenne d\'epend de
la situation exp\'erimentale que l'on envisage : une fonction
d'onde moyenne si celle-ci interf\`ere avec un \flqq oscillateur
local\frqq, c'est-\`a-dire une onde de r\'ef\'erence avec une
phase fix\'ee ; une fonction de coh\'erence de l'onde si l'on
\'etudie une figure d'interf\'erence moyenn\'ee sur un grand
nombre de r\'ealisations du syst\`eme ; ou encore une fonction
de corr\'elation d'ordre quatri\`eme (corr\'elations d'intensit\'e),
lorsque l'on s'int\'eresse par exemple aux propri\'et\'es 
de la figure d'interf\'erence cr\'e\'ee par une seule r\'ealisation
de l'exp\'erience (cette situation s'applique \`a l'interf\'erence
entre deux condensats de {\sc Bose--Einstein} ou encore aux
tavelures lumineuses d'un faisceau laser coh\'erent).

Les exemples que l'on vient de donner s'inscrivent essentiellement
dans un cadre statistique {\em ondulatoire}, o\`u les objets de base sont
des ondes (ou champs), comme les ondes de mati\`ere ou lumineuses.
Il existe \'egalement des probl\`emes o\`u ce sont des {\em particules}
qui sont diffus\'ees dans un milieu al\'eatoire. Dans ce cas, il 
n'a pas de sens de parler d'ondes ou d'interf\'erence : la description
naturelle des moyennes du syst\`eme passe simplement par une 
densit\'e spatiale 
ou une distribution des vitesses. L'on est habitu\'e \`a une
telle approche par la th\'eorie cin\'etique des gaz dont le point
de d\'epart est l'\'equation de {\sc Boltzmann} : cette \'equation
d\'ecrit comment la fonction de distribution d'un ensemble de
particules \'evolue en pr\'esence d'une force ext\'erieure d'une part,
et des collisions entre particules d'autre part. Un autre exemple,
un peu moins habituel, est celui du transfert radiatif : l'on
consid\`ere alors le transport de l'\'energie lumineuse dans
un milieu diffusant (et/ou absorbant), en d\'ecrivant cette
\'energie par sa distribution en fr\'equence, en position
et direction de propagation (la \flqq luminance\frqq). 
L'\'equation de base de cette th\'eorie ressemble \`a celle
de {\sc Boltzmann} et
s'appelle celle du transfert radiatif ; elle a \'et\'e
introduite au d\'ebut du si\`ecle et \'etudi\'ee abondamment
par {\sc Chandrasekhar} dans un contexte astrophysique,
pour mod\'eliser les atmosph\`eres des \'etoiles 
\cite{Chandrasekhar}. Les ingr\'edients de cette \'equation
sont les propri\'et\'es d'absorption et de diffusion du
milieu.  Pour des atomes dans un potentiel al\'eatoire,
une telle approche serait qualifi\'ee de \flqq classique\frqq,
en opposition \`a une approche \flqq quantique\frqq\ o\`u l'on
manipule des ondes de mati\`ere. Pourtant, comme on sait que
ce sont ces derni\`eres qui ont une r\'ealit\'e plus
fondamentale, l'on peut se poser la question du rapport entre
ces deux approches th\'eoriques : comment se justifie
l'approche corpusculaire \`a partir de l'approche ondulatoire ?
Ou encore : dans quelle limite peut-on se contenter d'une approche
corpusculaire, et o\`u vont se manifester des ph\'enom\`enes proprement
ondulatoires ?

\section*{Petit historique}
\addcontentsline{toc}{section}{Petit historique}

Ces questions ont \'et\'e poursuivies dans le contexte {\bf optique}
depuis les ann\'ees 1960 environ. Il s'agit de fonder
l'\'equation du transfert radiatif (description quasi-corpusculaire,
absence d'interf\'erences)
sur la th\'eorie de la diffusion multiple des ondes dans les milieux
al\'eatoires.  C'est maintenant que l'on
se rend compte du travail \'enorme des physiciens sovi\'etiques
dans ce domaine \cite{Kravtsov96}. Il est devenu clair que l'approche
du transfert radiatif est valable pour des champs lumineux dont
les profils spatiaux varient lentement \`a l'\'echelle de leur
longueur de coh\'erence (des champs \flqq quasi-uniformes\frqq).
Cette approche n'est plus valable, par exemple,
pour rendre compte de la r\'etro-diffusion exalt\'ee (``{\em coherent
backscattering\/}''), ou encore de la localisation (\flqq forte\frqq)
des ondes dans un milieu d\'esordonn\'e.
Dans ce rapport, nous allons nous appuyer abondamment sur l'analogie
avec la diffusion de la lumi\`ere. En particulier, nous allons
\'elaborer un peu les conditions de validit\'e de l'\'equation
du transfert radiatif.

Dans le domaine du transport des {\bf \'electrons} dans les solides,
le d\'eveloppement s'est poursuivi de fa\c con plut\^ot ind\'ependante,
semble-t-il. Le travail de {\sc P.-W. Anderson} sur la localisation
des \'electrons dans un r\'eseau d\'esordonn\'e 
(``{\em Absence of diffusion in certain random lattices\/}'',
\cite{Anderson58})
a \'et\'e d'un impact remarquable parce qu'il a donn\'e un
exemple o\`u les interf\'erences ont des cons\'equences
dramatiques pour le transport (une propri\'et\'e plut\^ot
\flqq macroscopique\frqq). Depuis, d'autres ph\'enom\`enes
d'origine ondulatoire ont \'et\'e \'etudi\'es, comme par exemple
les \flqq fluctuations de conductance universelles\frqq.
Un temps assez long s'est \'ecoul\'e jusqu'\`a ce que l'on prenne 
conscience que la localisation des \'electrons provient simplement
de leur caract\`ere ondulatoire, et que la localisation existe 
par cons\'equent
\'egalement pour les ondes lumineuses (voir \cite{vanTiggelen96}).
C'est dans ce contexte qu'a \'et\'e pr\'edite et observ\'ee
la r\'etro-diffusion exalt\'ee de la lumi\`ere, un autre
effet au-del\`a de l'\'equation standard du transfert radiatif.

En ce qui concerne les {\bf neutrons}, les recherches intenses
dans les r\'eacteurs nucl\'eaires ont rapidement conduit
\`a une th\'eorie du transport par analogie avec le transfert
radiatif. Des formalismes tr\`es divers ont \'et\'e utilis\'es, 
par physiciens et math\'ematiciens, 
pour la justifier \`a partir d'une approche microscopique
\cite{Overhauser59,Osborn65,Riska68,Bobker79}. La conclusion
actuelle semble que l'\'equation du transfert radiatif est
valable aux grandes \'echelles spatiales et pour des \'energies
pas trop basses.

\section*{Plan du rapport}
\addcontentsline{toc}{section}{Plan du rapport}

Ce rapport contient trois chapitres dont le {\bf premier} pr\'esente
l'\'equation du transfert radiatif pour la lumi\`ere d'un point de vue 
ph\'enom\'enologique. Nous \'etudions en particulier la limite
o\`u le milieu diffuseur favorise la diffusion vers l`avant
(les petits angles de diffusion). Dans ce r\'egime, l'on peut
reformuler le transfert radiatif \`a l'aide d'une \'equation de
{\sc Fokker--Planck}. 

Nous retrouverons cette approche au
{\bf second} chapitre qui est consacr\'e au point de vue classique du
transport d'atomes dans un potentiel al\'eatoire. Nous montrons
que c'est aux \'echelles spatiales grandes par rapport \`a la
longueur de corr\'elation du potentiel, que 
la distribution moyenne des atomes dans l'espace des phases
v\'erifie une \'equation de {\sc Fokker--Planck}. Nous
en calculons le tenseur de diffusion des vitesses. 
Aux temps encore plus longs (pour des atomes tr\`es lents),
la distribution spatiale atomique est d\'ecrite par une \'equation
de diffusion spatiale dont nous d\'eterminons le coefficient
de diffusion. 

Le {\bf troisi\`eme} chapitre \'etudie le point de vue ondulatoire :
la diffusion multiple des
ondes de mati\`ere dans un milieu al\'eatoire. Nous introduisons
les concepts statistiques pertinents ainsi que leurs \'equations
d'\'evolution. Le fil rouge du chapitre est de d\'etailler le
passage vers une approche du type transfert radiatif. Le probl\`eme
est pr\'esent\'e de fa\c con \`a couvrir aussi bien la diffusion
de la lumi\`ere que celle des atomes. Nous mettons en \'evidence
l'\'echelle spatiale minimale au-dessus de laquelle
l'\'equation du transfert radiatif est \'equivalente aux approches
ondulatoires statistiques.




\chapter{Transfert radiatif}
\label{s:radiatif}

Une petite introduction dans la th\'eorie du transfert radiatif.
La lumi\`ere y est d\'ecrite par la {\em luminance\/}. Nous
\'etudions un peu plus en d\'etail la situation d'une
diffusion \flqq piqu\'ee vers l'avant\frqq, o\`u l'\'equation
du transfert radiatif peut \^etre \'ecrite sous la forme
d'une \'equation de {\sc Fokker--Planck}. Nous verrons
au chapitre suivant qu'une \'equation identique appara\^{\i}t
pour le transport d'atomes. On illustre de cette fa\c con
le passage entre des th\'eories ondulatoire et corpusculaire.

\section{La luminance}
La quantit\'e centrale pour d\'ecrire la lumi\`ere dans
le transfert radiatif est
la {\em luminance\/} ${\cal I}({\bf r}, {\bf u})$ : elle
repr\'esente la puissance rayonn\'ee ${\rm d}{\cal F}$
par un \'el\'ement de surface ${\rm d}\sigma$
situ\'e \`a la position ${\bf r}$,
dans un \'el\'ement d'angle solide ${\rm d}\Omega$
autour de la direction caract\'eris\'ee par le vecteur
unitaire ${\bf u}$ (voir le chapitre~5.7 de {\sc Mandel} et
{\sc Wolf} \cite{MandelWolf} et le chapitre~4.3 de
{\sc Rytov, Kravtsov} et {\sc Tatarskii} \cite{Rytov})\thinspace:
\begin{equation}
{\rm d}{\cal F} =  {\cal I}({\bf r}, {\bf u}) {\rm d}\sigma \cdot
{\rm d}\Omega
	\label{eq:def-luminance}
\end{equation}
L'\'equation d'\'evolution pour la luminance est celle
du transfert radiatif\thinspace:
\begin{equation}
{\bf u}\cdot\nabla_{r} {\cal I} = 
\underbrace{ S({\bf r}, {\bf u}) }_{ \mbox{\scriptsize sources} }
 - \underbrace{ \beta({\bf r}) {\cal I} }_{ \mbox{\scriptsize att\'enuation} }
+ \frac{ 1 }{ \ell_{\rm ex}( {\bf r} ) } 
\int\limits_{4\pi} \underbrace{ {\rm d}\Omega' \,
p({\bf u}, {\bf u}'; {\bf r}) {\cal I}({\bf r}, {\bf u}')
}_{ \mbox{\scriptsize re-diffusion} }
	\label{eq:e-t-r}
\end{equation}
Cette \'equation traduit les processus suivants\thinspace:
\begin{itemize}
\item
l'\'energie lumineuse est transport\'ee dans la direction ${\bf u}$
le long des rayons g\'eom\'etriques ;
\item
la lumi\`ere peut \^etre \'emise par une source $S({\bf r}, {\bf u})$
dans la direction ${\bf u}$ ;
\item
elle peut \^etre att\'enu\'ee par de l'absorption
ou de la diffusion. Ce processus est caract\'eris\'e par
un coefficient $\beta({\bf r})$ ;
\item
la diffusion redistribue les directions de l'intensit\'e 
lumineuse. Ce processus est d\'ecrit par le libre parcours
moyen%
\footnote{%
L'indice \flqq ex\frqq\ indique qu'il s'agit de la distance
caract\'eristique pour l'extinction de la partie coh\'erente 
du champ lumineux.}
$\ell_{\rm ex}( {\bf r} )$ (\'egal \`a la section efficace
totale par unit\'e de volume) et la \flqq fonction
de phase\frqq\ $p({\bf u}, {\bf u}'; {\bf r})$ 
que l'on peut interpr\'eter comme une \flqq section efficace 
diff\'erentielle de diffusion\frqq\ pour un processus o\`u
un rayon venant d'une direction ${\bf u}'$ est diffus\'e dans
la direction ${\bf u}$. 
\end{itemize}

\section{Hypoth\`eses pour le transport radiatif}

\begin{enumerate}
\item
L'on d\'ecrit le milieu \`a une {\em \'echelle spatiale grande
devant la longueur d'onde optique\/} (pour pouvoir d\'efinir
une puissance rayonn\'ee dans une direction par unit\'e de
surface). Ceci implique que les coefficients $\beta({\bf r})$
et $f({\bf u}, {\bf u}'; {\bf r})$ et bien s\^ur la
luminance ${\cal I}({\bf r}, {\bf u})$ elle-m\^eme varient lentement,
en fonction de la position ${\bf r}$, \`a l'\'echelle 
de la longueur d'onde optique.
\item
La diffusion de la lumi\`ere est suppos\'ee {\em \'elastique\/}%
\thinspace: c'est seulement la direction de la lumi\`ere,
mais non pas sa fr\'equence qui est modifi\'ee par le milieu.
Dans cette hypoth\`ese, l'on peut formuler une \'equation du
transport radiatif pour chaque fr\'equence lumineuse
s\'epar\'ement. Il est cependant imm\'ediat
de rendre compte de la diffusion in\'elastique en introduisant
une section efficace in\'elastique et
une d\'ependance de la luminance de la fr\'equence.
\end{enumerate}

\section{Relation \`a une \'equation de \protect{\sc Fokker--Planck}}

Afin de comparer l'\'equation du transport radiatif~(\ref{eq:e-t-r})
au transport d'atomes, une petite modification\thinspace: nous allons en
d\'eduire une \'equation de type {\sc Fokker--Planck} que nous
retrouverons \'egalement pour les atomes.
A cette fin, nous supposons que la re-diffusion de la lumi\`ere
s'effectue surtout autour de la direction vers l'avant, c'est-%
\`a-dire que dans la fonction de phase
$f({\bf u}, {\bf u}'; {\bf r})$, le vecteur \flqq diffus\'e\frqq\ 
${\bf u}$ pointe dans une direction voisine du vecteur
\flqq incident\frqq\ ${\bf u}'$. Une telle situation correspond,
par exemple, \`a la diffusion par des inhomog\'en\'eit\'es d'indice,
lorsque celles-ci sont beaucoup plus grandes que la longueur d'onde.
Ce cas se pr\'esente fr\'equemment dans la diffusion de la lumi\`ere
par des tissus biologiques.

Dans la limite d'une diffusion piqu\'ee vers l'avant,
nous pouvons d\'evelopper la luminance ${\cal I}({\bf r}, {\bf u}')$
au troisi\`eme terme dans~(\ref{eq:e-t-r}) en fonction de
${\bf u}'$. Ceci faisant, l'on obtient l'expression suivante 
pour le terme de re-diffusion :
\begin{equation}
\mbox{re-diffusion} \to
\frac{ 1 }{ \ell_{\rm ex}( {\bf r} ) }
\, {\cal I}({\bf r}, {\bf u}) 
+ \sum_{\alpha} F_\alpha({\bf r}) \partial_{u_\alpha} 
{\cal I}({\bf r}, {\bf u}) 
+ \sum_{\alpha\beta} D_{\alpha\beta}({\bf r}) 
\partial_{u_\alpha} \partial_{u_\beta} 
{\cal I}({\bf r}, {\bf u}) 
	\label{eq:vers-FP}
\end{equation}
o\`u $u_{\alpha,\beta}$ d\'esignent les directions perpendiculaires
au vecteur ${\bf u}$. (Ce sont seulement ces directions-ci qui
interviennent dans la d\'eriv\'ee, \'etant donn\'e que ${\bf u}$
est un vecteur unitaire.) Dans~(\ref{eq:vers-FP}), 
$F_\alpha({\bf r})$ est la \flqq force moyenne\frqq\ et 
$D_{\alpha\beta}({\bf r})$ le tenseur de diffusion. Ils sont donn\'es par
\begin{eqnarray}
F_\alpha({\bf r}) &=& \frac{ 1 }{ \ell_{\rm ex}( {\bf r} ) }
\int\limits_{4\pi} \! {\rm d}\Omega' \,
(u'_\alpha - u_\alpha) \, p({\bf u}, {\bf u}'; {\bf r})
	\label{eq:def-force}\\
D_{\alpha\beta}({\bf r}) &=& \frac{ 1 }{ 2 \ell_{\rm ex}( {\bf r} ) }
\int\limits_{4\pi} \! {\rm d}\Omega' \,
(u'_\alpha - u_\alpha) (u'_\beta - u_\beta) 
\, p({\bf u}, {\bf u}'; {\bf r})
	\label{eq:def-coeff-diffusion}
\end{eqnarray}
Notons qu'une telle formulation du transfert radiatif a \'et\'e
\'etudi\'ee en d\'etail par {\sc G. C. Pomraning} \cite{Pomraning95},
dans le contexte de l'arr\^et d'un faisceau de particules \`a 
l'int\'erieur d'un solide diffusant. Le groupe de {\sc J. M. Luck}
l'a \'egalement utilis\'ee pour trouver des solutions analytiques
dans une g\'eom\'etrie planaire \cite{Luck96}.

Lorsque le milieu diffuseur est statistiquement
isotrope (la fonction de phase ne d\'epend que de
$\cos\theta \equiv {\bf u}\cdot{\bf u}'$), 
la force $F_\alpha( {\bf r} )$ 
s'annule et le tenseur $D_{\alpha\beta}( {\bf r} )$ est proportionnel 
au {\sc Kronecker} $\delta_{\alpha\beta}$\thinspace:
\[
\mbox{milieu isotrope}:\quad
D_{\alpha\beta}( {\bf r} )
= \frac{\delta_{\alpha\beta}}{4\ell_{\rm ex}( {\bf r} )} 
\langle \sin^2\theta \rangle_p 
\approx 
\frac{\delta_{\alpha\beta}}{4\ell_{\rm ex}( {\bf r} )} 
\langle \theta^2 \rangle_p 
\approx 
\frac{\delta_{\alpha\beta}}{2\ell^*_{\rm trans}( {\bf r} )} 
\]
o\`u $\ell^*_{\rm trans} = \ell_{\rm ex} 
/ ( 1 - \langle \cos\theta \rangle )$ est le libre parcours
moyen de transport. Nous constatons donc que
les \'el\'ements du tenseur de diffusion correspondent 
\`a la valeur quadratique moyenne de la d\'eviation angulaire
par unit\'e de longueur des rayons. Ceci appelle \`a une
interpr\'etation physique de l'\'equation de {\sc Fokker--Planck}~
(\ref{eq:vers-FP}).

\section{Mouvement diffusif des directions des rayons}

Introduisons les angles $\theta, \phi$ pour le vecteur unitaire
${\bf u}$ et supposons que le milieu diffusant est sans sources,
non-absorbant et isotrope.
Le tenseur de diffusion s'\'ecrit alors $D_{\alpha\beta} =
D \,\delta_{\alpha\beta}$ et
l'\'equation de {\sc F.--P.} pour la luminance devient
\begin{equation}
{\bf u} \cdot \nabla_r {\cal I}({\bf r}, \theta, \phi) =
D \left[
\frac{ 1 }{ \sin \theta } 
\frac{ \partial }{ \partial \theta }
\left( \sin\theta \frac{ \partial }{ \partial \theta } \right)
+ 
\frac{ 1 }{ \sin^2\theta } \frac{ \partial^2 }{ \partial\phi^2 } 
\right] 
{\cal I}({\bf r}, \theta, \phi)
	\label{eq:FP-sphere}
\end{equation}
Les d\'eriv\'ees angulaires correspondent au Laplacien sur la
sph\`ere. Leurs fonctions propres sont donc les polyn\^omes de
{\sc Legendre} $P_n(\cos\theta) {\rm e}^{ i m \phi}$. 
Pour une luminance qui
est initialement collimat\'ee autour de la direction $\theta = 0$
(l'axe $Oz$),
{\sc Frisch} donne alors la solution explicite suivante
(\cite{Frisch66}, \'eq.~(19.20))
\begin{eqnarray}
&& {\cal I}(z, \theta, \phi) = \frac{ 1 }{ 4\pi }
\sum\limits_{n = 0}^{\infty} ( 2 n + 1 )
P_n(\cos\theta) \, {\rm e}^{ - n ( n + 1 ) D z }
	\label{eq:solution-Frisch}\\
&& {\cal I}(z = 0, \theta, \phi) = \delta(\theta)
	\nonumber
\end{eqnarray}
On en d\'eduit que les structures angulaires fines (les harmoniques 
sph\'eriques sup\'erieurs avec $n$ grands) sont amorties plus rapidement
que les variations angulaires \flqq molles\frqq. Dans la limite $z \to \infty$
d'un milieu profond, il ne subsiste plus qu'une distribution angulaire
isotrope ($n = 0$). La distance caract\'eristique $L$ pour 
l'\flqq isotropisation\frqq\ de la luminance est donn\'ee par l'inverse
du coefficient de diffusion
\begin{equation}
L \sim \frac{ 1 }{ D } = \frac{ \ell^*_{\rm trans} }{ 2 }
	\label{eq:c-est-isotrope}
\end{equation}
Pour une fonction de phase tr\`es piqu\'ee vers l'avant, cette
distance (le libre parcours moyen de transport)
est beaucoup plus grande que la longueur d'att\'enuation
($\sim \ell_{\rm ex}$) de la partie collimat\'ee du faisceau lumineux.
La solution explicite~(\ref{eq:solution-Frisch}) permet donc
d'appr\'ecier la transition entre un \flqq transport balistique\frqq\ 
(distribution angulaire piqu\'ee) et un transport diffusif
(distribution angulaire isotrope). Ce qui lui manque, c'est
de prendre en compte le profil spatial de l'intensit\'e incidente
que l'on suppose en effet uniforme.

A titre d'exemple, consid\'erons l'\'evolution de la \flqq
directivit\'e\frqq\ de la luminance ; la solution~(\ref{eq:solution-Frisch}) 
donne
\begin{equation}
\langle \cos\theta \rangle = {\rm e}^{ - z / \ell^*_{\rm trans} }
	\label{eq:cos-theta-decroit}
\end{equation}
Pour des distances faibles par rapport \`a $\ell^*_{\rm trans}$, 
on trouve donc un \'elargissement diffusif
\begin{equation}
\langle \theta^2 \rangle \simeq 2 z / \ell^*_{\rm trans},
	\label{eq:c-est-diffusif}
\end{equation}
caract\'eristique pour une marche au hasard des directions de
la lumi\`ere sur la sph\`ere des vecteurs unitaires.
D'autre part, aux grandes distances $ z \gg \ell^*_{\rm trans}$, 
l'on a $\langle \cos\theta \rangle = 0 $\thinspace:
la distribution angulaire est devenue isotrope.

\paragraph{Remarque.}
L'analogie entre cette formulation du transfert radiatif et
le mouvement d'un ensemble de particules classique a fait
na\^{\i}tre l'id\'ee de \flqq simuler\frqq\ la diffusion
de la lumi\`ere par des techniques Monte Carlo
\cite{Feld94}. L'id\'ee est de repr\'esenter la luminance
comme la somme sur un grand nombre de trajectoires qui ont
subi des diffusions dans le milieu. Une telle approche
ressemble \`a la formulation de la m\'ecanique quantique
en termes d'une \flqq int\'egrale de chemins\frqq, et
l'on peut profiter de cette analogie pour faire un
calcul analytique approch\'e de l'int\'egrale
de trajectoires, en retenant, par exemple, seulement
la \flqq trajectoire la plus probable\frqq\ dans la
somme. Voir le papier de {\sc Perelman} {\em et al.\/}
\cite{Feld94} pour davantage de r\'ef\'erences.

\section{Conclusions}

Dans la limite des faibles angles de diffusion
$\langle \theta^2 \rangle \ll 1$, l'\'equation du
transfert radiatif se transforme d'une \'equation
int\'egro-diff\'erentielle en une \'equation aux d\'eriv\'ees
partielles. Elle ressemble \`a une \'equation de
{\sc Fokker--Planck} pour une distribution de probabilit\'e.
Une telle description est justifi\'ee lorsque la fonction
de phase, en fonction de l'angle de diffusion,
a une port\'ee beaucoup plus courte que la luminance
en fonction de la direction.

Les directions des rayons effectuent alors un mouvement diffusif
sur la sph\`ere des vecteurs unitaires qui est 
caract\'eris\'e par le tenseur de diffusion $D_{\alpha\beta}$.
Pour des milieux diffuseurs plus \'epais que $\sim 1 / D$,
la distribution angulaire du rayonnement devient isotrope.




\chapter{Transport d'atomes --- point de vue classique}
\label{s:classique}

Il s'agit maintenant de formuler le probl\`eme du mouvement
classique d'un ensemble d'atomes
dans un milieu al\'eatoire. Avant d'entreprendre
cette \'etude, nous allons pr\'eciser un mod\`ele pour le
milieu al\'eatoire\thinspace: il s'agit d'un {\em potentiel
al\'eatoire\/} tel qu'il appara\^{\i}t pour le d\'eplacement
lumineux dans un champ de tavelures lumineuses. Ensuite,
nous proposons une approche classique pour le
transport d'atomes : les atomes seront consid\'er\'es comme
des particules ponctuelles qui se d\'eplacent dans le
potentiel al\'eatoire. Au chapitre suivant, nous utiliserons
une approche ondulatoire o\`u l'on \'etudie la propagation des ondes
de mati\`ere dans le potentiel. Dans les deux cas, nous
allons faire le parall\`ele \`a la propagation de la
lumi\`ere.

\section{Mod\`ele pour un potentiel al\'eatoire}

Nous supposons les atomes plac\'es dans un champ de {\em
speckles} lumineux. Nous allons nous restreindre \`a un
champ lumineux monochromatique dont la fr\'equence
est d\'esaccord\'ee beaucoup par rapport \`a une fr\'equence
de r\'esonance atomique. Dans cette situation, le champ
\'electrique ${\bf E}({\bf r})$ (amplitude vectorielle 
complexe)
donne lieu \`a un {\em potentiel lumineux\/} proportionnel
au carr\'e du champ et inversement proportionnel au
d\'esaccord en fr\'equence $\omega_L - \omega_A$.
Nous nous limiterons dans un premier temps \`a un atome
avec un seul niveau interne dans l'\'etat fondamental,
donc \`a une transition atomique $J = 0 \to 1$. Le
potentiel lumineux $V({\bf r})$ prend alors la forme
\[
V({\bf r}) = \frac{ d^2 | {\bf E}({\bf r}) |^2 
}{ \hbar (\omega_L - \omega_A ) } ,
\]
il est donc proportionnel \`a l'intensit\'e lumineuse.
Dans cette expression, $d$ est l'\'el\'ement de matrice
de l'op\'erateur dipole \'electrique entre l'\'etat
fondamental et l'\'etat excit\'e. Une \'ecriture alternative
du potentiel lumineux qui fait directement intervenir
l'intensit\'e lumineuse $I({\bf r}) \equiv |{\bf E}({\bf r})|^2$
est 
\begin{equation}
V({\bf r}) = \frac{ \hbar \Gamma^2 }{ 8 (\omega_L - \omega_A ) }
\frac{ I({\bf r}) }{ I_{sat} }
	\label{eq:potentiel-lumineux}
\end{equation}
o\`u $I_{sat}$ est l'{\em intensit\'e de saturation\/}.
On d\`eduit de cette formule que l'ordre de grandeur
du potentiel lumineux est (tr\`es grossi\`erement) donn\'e 
par la largeur en \'energie $\hbar\Gamma$ de l'\'etat excit\'e.

Les tavelures lumineuses ont une structure spatiale
erratique dont l'\'echelle spatiale typique est 
la longueur d'onde lumineuse $\lambda_L = 2\pi / k_L = 
2 \pi c / \omega_L$.
A cause de leur complexit\'e, l'on peut tout au plus raisonner
en termes de leurs propri\'et\'es moyennes ou statistiques.
Nous allons supposer que ces propri\'et\'es sont enti\`erement
contenues dans la {\em fonction de corr\'elation de l'intensit\'e
lumineuse\/}\thinspace:
\begin{equation}
\langle V({\bf r}) V({\bf r}') \rangle = B^2
\langle I({\bf r}) I({\bf r}') \rangle
\equiv \bar{V}^2 \left[ 1 + g(|{\bf r} - {\bf r}'|) \right]
	\label{eq:def-pot-correlation}
\end{equation}
o\`u $B$ est le coefficient de proportionnalit\'e entre
l'intensit\'e et le potentiel lumineux,
$\bar{V}$ est la valeur moyenne du potentiel 
et $g(r)$ sa fonction de corr\'elation (sans dimension).
Nous supposons que les tavelures sont statistiquement
homog\`enes (les corr\'elations ne d\'ependent que de la
diff\'erence ${\bf r} - {\bf r}'$ entre deux positions)
et isotropes ($g$ ne d\'epend que du module du vecteur
diff\'erence). L'\'echelle de variation caract\'eristique
de la fonction de corr\'elation est la longueur d'onde
lumineuse\thinspace: \`a des positions plus distantes
que quelques $\lambda_L$, la fonction de corr\'elation
d\'ecro\^{\i}t rapidement vers z\'ero
\[
r \gg \lambda_L:\quad
g(r) \to 0.
\]
Un exemple particulier de fonction de corr\'elation pour
un champ de tavelures est donn\'e par\thinspace:
\[
g(r) = \left( 
\frac{ \sin k_L r }{ k_L r } 
\right)^2
\]
qui correspond \`a un champ lumineux avec une distribution
angulaire isotrope dans tout l'angle solide (voir 
{\sc Gori} \cite{Gori94} et {\sc Nussenzveig} \cite{Nussenzveig87a}).
Un autre exemple est une fonction de corr\'elation 
gaussienne\thinspace:
\[
g(r) = \exp\!\left( - \frac{ {\bf r}^2 }{ 2 \ell_c^2 } \right)
\]
qui correspond au champ lumineux rayonn\'e dans le champ
lointain par une source incoh\'erente non ponctuelle 
dont le profil spatial d'intensit\'e est gaussien.

\paragraph{Remarques.}
(i) Dans ce mod\`ele, le milieu al\'eatoire est
d\'ecrit par un {\em potentiel\/}, \`a la diff\'erence du
mouvement brownien, par exemple, o\`u l'on mod\'elise les
chocs que subit une particule immers\'e dans un fluide
par une force al\'eatoire et d\'ependante du temps. Dans
ce probl\`eme appara\^{\i}t par exemple un amortissement
de la vitesse atomique, en cons\'equence des chocs avec
le fluide environnant. Dans notre probl\`eme par contre,
l'\'energie de l'atome est conserv\'ee {\em a priori\/},
il peut tout au plus y avoir un transfert d'\'energie
entre \'energies potentielle et cin\'etique, ou bien
entre \'energie cin\'etique \flqq orient\'ee\frqq\ (correspondant 
\`a une distribution de vitesse collimat\'ee)
et \flqq diffuse\frqq\ (correspondant
\`a un mouvement diffus de l'ensemble d'atomes).

(ii) Si nous nous contentons de la fonction de corr\'elation
en deux points pour d\'ecrire le potentiel al\'eatoire, ceci
revient \`a supposer que ce dernier poss\`ede une {\em statistique
gaussienne\/}. C'est une hypoth\`ese commode pour le calcul,
mais qui peut s'av\'erer insuffisante dans certaines situations
exp\'erimentales (diffusion par les particules en suspension
liquide, par exemple). {\sc V.~M. Finkel'berg} discute un potentiel
avec une statistique plus g\'en\'erale et donne les
formules n\'ecessaires pour calculer les fonctions de
corr\'elation \cite{Finkelberg67}.

\section{Distribution dans l'espace des phases}

La quantit\'e de base pour d\'ecrire le mouvement d'un
ensemble d'atomes classiques est sa {\em densit\'e de
probabilit\'e dans l'espace des phases\/} $f({\bf r}, {\bf v})$
qui donne le nombre d'atomes ${\rm d}N$ se trouvant autour
de la position ${\bf r}$ et ayant une vitesse ${\bf v}$\thinspace:
\begin{equation}
{\rm d}N = f({\bf r}, {\bf v}) \, {\rm d}{\bf r} \, {\rm d}{\bf v}
	\label{eq:def-f-de-rv}
\end{equation}
Nous notons l'analogie \`a la luminance ${\cal I}({\bf r}, {\bf s})$
d\'efinie \`a l'\'equation~(\ref{eq:def-luminance}). A la diff\'erence
de cette derni\`ere, la distribution dans l'espace des phases
d\'epend \`a la fois de la direction et du module de la vitesse  
atomique. Ceci traduit le fait qu'en g\'en\'eral, les atomes
peuvent avoir une vitesse (une \'energie cin\'etique) quelconque.
En outre, lors du mouvement dans le potentiel al\'eatoire,
l'\'energie cin\'etique des atomes est modifi\'ee.
A la diff\'erence de la lumi\`ere,
la diffusion d'atomes n'est donc pas monochromatique\thinspace:
la fr\'equence de l'onde de {\sc de Broglie} change.
Nous pouvons tout au plus esp\'erer de retrouver une conservation
d'\'energie cin\'etique \flqq en moyenne\frqq\ lorsque le potentiel
al\'eatoire a une valeur moyenne spatialement constante. Cette loi de 
conservation serait cependant approch\'ee et seulement
valable \`a une \'echelle spatiale grande par rapport \`a 
l'\'echelle de variation des tavelures (la longueur d'onde lumineuse).

\section{D\'erivation d'une \'equation de Fokker--Planck}

\subsection{Approche heuristique}

Nous partons de l'\'equation de {\sc Liouville} pour la
distribution $f({\bf r}, {\bf v}; t)$ dans l'espace des
phases\thinspace:
\begin{equation}
\partial_t f({\bf r}, {\bf v}; t) + {\bf v}\cdot\nabla_r f
+ \frac{ 1 }{ M } {\bf F}( {\bf r} )\cdot\nabla_v f = 0 ,
	\label{eq:Liouville}
\end{equation}
o\`u ${\bf F}({\bf r}) = - \nabla_r V({\bf r})$ est la force
qui agit sur l'atome.%
\footnote{
En \'ecrivant~(\ref{eq:Liouville}), nous n\'egligeons les 
interactions entre atomes.}
Un moyen simple d'int\'egrer l'\'equation~%
(\ref{eq:Liouville}) est d'utiliser la conservation du nombre
d'atomes le long des trajectoires classiques. Les trajectoires
dans l'espace des phases $( {\bf r}(t), {\bf v}(t) )$ sont
les solutions des \'equations du mouvement classiques, et
la conservation du nombre de particules s'\'ecrit\thinspace:
\begin{equation}
f({\bf r}(t), {\bf v}(t); t) = \mbox{const.}
	\label{eq:nb-const}
\end{equation}
Cette expression permet de calculer la fonction de distribution
\`a des instants ult\'erieurs si l'on se la donne \`a un
instant initial.

Utilisons maintenant l'\'equation~(\ref{eq:nb-const})
pour deux instants de temps voisins $t$ et $t - \Delta t:$ 
\begin{equation}
f({\bf r}_0, {\bf v}_0; t) = 
f({\bf r}(t - \Delta t), {\bf v}(t - \Delta t); t - \Delta t)
	\label{eq:ft-et-ft+dt}
\end{equation}
de sorte qu'entre $t - \Delta t$ et $t$,
les trajectoires classiques sont de la forme
(avec $0 \le \tau \le \Delta t$)
\begin{eqnarray}
{\bf r}_0 &\simeq& {\bf r}(t - \tau) 
+ \tau {\bf v}_0 \nonumber\\
{\bf v}_0 &\simeq& {\bf v}(t - \tau) 
+ \frac{ 1 }{ M }
\int_0^\tau\!{\rm d}\tau' {\bf F}[{\bf r}(t - \tau')] 
	\label{eq:traj-droites}
\end{eqnarray}
Nous supposons donc l'intervalle $\Delta t$ suffisamment court
et le potentiel al\'eatoire suffisamment faible
pour que les trajectoires ne soient pas fortement courb\'ees
sous l'influence du potentiel al\'eatoire.
En admettant en outre que la fonction de distribution $f({\bf r},
{\bf v}; t)$ varie lentement \`a l'\'echelle des d\'eplacements~%
(\ref{eq:traj-droites}), nous pouvons la d\'evelopper \`a
l'\'equation~(\ref{eq:ft-et-ft+dt}), pour trouver
\begin{eqnarray}
&& f({\bf r}_0, {\bf v}_0; t) - f({\bf r}_0, {\bf v}_0; t - \Delta t) =
\nonumber\\
&& - \, \Delta t \, {\bf v}_0 \cdot \nabla_r f(t - \Delta t)
- ( {\bf v}(t - \Delta t) - {\bf v}_0 )
\cdot \nabla_v f(t - \Delta t) \, + 
\nonumber\\
&& \qquad + \, \frac 12 \sum_{ij} ( {\bf v}(t - \Delta t) - {\bf v}_0 )_i
( {\bf v}(t - \Delta t) - {\bf v}_0 )_j 
\partial_{v_i} \partial_{v_j} f(t - \Delta t) . 
	\label{eq:Liouville-ordre-2}
\end{eqnarray}
Nous allons maintenant prendre la valeur moyenne de cette
\'equation par rapport au potentiel al\'eatoire (qui appara\^{\i}t
dans la vitesse ${\bf v}(t - \Delta t)$). Pour factoriser
la valeur moyenne de la troisi\`eme ligne de~(\ref{eq:Liouville-ordre-2}),
nous supposons que le long des trajectoires, la force al\'eatoire 
et la fonction de distribution forment
un processus {\em markovien}\thinspace: 
la fonction de distribution \`a l'instant $t - \Delta t$ n'est
alors pas corr\'el\'ee avec les valeurs de la force
pour $t - \tau > t - \Delta t$.
Cette hypoth\`ese semble raisonnable si pendant l'intervalle
de temps $\Delta t$, l'atome traverse un grand nombre de longueurs
de corr\'elation du potentiel\thinspace:
\begin{equation}
v \Delta t \gg \lambda_L
	\label{eq:condition-Markov}
\end{equation}
Il s'ensuit que cette approche ne peut ni d\'ecrire des atomes tr\`es
lents, ni rendre compte de leur mouvement \`a une \'echelle de
temps comparable au temps entre deux \flqq collisions\frqq.

Avec l'hypoth\`ese de {\sc Markov}, la moyenne sur les r\'ealisations
du milieu al\'eatoire fait appara\^{\i}tre l'int\'egrale de la
fonction de corr\'elation de la force dans le dernier terme de
l'\'equation~(\ref{eq:Liouville-ordre-2}). Celui-ci devient alors
\begin{equation}
\frac{ 1 }{ 2 M^2 } \int\limits_0^{\Delta t} 
\! {\rm d}\tau'_1 {\rm d}\tau'_2 \sum_{ij} \,
\left\langle F_i[ {\bf r}_0 - \tau'_1 {\bf v}_0 ]
F_j[ {\bf r}_0 - \tau'_2 {\bf v}_0 ] \right\rangle
\partial_{v_i} \partial_{v_j} \langle f \rangle
	\label{eq:force-correlation}
\end{equation}
Par cons\'equent, le coefficient de diffusion de la vitesse
est donn\'e par l'expression 
\begin{equation}
D_{ij}({\bf v}) = 
\lim_{\lambda_L / v \ll \Delta t \atop \Delta t \ll T}
\frac{ 1 }{ 2 M^2 \Delta t} \int\limits_0^{\Delta t} 
\! {\rm d}\tau'_1 {\rm d}\tau'_2 \,
G_{ij}[ (\tau'_1 - \tau'_2) {\bf v} ]
	\label{eq:coeff-diffusion}
\end{equation}
o\`u $T \gg \lambda_L / v$ est l'\'echelle de temps 
\`a laquelle l'on regarde la fonction de distribution moyenne
(beaucoup plus grande que le temps entre deux collisions), et
o\`u $G_{ij}({\bf r})$ est le tenseur de corr\'elations de la force
al\'eatoire
\begin{equation}
G_{ij}({\bf r}) \equiv 
\left\langle F_i( {\bf 0} ) F_j( {\bf r} ) \right\rangle.
	\label{eq:def-corr-force}
\end{equation}

Avant de calculer en d\'etail le coefficient de diffusion
de la vitesse $D_{ij}({\bf v})$, donnons 
l'\'equation de {\sc Fokker--Planck} que l'on trouve par
cette proc\'edure. Il y appara\^{\i}t une complication
parce que le coefficient de diffusion d\'epend de la vitesse. 
Nous suivons l'article de {\sc Hodapp} {\em et al.}
\cite{Dalibard95} (qui suit le livre de {\sc van Kampen}
\cite{vanKampen}) pour la forme de l'\'equation de 
{\sc F.-P.}\thinspace:
\begin{equation}
\partial_t \langle f({\bf r}, {\bf v}; t) \rangle 
+ {\bf v}\cdot\nabla_r \langle f \rangle =
\sum_{ij} \partial_{v_i} 
\left\{ - \frac{ F^{d}_i({\bf v}) }{ M } \langle f \rangle 
+ D_{ij}({\bf v}) \partial_{v_j} \langle f \rangle 
\right\}
+ \frac{ \partial_{v_i} F^{d}_i({\bf v}) }{ M } \langle f \rangle 
	\label{eq:Fokker-Planck}
\end{equation}
o\`u la force de d\'erive $F^{d}_i({\bf v})$ ainsi que le dernier terme
proviennent du r\'e-arrangement
des termes sous les d\'eriv\'ees $\partial_{v_i}$. La force
de d\'erive est donn\'ee par
\begin{equation}
F^{d}_i({\bf v}) = M \sum_{j} \partial_{v_j} D_{ij}({\bf v}).
	\label{eq:def-friction}
\end{equation}

\paragraph{Calcul du coefficient de diffusion.}
Nous donnons d'abord la fonction de corr\'elation de la force
al\'eatoire~(\ref{eq:def-corr-force})\thinspace:
\begin{equation}
G_{ij}({\bf r}) 
= \bar{V}^2 \left[
\frac{ 1 }{ r } \frac{ {\rm d}g }{ {\rm d}r }
\left( \frac{ r_i r_j }{ r^2 } - \delta_{ij} \right)
- \frac{ {\rm d}^2g }{ {\rm d}r^2 } \frac{ r_i r_j }{ r^2 } 
\right] ,
\qquad r \equiv |{\bf r}|,
	\label{eq:resultat-corr-force}
\end{equation}
r\'esultat que l'on d\'eduit de la d\'efinition \'el\'ementaire
de la force comme le gradient du potentiel, et en \'echangeant
l'ordre de la moyenne statistique et du processus limite
pour le gradient.  Ce calcul suppose que 
le potentiel al\'eatoire admet une d\'eriv\'ee,
hypoth\`ese qui semble justifi\'ee pour un champ de {\em speckles\/}. 

Pour calculer le coefficient de diffusion~(\ref{eq:coeff-diffusion}), 
passons aux variables d'int\'egration $(\tau'_1 + \tau'_2) / 2,
\, \tau'_2 - \tau'_1$. 
Pour l'int\'egration sur la diff\'erence,
nous utilisons l'\'equation~(\ref{eq:condition-Markov})\thinspace:
l'intervalle $\Delta t$ est beaucoup plus
long que le temps de corr\'elation du tenseur de corr\'elations
$G_{ij}[ (\tau'_1 - \tau'_2) {\bf v} ]$. Par cons\'equent, celui-ci
s'annule aux bornes d'int\'egration $\pm (\tau'_1 + \tau'_2)/2$ 
que nous pouvons donc remplacer par $\pm\infty$. 
L'int\'egrande ne d\'epend alors plus de la
somme, et l'int\'egrale est proportionnelle \`a $\Delta t$.
Nous trouvons ainsi le r\'esultat\thinspace:
\begin{eqnarray}
D_{ij}({\bf v}) &=& \frac{ 1 }{ 2 M^2 v } \int\limits_{-\infty}^{+\infty}
\! {\rm d}r \, G_{ij}( r {\bf v} / v )
= \frac{ K }{ v } \left( \delta_{ij} - \frac{ v_i v_j }{ v^2 } \right)
	\label{eq:resultat-Dij}\\
\mbox{avec}\quad
K &=& \frac{ \bar{V}^2 }{ M^2 } 
\int\limits_0^\infty \! \frac{ {\rm d}r }{ r } 
\left( - \frac{ {\rm d}g }{ {\rm d}r } \right) > 0.
	\label{eq:def-K}
\end{eqnarray}
Le deuxi\`eme terme dans~(\ref{eq:resultat-corr-force})
(qui fait intervenir la deuxi\`eme d\'eriv\'ee de la
fonction de corr\'elation) ne contribue pas au coefficient
de diffusion parce que son int\'egrale radiale s'annule.
Pour les deux fonctions de corr\'elations propos\'ees ci-dessus,
la constante $K$ prend les valeurs suivantes\thinspace:
\begin{eqnarray}
\mbox{lumi\`ere isotrope}: \quad K &=& \frac{ \pi k_L }{ 3 }
\frac{ \bar{V}^2 }{ M^2 } , \\
\mbox{{\em speckles\/} gaussiens}: \quad K &=& \sqrt{ \frac{ \pi }{ 2 } }
\frac{ 1 }{ \ell_c } \frac{ \bar{V}^2 }{ M^2 } .
\end{eqnarray}

\paragraph{Force de friction.}
Du tenseur de diffusion~(\ref{eq:resultat-Dij}) nous d\'eduisons
que la force de d\'erive
${\bf F}^{d}({\bf v})$ dans l'\'equation de {\sc F.--%
P.}~(\ref{eq:Fokker-Planck}) est en fait une \flqq force de friction\frqq\ 
avec
\begin{equation}
{\bf F}^{d}({\bf v}) = - 2 M K \frac{ {\bf v} }{ v^3 },
\qquad \nabla_v \cdot {\bf F}^{d} \equiv 8 \pi M K \delta({\bf v})
	\label{eq:expression-force-d}
\end{equation}
Puisque notre approche n'est pas valable pour des atomes \`a 
vitesse nulle [voir la condition~(\ref{eq:condition-Markov})], 
nous allons n\'egliger le terme proportionnel \`a 
$\nabla_v \cdot {\bf F}^{d} \propto \delta({\bf v}) \langle f \rangle$ 
dans l'\'equation de F.--P.~(\ref{eq:Fokker-Planck}).

\subsection{Discussion physique}

\paragraph{Amortissement de la vitesse moyenne.}
On d\'eduit facilement de l'\'equation de {\sc Fokker--%
Planck}~(\ref{eq:Fokker-Planck}) l'\'equation d'\'evolution
pour la vitesse moyenne
(apr\`es deux int\'egrations par parties et en n\'egligeant les
termes de bord)\thinspace:
\begin{eqnarray}
\frac{ {\rm d} \langle v_k \rangle }{ {\rm d}t } 
& = &
\frac{ {\rm d} }{ {\rm d}t } 
\int \!{\rm d}{\bf r} {\rm d}{\bf v} 
\, v_k \langle f({\bf r}, {\bf v}; t) \rangle
	\nonumber\\
&= &
\int \!{\rm d}{\bf r} {\rm d}{\bf v} 
\left[ \frac{ F^{d}_k({\bf v}) }{ M }
 + \sum_j \partial_{v_j} D_{jk}({\bf v}) \right]
\langle f({\bf r}, {\bf v}; t) \rangle
	\label{eq:amortissement-v-general}
\\
& = &
- 4 K \int \!{\rm d}{\bf r} {\rm d}{\bf v} 
\, \frac{ v_k }{ v^3 } \langle f({\bf r}, {\bf v}; t) \rangle
	\label{eq:amortissement-v}
\end{eqnarray}
La valeur moyenne de la vitesse diminue donc sous l'effet de
la force de friction et de la diffusion. Ceci se produit \`a une
\'echelle de temps caract\'eristique $T \sim v^3 / K$, qui d\'epend
du module de la vitesse initiale. Pour que notre th\'eorie soit
valable, il faut que ce temps soit beaucoup plus grand que
l'intervalle entre deux collisions. Cette condition se
traduit par l'in\'egalit\'e\thinspace:
\begin{equation}
\frac{ T }{ \lambda_L / v } \sim 
\left( \frac{ M v^2 }{ \bar{V} } \right)^2 \gg 1
	\label{eq:deux-temps}
\end{equation}
Il faut donc que le potentiel al\'eatoire soit en moyenne
beaucoup plus faible que l'\'energie cin\'etique des atomes.
Dans ce r\'egime, l'amortissement de la vitesse est 
lente \`a l'\'echelle des collisions avec les {\em speckles.} 

\paragraph{Elargissement diffusif des composantes de vitesse
transverses.}
Nous nous attendons \`a ce que l'amortissement de la 
vitesse moyenne s'accompagne d'une \flqq isotropisation\frqq\ 
de la distribution
angulaire de la vitesse parce qu'il semble peu probable que la
distribution puisse \^etre \flqq refroidie\frqq\ par le potentiel al\'eatoire.

Nous rappelons que le coefficient de diffusion dans l'\'equation
de {\sc F.--P.} donne la valeur quadratique moyenne
du changement de la vitesse\thinspace:
\begin{equation}
\langle \Delta v_i(t) \Delta v_j(t) \rangle
\simeq 2 D_{ij}[{\bf v}(0)] t.
\end{equation}
En utilisant le tenseur de diffusion~(\ref{eq:resultat-Dij}), 
nous constatons que les composantes parall\`eles \`a la 
vitesse initiale ${\bf v}(0)$ ne subissent aucune diffusion%
\footnote{%
Pr\'ecisons que $\Delta v_\Vert$ d\'enote la partie fluctuante
de la vitesse. C'est la vitesse moyenne qui est amortie par la 
force de friction ${\bf F}^{d}( {\bf v} )$.}
$\langle \Delta v^2_\Vert( t ) \rangle = 0$. Par contre,
les composantes perpendiculaires \`a 
${\bf v}(0)$ s'\'elargissent de fa\c{c}on diffusive
\begin{equation}
\left\langle {\bf v}^2_\perp (t) \right\rangle \simeq 
\frac{ 4 K }{ v(0) } t.
\end{equation}
La distribution des vitesses devient isotrope lorsque les composantes
transverses sont du m\^eme ordre de grandeur que la vitesse initiale.
Nous trouvons que ce processus a lieu sur la m\^eme \'echelle de temps
$T \sim K / v^3(0)$ que l'amortissement de la
valeur moyenne de la vitesse.

\paragraph{Variation de l'\'energie cin\'etique moyenne.}
Calculons maintenant une \'equation d'\'evolution pour la
valeur moyenne de l'\'energie cin\'etique des atomes. L'\'equation
de {\sc F.--P.}~(\ref{eq:Fokker-Planck}) donne, apr\`es
un calcul similaire \`a celui de l'\'eq.(\ref{eq:amortissement-v-general}),
\begin{eqnarray}
&& \frac{ {\rm d} }{ {\rm d}t }
\left\langle \frac{ M }{ 2 } {\bf v}^2 \right\rangle
= \nonumber\\
&& \quad = 
\int {\rm d}{\bf r} {\rm d}{\bf v} 
\left[ {\bf v}\cdot{\bf F}^{d}({\bf v}) + M 
\sum_{ij} \partial_{v_i} \left( D_{ij}({\bf v}) v_j \right) \right]
\langle f({\bf r}, {\bf v}; t) \rangle
	\label{eq:derivee-energie-cinetique}
\\
&& \quad = - 2 K M
\int {\rm d}{\bf r} {\rm d}{\bf v} 
\frac{ 1 }{ v }
\langle f({\bf r}, {\bf v}; t) \rangle.
	\label{eq:refroidissement?}
\end{eqnarray}
L'\'energie cin\'etique moyenne diminue donc sous l'influence
de la force de friction [le premier terme en crochets de~%
(\ref{eq:derivee-energie-cinetique})]. 
Ceci appara\^{\i}t \'egalement dans l'\'equation du mouvement
pour la vitesse sous l'influence de la friction\thinspace:
\begin{eqnarray}
&& \frac{ {\rm d} }{ {\rm d}t } {\bf v} = 
{\bf F}^{d}( {\bf v} ) = - 2 K \frac{ {\bf v} }{ v^3 }
\\
&& \Longrightarrow
{\bf v}(t) = \frac{ {\bf v}(0) }{ v(0) } 
\left[ v^3(0) - 6 K t \right]^{1/3},
\quad \left( 0 < t < v^3(0) / 6 K \right)
	\label{eq:solution-v-moyenne}
\end{eqnarray}
Toutes les vitesses sont donc amorties radialement et
deviennent nulles en un {\em temps fini} 
$T_{v(0)} = v^3(0) / 6 K$ de l'ordre de l'\'echelle de
temps caract\'eristique $T$ trouv\'ee ci-dessus.
 
Cependant, nous trouvons ici des r\'esultats en contradiction 
avec l'intuition que le potentiel al\'eatoire conserve
en moyenne l'\'energie cin\'etique.
Cette contradiction nous am\`ene \`a la remarque suivante\thinspace:

\paragraph{D\'emonstration \flqq malhonn\^ete\frqq.}
D'apr\`es {\sc Keller} (cit\'e dans \cite{Frisch66}),
une telle d\'emonstration de l'\'equation de {\sc F.--%
P.} pour la fonction de distribution \linebreak moyenne est
\flqq malhonn\^ete\frqq. Nous faisons en effet des hypoth\`eses sur
la nature markovienne de la force al\'eatoire qui ne sont
justifi\'ees que dans un certain r\'egime de param\`etres
ou \`a certaines \'echelles de vitesse et de position
(grande devant la longueur de corr\'elation). Nous verrons
au prochain paragraphe que l'on peut trouver l'\'equation
de {\sc F.--P.} par une proc\'edure diff\'erente et mieux
justifi\'ee d'apr\`es {\sc Frisch}.  Elle aura en outre la
vertu de donner une \'equation de {\sc F.--P.} \flqq physiquement
acceptable\frqq\ qui conserve l'\'energie cin\'etique moyenne.

\section{D\'emonstrations plus rigoureuses}

\subsection{M\'ethode p\'edestre}

La fa\c{c}on la plus simple d'\'eviter la contradiction du
refroidissement dans un potentiel al\'eatoire consiste \`a
renverser une partie de l'argumentation pr\'ec\'edente
pour d\'eduire l'\'equation de {\sc Fokker--Planck}.
(C'est ainsi que {\sc Frisch} la trouve habituellement.)
D\'efinissons d'abord les valeurs moyennes
\begin{eqnarray}
F^{d}_i({\bf v}) &:=& M 
\lim_{\Delta t \to 0}
\left.\frac{ \left\langle v_i(t + \Delta t) - v_i(t) \right\rangle }{
\Delta t} \right|_{ \mbox{${\bf v}(t) = {\bf v}$} } 
	\label{eq:force-Frisch}\\
D_{ij}({\bf v}) & := &
\lim_{\Delta t \to 0}
\left.\frac{ 
\left\langle \Delta v_i(t + \Delta t) \Delta v_j(t + \Delta t) \right\rangle }{
2 \Delta t} \right|_{ \mbox{${\bf v}(t) = {\bf v}$} } 
	\label{eq:diff-Frisch}
\end{eqnarray}
L'\'equation de {\sc F.--P.} est alors donn\'ee par
\begin{equation}
\partial_t \langle f({\bf r}, {\bf v}; t) \rangle 
+ {\bf v}\cdot\nabla_r \langle f \rangle =
\sum_{ij} \partial_{v_i} 
\left\{ - \frac{ F^{d}_i({\bf v}) }{ M } \langle f \rangle 
+ D_{ij}({\bf v}) \, \partial_{v_j} \langle f \rangle 
\right\} ,
	\label{eq:FP-correcte}
\end{equation}
o\`u la force de d\'erive ${\bf F}^{d}({\bf v})$ et le tenseur de
diffusion $D_{ij}({\bf v})$ sont donn\'es par (\ref{eq:force-Frisch}) 
et~(\ref{eq:diff-Frisch}).

Pour calculer la d\'erive et le tenseur de diffusion
dans un potentiel al\'eatoire faible, nous pouvons
int\'egrer le gradient du potentiel le long d'une trajectoire droite,
comme \`a l'\'equation~(\ref{eq:traj-droites}).
De cette fa\c{c}on, l'on trouve que 
\begin{itemize}
\item
la force de d\'erive
$F^{d}_i({\bf v})$ {\em s'annule}
(en moyenne, le potentiel al\'eatoire ne donne pas de transfert de
vitesse), et
\item
un r\'esultat identique \`a (\ref{eq:resultat-Dij}) pour
le tenseur de diffusion. 
\end{itemize}
Il reste cependant vrai que la valeur moyenne de la vitesse
est amortie\thinspace: l'\'equation~(\ref{eq:amortissement-v-general})
est encore correcte et la seule diff\'erence par rapport au resultat~%
(\ref{eq:amortissement-v}) est un facteur num\'erique qui change
(remplacer $4 K$ par $2 K$). 
Et finalement, pour l'\'energie cin\'etique moyenne, l'\'equation~%
(\ref{eq:derivee-energie-cinetique}) reste vraie (elle ne d\'epend
que de la forme de l'\'equation de {\sc F.--P.}) et donne
un r\'esultat nul parce que le tenseur de diffusion est 
\flqq orthogonal\frqq\ \`a la vitesse, 
$\sum_j D_{ij}({\bf v}) v_j = 0$. Nous obtenons donc le 
r\'esultat attendu que {\em l'\'energie
cin\'etique moyenne n'est pas modifi\'ee par le potentiel
al\'eatoire.\/} Ce qui se produit c'est bien une redistribution de
la vitesse d'une composante collimat\'ee vers une composante
diffuse.

\subsection{D\'eveloppement multi-\'echelles de {\sc Ryzhik},
{\sc Papanicolaou} et {\sc Keller}}

\paragraph{Id\'ee du d\'eveloppement.}
Nous allons pr\'esenter ici une m\'ethode alternative pour
\'etablir l'\'equation de {\sc Fokker--Planck}. 
Elle met en relief le fait
que le potentiel al\'eatoire $V({\bf r})$ d'une part, et la
fonction de distribution moyenne $\langle f( {\bf r}, {\bf v}; t )
\rangle$ d'autre part, varient sur des \'echelles spatiales
tr\`es diff\'erentes. Il s'agit alors d'\'eliminer
les quantit\'es avec des variations spatiales \flqq rapides\frqq\ 
et de formuler une \'equation ferm\'ee pour les quantit\'es moyennes
avec une variation spatiale \flqq lente\frqq.
{\sc Ryzhik}, {\sc Papanicolaou} et {\sc Keller} \cite{Keller96}
introduisent
\`a cet effet le rapport entre les \'echelles caract\'eristiques
\flqq rapides\frqq\ et \flqq lentes\frqq, ce qui revient pour notre situation
\`a introduire le param\`etre
\begin{equation}
\epsilon = \frac{ \lambda_L }{ L } 
	\label{eq:def-epsilon-Keller}
\end{equation}
Ils consid\`erent ensuite un potentiel al\'eatoire faible
proportionnel \`a $\sqrt{\epsilon}$. Le calcul consiste
\`a effectuer un d\'eveloppement asymptotique pour
$\epsilon \ll 1$. 

Cette m\'ethode semble d'abord se limiter \`a une situation
o\`u existe une relation entre la force du potentiel, d'une part, 
et la port\'ee des corr\'elations, d'autre part. Cependant, 
l'on peut aussi la voir sous un autre point de vue et choisir
ind\'ependamment la force du potentiel et la longueur de
corr\'elation $\lambda_L$\thinspace: {\em l'approche de {\sc Ryzhik}
fixe alors l'\'echelle $L$ \`a laquelle on regarde la
fonction de distribution moyenne.\/} La discussion pr\'ec\'edente
confirme que ce point de vue est adapt\'e \`a notre 
probl\`eme\thinspace:
plus pr\'ecis\'ement, nous pouvons estimer que
l'\'echelle spatiale caract\'eristique $L$ de 
la fonction de distribution moyenne est \'egale \`a 
la distance parcourue n\'ecessaire pour que la diffusion 
ait rendu isotrope la distribution angulaire de la vitesse%
\thinspace:
\begin{equation}
L \sim v T \sim \frac{ v^4 }{ K } 
\sim \lambda_L \left( \frac{ M v^2 }{ \bar{V} } \right)^2
	\label{eq:longueur-caracteristique}
\end{equation}
Cet argument permet d'\'etablir le lien suivant entre les
petits param\`etres de notre probl\`eme :
\begin{equation}
\epsilon = \left( \frac{ \bar{V} }{ Mv^2 } \right)^2 ,
	\label{eq:def-epsilon}
\end{equation}
et la condition $\epsilon \ll 1$ est bien celle d'un potentiel
al\'eatoire faible que nous avons rencontr\'ee et utilis\'ee
ci-dessus. Lorsque {\sc Ryzhik} {\em et al.\/} parlent d'un
\flqq potentiel al\'eatoire proportionnel \`a $\sqrt{\epsilon}$\frqq,
ce nombre donne donc l'ordre de grandeur de $V( {\bf r} )$ en 
unit\'es de l'\'energie cin\'etique des atomes (la seule
\'echelle d'\'energie dans le probl\`eme).

Notons que le d\'eveloppement multi-\'echelles est \`a manipuler
avec une certaine pr\'ecaution\thinspace: il faut identifier d'avance
les \'echelles caract\'eristiques spatiales ainsi que les
\'energies typiques du probl\`eme et regrouper ces param\`etres
dans une seule petite quantit\'e $\epsilon$.

\paragraph{D\'eveloppement.}
En fonction du petit param\`etre $\epsilon$, la force al\'eatoire
est proportionnelle \`a
\begin{equation}
| {\bf F} | \sim \frac{ \bar{V} }{ \lambda_L } \propto 
\frac{ \sqrt{\epsilon} }{ \epsilon } = \frac{ 1 }{ \sqrt{\epsilon} }
	\label{eq:force-et-epsilon}
\end{equation}
Par cons\'equent, {\sc Ryzhik} {\em et al.} \'ecriraient
l'\'equation de {\sc Liouville}~(\ref{eq:Liouville})
sous la forme
\begin{equation}
\partial_t f({\bf r}, {\bf v}; t) + {\bf v}\cdot\nabla_r f
+ \frac{ 1 }{ M \sqrt{ \epsilon} } \tilde{{\bf F}}\cdot\nabla_v f = 0 ,
	\label{eq:Liouville-Keller}
\end{equation}
o\`u $\tilde{{\bf F}}$ est une quantit\'e \flqq de l'ordre unit\'e\frqq.
La fonction de distribution est alors d\'evelopp\'ee en une
s\'erie de puissances 
\begin{eqnarray}
f( {\bf r}, {\bf v}; t ) & = &
f^{(0)}( {\bf r}, {\bf v}; t ) \, + 
	\nonumber\\
&& + \, \sqrt{\epsilon}
f^{(1)}( {\bf r}, {\mbf \rho}, {\bf v}; t ) + \epsilon
f^{(2)}( {\bf r}, {\mbf \rho}, {\bf v}; t ) + \ldots
	\label{eq:developpement-Keller}
\end{eqnarray}
o\`u ${\bf r}, {\mbf \rho}$ d\'enotent les variables spatiales
\flqq lentes\frqq\ et \flqq rapides\frqq. On fait l'hypoth\`ese que 
le premier terme $f^{(0)}$ est ind\'ependant de la variable
\flqq rapide\frqq\ ${\mbf \rho}$ et que c'est le seul
\`a donner un r\'esultat non nul apr\`es moyenne statistique 
sur le potentiel. Les termes suivants $f^{(1,2)}$ d\'ependent
\`a la fois de la variable rapide ${\mbf \rho}$ et de la variable
\flqq lente\frqq\ ${\bf r}$. 

La diff\'erence en \'echelle entre ces deux variables
se traduit par l'\'ecriture suivante de l'op\'erateur gradient spatial
dans~(\ref{eq:Liouville-Keller})
\begin{equation}
\nabla_r \mapsto \nabla_r + \frac{ 1 }{ \epsilon } \nabla_\rho
	\label{eq:gradient-deux-echelles}
\end{equation}
Nous pouvons maintenant ins\'erer le d\'eveloppement de la fonction
de distribution~(\ref{eq:developpement-Keller}) dans
l'\'equation de {\sc Liouville}~(\ref{eq:Liouville-Keller})
et s\'eparer les diff\'erents ordres en $\epsilon$. 
Les deux ordres les plus bas donnent alors\thinspace:
\begin{eqnarray}
\mbox{ordre ${\cal O}( 1/\sqrt{\epsilon} )$}: \quad
0 & = &
{\bf v}\cdot\nabla_\rho f^{(1)} + \tilde{{\bf F}}\cdot\nabla_v f^{(0)}
	\label{eq:exprimer-f1}\\
\mbox{ordre ${\cal O}( 1 )$}: \quad
0 & = &
\left( \partial_t + {\bf v}\cdot\nabla_r \right) f^{(0)} 
+ {\bf v}\cdot\nabla_\rho f^{(2)} +
\tilde{{\bf F}}\cdot\nabla_v f^{(1)}
	\label{eq:equation-f0}
\end{eqnarray}
La premi\`ere \'equation permet d'exprimer la \flqq partie fluctuante\frqq\ 
$f^{(1)}$ de
la distribution en fonction de la force al\'eatoire et de
la distribution \flqq lente\frqq\ $f^{(0)}$. La deuxi\'eme \'equation permet
de trouver une \'equation ferm\'ee pour la distribution moyenne
$\langle f \rangle = \langle f^{(0)} \rangle$ en ins\'erant
la solution de la premi\`ere \'equation {\em dans la
valeur moyenne de la deuxi\`eme \'equation\/}\thinspace:
le terme $f^{(2)}$ en dispara\^{\i}t en effet apr\`es la moyenne.

L'\'equation~(\ref{eq:exprimer-f1}) peut se r\'esoudre par
une transformation de {\sc Fourier}, et l'on trouve 
\begin{equation}
\tilde{ f }^{(1)}( {\bf r}, {\mbf \kappa}, {\bf v}; t ) =
- \frac{ 1 }{ {\rm i} {\bf v} \cdot {\mbf \kappa} + \eta }
\int\!{\rm d}{\mbf \rho} 
\exp( - {\rm i} {\mbf \kappa} \cdot {\mbf \rho} )
\tilde{ {\bf F} }({\mbf \rho}) \cdot \nabla_v 
f^{(0)}( {\bf r}, {\bf v}; t ) ,
	\label{eq:solution-f1}
\end{equation}
o\`u $\eta \to 0^+$ est un param\`etre de r\'egularisation.
Nous ins\'erons ce
r\'esultat en~(\ref{eq:equation-f0}) et trouvons apr\`es
la moyenne une \'equation de {\sc Fokker--Planck}
\begin{equation}
\left( \partial_t + {\bf v}\cdot\nabla_r \right) \langle f \rangle
= \sum_{ij} \partial_{v_i} D_{ij}( {\bf v} ) \partial_{v_j}
\langle f \rangle ,
	\label{eq:FP-Keller}
\end{equation}
o\`u le tenseur de diffusion est donn\'e par
\begin{equation}
D_{ij}({\bf v}) = \frac{ 1 }{ M^2 }
\int\!{\rm d}{\mbf \rho}' 
\langle F_i({\mbf \rho}) \, F_j({\mbf \rho}') \rangle
\int\!\frac{ {\rm d}{\mbf \kappa} }{ (2\pi)^3 }
\frac{ 
\exp{ {\rm i} {\mbf \kappa}\cdot( {\mbf \rho} - {\mbf \rho}' ) } }%
{ {\rm i} {\bf v}\cdot{\mbf \kappa} + \eta }
	\label{eq:def-Dij-Keller}
\end{equation}
Comme le potentiel al\'eatoire est statistiquement homog\`ene,
le tenseur de diffusion~(\ref{eq:def-Dij-Keller})
ne d\'epend pas de ${\mbf \rho}$. En outre, l'int\'egrale
sur ${\mbf \kappa}$ peut se calculer%
\footnote{%
\label{fn:delta-transverse}%
Le r\'esultat est
\[
\int\!\frac{ {\rm d}{\mbf \kappa} }{ (2\pi)^3 }
\frac{ \exp( - {\rm i} {\mbf \kappa} \cdot {\bf r} ) 
}{ {\rm i} {\bf v}\cdot{\mbf \kappa} + \eta }
= \left\{
\begin{array}{l}
\delta({\bf R}_\perp) / v \quad
\mbox{si ${\bf r}\cdot{\bf v} < 0$,}\\
0 \quad \mbox{sinon.}
\end{array}
\right.
\]
o\`u ${\bf R}_\perp$ d\'enote les composantes de ${\bf r}$ 
perpendiculaires au vecteur ${\bf v}$.%
} et l'on trouve [en utilisant~(\ref{eq:def-corr-force})]
\begin{equation}
D_{ij}({\bf v}) = \frac{ 1 }{ M^2 v } \int\limits_{-\infty}^{0}
\! {\rm d}r \, G_{ij}( r {\bf v}/v )
	\label{eq:resultat-Dij-Keller}
\end{equation}
L'on trouve alors un r\'esultat identique \`a~(\ref{eq:resultat-Dij})
si la fonction de corr\'elation $G_{ij}({\bf r})$ ne d\'epend pas
du signe de ${\bf r}$ (c'est l'invariance par parit\'e). En plus,
le tenseur de diffusion se trouve \flqq \`a la bonne position\frqq\ dans
l'\'equation de {\sc F.--P.}~(\ref{eq:FP-Keller}) (entre les
d\'eriv\'ees $\partial_{v_i}$) de sorte que celle-ci est
compatible avec la conservation en moyenne de l'\'energie.

\paragraph{Remarque.}
Nous constatons que le d\'eveloppement multi-\'echelles
conduit de fa\c{c}on g\'en\'erique \`a une \'equation de
{\sc F.--P.} \`a partir de l'\'equation de {\sc Liouville}.
Ce type de description du transport d'atomes dans un potentiel al\'eatoire
appara\^{\i}t donc comme l'outil standard dans le point de vue classique.

\subsection{Raisonnement formel de {\sc Frisch}}
\paragraph{Id\'ee.}
Cette approche proc\`ede par analogie avec les \'equations
de propagation des ondes en milieu al\'eatoire, en les formulant
d'une fa\c{c}on assez g\'en\'erale. L'on identifie des 
\flqq op\'erateurs de {\sc Liouville}\frqq\ ${\cal L}_0$ et ${\cal L}_1$
dont le deuxi\`eme contient la partie al\'eatoire de 
l'\'equation de propagation\thinspace:
\begin{equation}
\partial_t A(t) = {\cal L}_0 A(t) + {\cal L}_1(t) A(t) ,
	\label{eq:equation-generale}
\end{equation}
o\`u $A(t)$ d\'esigne un \flqq champ\frqq\ quelconque.
Dans l'\'equation de {\sc Liouville} proprement dite~(\ref{eq:Liouville}),
nous avons les correspondances suivantes
\begin{eqnarray}
A(t) &\longleftrightarrow& f({\bf r}, {\bf v}; t)
	\nonumber\\
{\cal L}_0 &\longleftrightarrow& - {\bf v}\cdot\nabla_r 
	\nonumber\\
{\cal L}_1(t) &\longleftrightarrow& - \frac{ 1 }{ M } {\bf F}\cdot\nabla_v 
	\label{eq:analogie-classique}
\end{eqnarray}
Pour l'\'equation de propagation des ondes de {\sc Schr\"odinger}, 
l'on aurait par exemple
\begin{eqnarray}
A(t) &\longleftrightarrow& \psi({\bf r}; t)
	\nonumber\\
{\cal L}_0 &\longleftrightarrow& 
\frac{ {\rm i}\hbar }{ 2 M } \nabla^2_r 
	\nonumber\\
{\cal L}_1(t) &\longleftrightarrow& 
- \frac{ {\rm i} }{ \hbar } V({\bf r})
	\label{eq:analogie-quantique}
\end{eqnarray}

{\sc Frisch} montre alors que l'on peut \'ecrire des \'equations
s\'epar\'ees pour la valeur moyenne $\langle A(t) \rangle$
et la partie fluctuante $\delta A(t) = A(t) - \langle A(t) \rangle$
du champ\thinspace:
\begin{equation}
\begin{array}{rcl}
\partial_t \langle A(t) \rangle & = & {\cal L}_0 \langle A(t) \rangle 
+ \langle {\cal L}_1(t) \delta A(t) \rangle 
	\\
\partial_t \delta A(t) & = & {\cal L}_0 \delta A(t) 
+ \delta \left\{
{\cal L}_1(t) \left[ \left\langle A(t) \right\rangle 
+ \delta A(t) \right] \right\},
\end{array}
	\label{eq:2eq-Frisch}
\end{equation}
La deuxi\`eme \'equation peut \^etre r\'esolue par un 
d\'eveloppement it\'eratif, au moins de fa\c{c}on formelle.
Sa solution exprime donc la partie fluctuante du champ
en fonction du champ moyen et du Liouvillien al\'eatoire.
En la reportant dans la premi\`ere \'equation, l'on trouve
une \'equation pour le champ moyen. Dans le cas de la 
propagation des ondes, l'\'equation r\'esultante est 
l'\'equation de {\sc Dyson} pour le champ moyen $\langle \psi \rangle$.

\paragraph{Approximation pour le champ fluctuant.}
Il est cependant difficile de r\'esoudre explicitement
pour la partie fluctuante du champ. Ce que {\sc Frisch} appelle
l'{\em approximation de r\'egularisation au premier ordre\/}
revient alors \`a ne retenir que le premier terme de l'it\'eration.%
\footnote{%
Notons que dans le contexte de la propagation des ondes,
cette approximation est celle de {\sc Bourret} ou encore la
{\em ladder approximation\/} pour l'op\'erateur
de masse. Dans d'autres contextes, on l'appelerait 
\flqq th\'eorie effective \`a basse \'energie ou \`a grande
\'echelle\frqq\ (physique des particules) ou encore 
\flqq \'elimination adiabatique des variables rapides\frqq\ 
(atome pilot\'e par un champ \`a grand d\'esaccord :
r\'eduction des \'equations de {\sc Bloch} optiques
\`a la population de l'\'etat fondamental...)
}
L'on a alors la solution
\begin{equation}
\delta A(t) \simeq \int\limits_{t_0}^t \! {\rm d}t' \,
\exp{\left[ {\cal L}_0 (t - t') \right]}
\left\{ {\cal L}_1(t') \langle A(t') \rangle \right\} .
	\label{eq:solution-Bourret}
\end{equation}
Dans l'approche de {\sc Ryzhik} {\em et al.\/},
une telle approximation arrive naturellement parce que la
partie fluctuante est calcul\'ee par un d\'eveloppement
perturbatif\thinspace: \`a l'ordre le plus bas, c'est 
le Liouvillien fluctuant et le champ moyen
(d'ordre z\'ero) qui d\'eterminent les fluctuations du champ 
(voir l'\'equation~(\ref{eq:exprimer-f1})).

Avec cette approximation,
l'\'equation d'\'evolution pour le champ moyen devient 
\begin{equation}
\partial_t \langle A(t) \rangle = {\cal L}_0 \langle A(t) \rangle 
+ \int\limits_{t_0}^t \! {\rm d}t' \,
\left\langle {\cal L}_1(t) 
\exp{\left[ {\cal L}_0 (t - t') \right]}
{\cal L}_1(t') \right\rangle 
\langle A(t') \rangle .
	\label{eq:Dyson-Frisch}
\end{equation}
o\`u nous voyons appara\^{\i}tre sous l'int\'egrale la fonction
de corr\'elation du Liouvillien fluctuant, prise pour les instants
$t$ et $t'$. Entre ces instants, l'exponentielle 
$\exp{\left[ {\cal L}_0 (t - t') \right]}$ d\'ecrit le mouvement
\flqq libre\frqq\ du syst\`eme sous la seule influence du Liouvillien
r\'egulier.

\paragraph{Application \`a notre probl\`eme.}
En utilisant les correspondances~(\ref{eq:analogie-classique}),
le deuxi\`eme terme \`a l'\'equation~(\ref{eq:Dyson-Frisch}) 
pour le champ moyen s'\'ecrit
\begin{equation}
\sum_{ij} \partial_{v_i}
\int\limits_{t_0}^t \! {\rm d}t' \,
\left\langle F_i({\bf r})
F_j( {\bf r} + (t' - t){\bf v} ) \right\rangle 
\partial_{v_j}
\langle f({\bf r} + (t' - t){\bf v}, {\bf v}; t') \rangle .
	\label{eq:eqdiff-Frisch}
\end{equation}
o\`u ${\bf r} + (t' - t){\bf v}$ d\'ecrit le \flqq vol libre\frqq\ aux 
instants $t'$ ant\'erieurs \`a $t$.
Nous con\-sta\-tons que l'\'equation pour la fonction de distribution
moyenne {\em n'est pas \'equivalente\/} \`a
l'\'equation de {\sc F.--P.}~(\ref{eq:FP-Keller}) parce qu'elle
fait intervenir une int\'egrale sur $\langle f \rangle$ \`a des
instants ant\'erieurs.
C'est seulement lorsque nous invoquons une
{\em hypoth\`ese suppl\'ementaire\/} que nous retrouvons la 
th\'eorie pr\'ec\'edente\thinspace: cette l'hypoth\`ese \'etant
que la port\'ee de la fonction de corr\'elation de la force al\'eatoire
(en fonction de $(t' - t){\bf v}$)
est beaucoup plus courte que l'\'echelle spatiale caract\'eristique 
pour la fonction de distribution moyenne. Dans cette
situation (que nous pourrions appeler \flqq markovienne\frqq) l'on
peut sortir la fonction de distribution moyenne de l'int\'egrale
sur $t'$. 
En rempla\c{c}ant ensuite la borne inf\'erieure $t_0$ de l'int\'egrale
par $-\infty$, l'on trouve l'\'equation de 
{\sc Fokker--Planck}~(\ref{eq:FP-Keller}).

Notons que l'approche de {\sc Frisch} donne elle aussi un
tenseur de diffusion \`a la bonne position (entre les d\'eriv\'ees
$\partial_{v_i}$).

\section{Conclusions}

Dans le point de vue classique, le transport d'atomes dans un potentiel
al\'eatoire faible est d\'ecrit
de fa\c{c}on naturelle par une \'equation de {\sc Fokker--Planck}, et
non pas par une \'equation de transfert radiatif.
L'\'equation de {\sc F.--P.} fait seulement intervenir la fonction de
distribution au m\^eme instant (elle est locale dans le temps)
lorsque les fluctuations du potentiel
al\'eatoire ont une \'echelle spatiale beaucoup plus courte que
la fonction de distribution moyenne. Le transport est alors d\'ecrit
par une \'equation lin\'eaire aux d\'eriv\'ees partielles.
Son tenseur de diffusion est reli\'e au tenseur de corr\'elation
de la force al\'eatoire, il d\'epend de la vitesse des atomes. 
L'\'equation de {\sc F.--P} permet de montrer que
dans le potentiel al\'eatoire, la valeur moyenne de la vitesse 
d\'ecro\^{\i}t parce que la distribution angulaire des vitesses
devient isotrope. Le temps caract\'eristique pour
ce processus diffusif est plus long que le temps n\'ecessaire
pour traverser une longueur de corr\'elation du potentiel
al\'eatoire. A cette \'echelle de temps, la distribution
angulaire d'un faisceau d'atomes initialement bien
collimat\'e, s'\'elargit pour devenir isotrope, tout
en conservant son \'energie cin\'etique moyenne.

Nous avons constat\'e que le tenseur de diffusion d\'epend
de la vitesse. Il faut alors faire attention \`a sa position
dans l'\'equation de {\sc F.--P.}  Il y a plusieurs m\'ethodes
pour trouver la \flqq bonne \'equation\frqq: un retour \`a la
signification statistique de l'\'equation de {\sc Fokker--Planck}
est d\'ej\`a fructueux (notre \flqq m\'ethode p\'edestre\frqq) ;
des approches plus formelles sont soit
le d\'eveloppement \`a plusieurs \'echelles (m\'ethode
de {\sc Ryzhik, Papanicolaou} et {\sc Keller}), soit 
l'analogie \`a l'approximation de {\sc Bourret} de la th\'eorie
de propagation des ondes en milieu al\'eatoire 
(m\'ethode de {\sc Frisch}).

Finalement, une remarque sur les approches
du type {\sc F.--P.} et transfert radiatif : nous pouvons
attribuer leur diff\'erence aux descriptions
classique ou ondulatoire du mouvement.
En effet, la th\'eorie classique part de l'\'equation de {\sc Liouville}
o\`u interviennent des d\'eriv\'ees de la fonction de
distribution dans l'espace des phases. L'\'equation de {\sc F.--P.}
h\'erite alors de cette m\^eme structure. Dans l'approche
ondulatoire, la diffusion par le potentiel al\'eatoire est
d\'ecrite par des int\'egrales de diffusion 
(voir l'\'equation du transfert radiatif~%
(\ref{eq:e-t-r})) parce que pour la diffusion d'une onde,
tout potentiel, aussi faible soit-il, peut donner lieu \`a
des grands angles de diffusion avec une probabilit\'e non
nulle (bien que faible). Dans une approche classique, par
contre, la vitesse change toujours de fa\c{c}on continue,
et ce changement tend vers z\'ero lorsque le potentiel est faible.
Les deux descriptions se rejoignent si la longueur d'onde est
beaucoup plus petite que la taille des inhomog\'en\'eit\'es du
potentiel : la diffusion est alors piqu\'ee vers l'avant,
et l'on peut passer du transfert radiatif \`a l'\'equation
de {\sc F.--P.} (voir au chapitre~\ref{s:radiatif}).

\section{Exemples de transport classique}

Pour finir ce chapitre sur le transport classique d'atomes,
nous donnons quel\-ques exemples o\`u l'on peut trouver des
solutions explicites \`a l'\'equation de {\sc Fokker--Planck}.
Il s'agit d'une part de l'\'elargissement angulaire de la distribution
des vitesses et d'autre part de la diffusion spatiale.

Notre point de d\'epart est l'\'equation de {\sc F.--P}
suivante
\begin{equation}
\left( \partial_t + {\bf v}\cdot\nabla_r \right) f( {\bf r}, {\bf v} ; t )
= \frac{ K }{ v^3 } \triangle_v^{(T)} f
	\label{eq:rappel-FP}
\end{equation}
o\`u la fonction de distribution moyenn\'ee dans l'espace des
phases est not\'ee $f$, pour simplifier, et o\`u $\triangle_v^{(T)}$
est l'op\'erateur diff\'erentiel Laplacien transverse. En fonction
des angles $\theta, \phi$ du vecteur ${\bf v}$, il est donn\'e
par l'expression en crochets de l'\'equation~(\ref{eq:FP-sphere}).
La constante $K$ dans~(\ref{eq:rappel-FP}) est proportionnelle
\`a $\bar{V}^2$, elle est donn\'ee par~(\ref{eq:def-K}).

\subsection{Diffusion des vitesses}

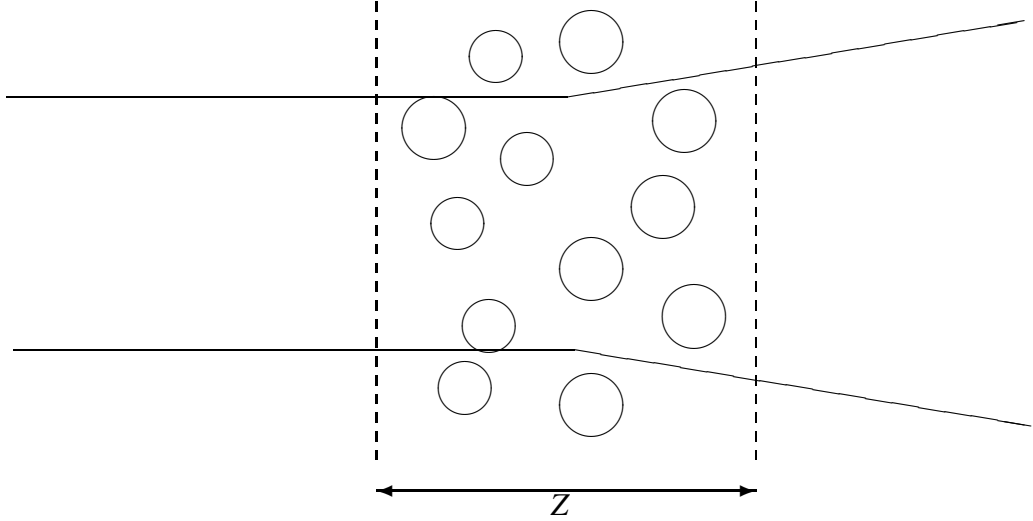
\begin{figure}
\begin{center}
\setlength{\unitlength}{0.012500in}%
\begingroup\makeatletter\ifx\SetFigFont\undefined
\def\x#1#2#3#4#5#6#7\relax{\def\x{#1#2#3#4#5#6}}%
\expandafter\x\fmtname xxxxxx\relax \def\y{splain}%
\ifx\x\y   
\gdef\SetFigFont#1#2#3{%
  \ifnum #1<17\tiny\else \ifnum #1<20\small\else
  \ifnum #1<24\normalsize\else \ifnum #1<29\large\else
  \ifnum #1<34\Large\else \ifnum #1<41\LARGE\else
     \huge\fi\fi\fi\fi\fi\fi
  \csname #3\endcsname}%
\else
\gdef\SetFigFont#1#2#3{\begingroup
  \count@#1\relax \ifnum 25<\count@\count@25\fi
  \def\x{\endgroup\@setsize\SetFigFont{#2pt}}%
  \expandafter\x
    \csname \romannumeral\the\count@ pt\expandafter\endcsname
    \csname @\romannumeral\the\count@ pt\endcsname
  \csname #3\endcsname}%
\fi
\fi\endgroup
\begin{picture}(430,215)(285,505)
\thinlines
\put(477,558){\circle{20}}
\put(464,667){\circle{26}}
\put(530,703){\circle{26}}
\put(530,551){\circle{26}}
\put(490,697){\circle{20}}
\put(569,670){\circle{26}}
\put(530,608){\circle{26}}
\put(474,627){\circle{20}}
\put(487,584){\circle{20}}
\put(573,588){\circle{26}}
\put(560,634){\circle{26}}
\put(503,654){\circle{20}}
\special{ps: gsave 0 0 0 setrgbcolor}\put(288,574){\line( 1, 0){235}}
\put(523,574){\line( 6,-1){191.027}}
\special{ps: grestore}\special{ps: gsave 0 0 0 setrgbcolor}\put(285,680){\line( 1, 0){235}}
\put(520,680){\line( 6, 1){191.189}}
\special{ps: grestore}\thicklines
\put(440,515){\vector(-1, 0){  0}}
\put(440,515){\vector( 1, 0){159}}
\thinlines
\multiput(599,720)(0.00000,-7.83673){25}{\line( 0,-1){  3.918}}
\multiput(440,528)(0.00000,7.83673){25}{\line( 0, 1){  3.918}}
\put(513,505){\makebox(0,0)[lb]{\smash{\SetFigFont{12}{14.4}{it}Z}}}
\end{picture}
\end{center}
\caption[fig:slab]{Transport d'un faisceau d'atomes \`a travers
une tranche de tavelures. Illustration de la g\'eom\'etrie.}
\label{fig:slab}
\end{figure}

Etudions d'abord l'\'evolution d'un jet d'atomes collimat\'e
qui traverse un champ de tavelures. Dans un premier temps, supposons
que les tavalures occupent une tranche d'\'epaisseur $Z$ et que
le jet a un profil spatial uniforme (voir la figure~\ref{fig:slab}). 
En nous pla\c cant dans le r\'egime stationnaire, la fonction de 
distribution ne d\'epend que de la distance $z$ et de la vitesse
${\bf v}$. Nous notons \'egalement que dans l'\'equation de
{\sc Fokker--Planck}~(\ref{eq:rappel-FP}), le module de la vitesse
$v$ intervient seulement comme param\`etre. Nous pouvons donc
raisonner pour une classe de vitesses donn\'ee. A l'entr\'ee du
champ de {\em speckles\/}, la fonction de distribution est
de la forme
\begin{equation}
f( z = 0; {\bf v} ) = f_i(v) \delta(\theta)
= \frac{ f_i(v) }{ 4\pi }
\sum_{n = 0}^{\infty} (2n+1) P_n(\cos\theta)
	\label{eq:f-init}
\end{equation}
o\`u $P_n(\cos\theta)$ sont les polyn\^omes de {\sc Legendre}
qui permettent de repr\'esenter la fonction $\delta$ angulaire
(voir aussi~(\ref{eq:solution-Frisch}) ; $\theta$ d\'enote l'angle
par rapport \`a la direction initiale du faisceau). 
En introduisant les
coefficients $f_n(z)$ pour le d\'eveloppement en harmoniques
sph\'eriques de la fonction
de distribution, nous trouvons l'\'equation suivante :
\begin{equation}
\sum_n \cos\theta \frac{ {\rm d}f_n }{ {\rm d}z } ( 2 n + 1 )
P_n(\cos\theta) =
-\frac{ K }{ v^4 }
\sum_n n ( n + 1 ) f_n( z ) ( 2 n + 1 )
P_n(\cos\theta) 
	\label{eq:FP-Legendre}
\end{equation}
Lorsque l'on projette cette \'equation sur un polyn\^ome de
{\sc Legendre} $P_m(\cos\theta)$, le $\cos\theta$ au membre gauche 
introduit un couplage entre les polyn\^omes voisins 
$P_{m\pm1}(\cos\theta)$. L'on trouve alors un syst\`eme lin\'eaire 
infini d'\'equations diff\'erentielles ordinaires que l'on peut 
r\'esoudre, par exemple, avec des techniques de diagonalisation
de matrices. Nous allons nous contenter d'une discussion qualitative
du r\'egime o\`u le champ de {\em speckles\/} a une \'epaisseur faible.
Plus pr\'ecis\'ement, nous supposons que la distribution angulaire
reste assez bien collimat\'ee de sorte que nous pouvons
remplacer le $\cos\theta$ dans~(\ref{eq:FP-Legendre}) par l'unit\'e.
Dans cette approximation, les diff\'erentes harmoniques 
sph\'eriques se d\'ecouplent,
et l'on retrouve la solution~(\ref{eq:solution-Frisch})
donn\'ee par {\sc Frisch}
\begin{equation}
f_n( Z ) = {\rm e}^{ - K n (n+1) Z / v^4 } f_i(v)
	\label{eq:solution-Frisch-2}
\end{equation}
Comme nous l'avons vu pour le transfert radiatif piqu\'e vers l'avant
(chap.~\ref{s:radiatif}), les harmoniques \'elev\'es (les structures
angulaires $\Delta\theta$ fines) disparaissent sur une \'epaisseur 
caract\'eristique donn\'ee par
\begin{equation}
\frac{ 1 }{ \Delta\theta^2 } \frac{ K Z }{ v^4 } \sim 1
\Longrightarrow
\Delta\theta \sim \left( \frac{ K Z }{ v^4 } \right)^{1/2}
	\label{eq:delta-theta}
\end{equation}
La solution~(\ref{eq:solution-Frisch-2}) est valable tant que cette
largeur angulaire reste faible. L'\'equation~(\ref{eq:delta-theta})
permet de traduire ce crit\`ere par une \'epaisseur maximale
du milieu al\'eatoire : celle-ci est \'evidemment 
beaucoup plus grande que la longueur de corr\'elation (la
longueur d'onde optique).

Nous passons maintenant \`a un deuxi\`eme exemple o\`u le potentiel
lumineux est \flqq branch\'e\frqq\ pendant un temps $T$ fini.
Prenons encore un jet d'atomes collimat\'e \`a l'instant $t=0$
et spatialement uniforme. Nous supposons \'egalement que le potentiel
lumineux (lorsqu'il est branch\'e) remplit tout l'espace.
La fonction de distribution ne d\'epend alors que de $t$ et ${\bf v}$.
Nous introduisons les coefficients $f_n( t )$ d'un d\'eveloppement
en polyn\^omes de {\sc Legendre}, et~(\ref{eq:FP-Legendre}) devient
\begin{equation}
\sum_n \frac{ {\rm d}f_n }{ {\rm d}t } ( 2 n + 1 )
P_n(\cos\theta) =
- \frac{ K }{ v^3 }
\sum_n n ( n + 1 ) f_n( z ) ( 2 n + 1 )
P_n(\cos\theta) 
	\label{eq:FP-Legendre-2}
\end{equation}
Dans cette \'equation, les harmoniques sph\'eriques
se d\'ecouplent exactement, et nous trouvons une
solution formellement identique \`a~(\ref{eq:solution-Frisch-2}) :
\begin{equation}
f_n( T ) = {\rm e}^{ - K n (n+1) T / v^3 } f_i(v)
	\label{eq:solution-Frisch-2b}
\end{equation}
De nouveau, nous identifions un temps d'interaction caract\'eristique
$T_{\rm iso} \sim v^3 / K$
au bout duquel la distribution angulaire devient isotrope. Si l'on
sait r\'esoudre des d\'etails angulaires $\Delta\theta$ plus fins, 
l'effet du potentiel al\'eatoire se fait voir d\'ej\`a \`a des
temps plus courts, de l'ordre de $\Delta\theta^2 \times T_{\rm iso}$.
 
Notons finalement que le temps caract\'eristique $T_{\rm iso}$
d\'epend \`a la fois de la vitesse des atomes et du
potentiel al\'eatoire (qui appara\^{\i}t dans la constante $K$). 
L'on peut donc 
faire varier ces deux param\`etres pour explorer une large
gamme de temps d'interaction \flqq effectifs\frqq, m\^eme si pour 
des raisons exp\'erimentales (effet de la gravit\'e, taille de
la chambre \`a vide), le temps r\'eel est limit\'e.

\subsection{Diffusion spatiale}

En derni\`ere application du transport d'atomes, nous allons
\'etudier le r\'egime des temps tr\`es longs, o\`u la 
distribution des vitesses est d\'ej\`a devenue isotrope.
Les atomes continuent cependant d'\'evoluer \`a cause d'un
gradient spatial dans la densit\'e atomique.  Dans ce r\'egime,
la distribution spatiale $F({\bf r}; t)$ des atomes v\'erifie
une \'equation de la diffusion que nous allons d\'eterminer
maintenant.

A cet effet, nous \'ecrivons la distribution dans l'espace
des phases comme
la somme d'une partie isotrope $F({\bf r}; t)$ et d'un terme
\flqq dipolaire\frqq\ ${\bf v}\cdot{\bf j}({\bf r}; t)$
\begin{equation}
f({\bf r}, {\bf v}; t) = F({\bf r}; t) + {\bf v}\cdot{\bf j}({\bf r}; t)
	\label{eq:approx-P1}
\end{equation}
(Cette approximation est appel\'ee \flqq $P_1$\frqq parce 
qu'elle se contente de $P_{0,1}(\cos\theta)$ dans le d\'eveloppement 
en harmoniques sph\'eriques.)
En projetant l'\'equation de {\sc Fokker--Planck} sur les
harmoniques sph\'eriques d'ordre $n=0,1$, l'on trouve 
\begin{eqnarray}
\partial_t F( {\bf r} ; t)
+ \int\frac{ {\rm d}\Omega_v }{ 4\pi }
\nabla_r\cdot{\bf v} [ {\bf v}\cdot{\bf j}( {\bf r}; t ) ]
& = & 0
	\label{eq:FP-P0}\\
\partial_t \,{\bf j}( {\bf r}; t )
+ \nabla_r F( {\bf r}; t ) 
& = &
\frac{ 2 K }{ v^3 } {\bf j}( {\bf r}; t )
	\label{eq:FP-P1}
\end{eqnarray}
En nous pla\c cant \`a des temps longs par rapport au temps
caract\'eristique d'\'evolution $T_{\rm iso}$ de la distribution
angulaire, nous pouvons nous contenter de la solution stationnaire
de l'\'equation~(\ref{eq:FP-P1}) : elle relie le \flqq flux
diffusif\frqq\ ${\bf j}( {\bf r} ; t )$ au gradient de la densit\'e spatiale
(loi de {\sc Fick})
\begin{equation}
{\bf j}( {\bf r}; t ) = - \frac{ v^3 }{ 2 K } \nabla_r F( {\bf r}; t )
	\label{eq:Fick}
\end{equation}
En reportant ce r\'esultat \`a l'\'equation~(\ref{eq:FP-P1}) et 
en calculant l'int\'egrale angulaire, nous trouvons alors une \'equation
de diffusion spatiale
\begin{equation}
\partial_t F( {\bf r}; t ) - \frac{ v^5 }{ 6 K } 
\nabla_r^2 F( {\bf r}; t ) = 0
	\label{eq:diffusion-spatiale}
\end{equation}
Le coefficient de diffusion (spatiale) est donc donn\'e par
\begin{equation}
D_{\rm sp} = \frac{ v^5 }{ 6 K } .
	\label{eq:estimation-Dsp}
\end{equation}

\paragraph{Discussion.}
Une situation physique int\'eressante correspond \`a l'expansion
d'un nuage initialement confin\'e et plong\'e dans un potentiel 
al\'eatoire.
Apr\`es un temps $\Delta t$, les atomes ont diffus\'e \`a travers
une distance quadratique moyenne 
\begin{equation}
\Delta{\bf r}^2 \simeq 6 D_{\rm sp} \Delta t
= \frac{ v^5 \Delta t }{ K }
\sim v^2 T_{\rm iso} \Delta t 
\ll ( v \Delta t )^2
	\label{eq:c-est-pas-rapide}
\end{equation}
A premi\`ere vue, il semble que les atomes diffusent tr\`es
rapidement (d\'ependance en $v^5$). Mais la deuxi\`eme formulation
dans~(\ref{eq:c-est-pas-rapide}) nous rappelle que,
lorsque le temps
$\Delta t$ est plus long que le temps caract\'eristique
$T_{\rm iso}$ de l'\'evolution angulaire,
la diffusion est au contraire beaucoup plus lente que le vol libre.

Il reste \`a remarquer que pour un temps d'exp\'erience donn\'e,
ce sont les atomes les plus lents qui \flqq ont le temps\frqq\
d'entrer dans le r\'egime diffusif (leur $T_{\rm iso} \propto v^3$ est
compris dans la dur\'ee de l'exp\'erience). Les atomes plus rapides
suivent par contre un transport balistique avec $\Delta {\bf r}^2
\simeq ({\bf v} \Delta t)^2$, 
ils n'ont en outre pas encore \flqq oubli\'e\frqq\
leur distribution angulaire initiale. C'est ainsi qu'ils ont parcouru
la distance la plus grande (et que leur distribution en position
permet de remonter \`a la distribution des vitesses initiale).
Nous constatons donc qu'il
est possible d'observer simultan\'ement plusieurs comportement
du transport, pour chacune des classes de vitesses (radiales)
des atomes. Le comportement diffusif appara\^{\i}t seulement pour
les classes les plus lentes, il va se manifester,
\`a la fin de l'exp\'erience, au centre de la distribution spatiale.




\chapter{Transport d'atomes --- point de vue ondulatoire}
\label{s:quantique}

Un bref aper\c cu d'approches ondulatoires pour mod\'eliser
le transport d'atomes dans un milieu al\'eatoire. Un grand
nombre de r\'esultats sont bien connus dans le contexte optique
(\flqq diffusion multiple de la lumi\`ere\frqq). Nous
montrerons en particulier comment l'on obtient une
\'equation du transfert radiatif \`a partir de la
description ondulatoire. C'est un exemple particulier
du passage du \flqq microscopique\frqq\ vers le
\flqq macroscopique\frqq.

\section{Introduction}

Nous nous pla\c cons maintenant dans un contexte ondulatoire
o\`u les atomes sont d\'ecrits, au niveau le plus fondamental
de la th\'eorie, par une fonction d'onde $\psi( {\bf r}, t)$.
Son \'equation d'\'evolution est l'\'equation de {\sc 
Schr\"odinger} :
\begin{equation}
{\rm i} \hbar \partial_t \psi( {\bf r}, t ) +
\frac{ \hbar^2 }{ 2 M } \nabla^2 \psi = V( {\bf r} ) \, \psi
	\label{eq:Schroedinger-ici}
\end{equation}
o\`u $V( {\bf r} )$ est le potentiel al\'eatoire.
Nous nous int\'eressons \`a des {\em propri\'et\'es moyennes\/}
des atomes lorsque l'on ignore presque tout du potentiel
al\'eatoire, sauf sa valeur moyenne (que nous supposons
\'egale \`a z\'ero, avec une red\'efinition convenable de
l'\'energie%
\footnote{%
Pour des atomes dans un potentiel al\'eatoire \flqq profond\frqq\
(qui peut devenir plus grand que l'\'energie cin\'etique
atomique), cette red\'efinition de l'\'energie implique
que la fonction d'onde moyenne est une onde \flqq localis\'ee\frqq :
son vecteur d'onde $k_0$ est imaginaire parce que $\langle V \rangle
> E_{i}$. Cette situation n'a pas d'analogue lumineux, et
il serait int\'eressant d'en explorer les cons\'equences
d'un point de vue th\'eorique. Mais l'on n'aura pas le
temps de le faire ici...}%
) et sa fonction de corr\'elation
$\langle V({\bf r}_1 ) \, V({\bf r}_2 )\rangle$.
Les quantit\'es int\'eressantes dans le contexte de cette
approche statistique sont, d'une part, la {\em valeur moyenne\/}
$\langle \psi( {\bf r}) \rangle$
de la fonction d'onde et, d'autre part,
sa {\em fonction de corr\'elation\/}
$\langle \psi({\bf r}_1 ) \, \psi^*({\bf r}_2 )\rangle$ que nous
appelerons \'egalement la fonction de coh\'erence.
Notons que la premi\`ere de ces quantit\'es n'est pas
forc\'ement nulle lorsque l'on se trouve au voisinage d'une
\flqq source coh\'erente\frqq\ d'atomes. La fonction d'onde moyenne
d\'ecrit alors la partie \flqq coh\'erente\frqq\ de l'onde atomique
dont la phase est suffisamment bien d\'efinie pour qu'elle 
puisse interf\'erer avec une onde de r\'ef\'erence. La
fonction de coh\'erence, quant \`a elle, d\'ecrit la
distribution en position et en impulsion de l'ensemble
atomique, comme nous verrons dans la repr\'esentation de
{\sc Wigner}. Elle d\'ecrit \'egalement le contraste des
franges lorsque l'on fait interf\'erer les ondes rayonn\'ees
par deux points \flqq du m\^eme ensemble\frqq\ 
(en opposition \`a l'interf\'erence
avec une onde de r\'ef\'erence dont il \'etait question pour
la fonction d'onde moyenne).

Dans la suite de cette introduction, 
nous allons fixer quelques notations.
Quant au potentiel al\'eatoire, nous introduisons sa
densit\'e spectrale (spatiale)
\begin{equation}
S( {\bf q} ) = \int\!{\rm d}{\bf r} \,
\langle V({\bf r} + {\bf r}_0 ) \, V({\bf r}_0 ) \rangle
\exp{(- {\rm i} {\bf q}\!\cdot\!{\bf r} ) }
	\label{eq:def-densite-spectrale}
\end{equation}
qui est la transform\'ee de {\sc Fourier} de la fonction
de corr\'elation du potentiel. Dans cette d\'efinition,
nous supposons que le potentiel al\'eatoire est 
(statistiquement) homog\`ene, c'est-\`a-dire que sa
fonction de corr\'elation en deux points ${\bf r}_{1,2}$
ne d\'epend que de la diff\'erence ${\bf r}_2 - {\bf r}_1$.
La port\'ee de la densit\'e 
spectrale~(\ref{eq:def-densite-spectrale}) en fonction
du vecteur d'onde ${\bf q}$ est de l'ordre de l'inverse
$1 / \ell_c$ de la longueur de corr\'elation.

Quant \`a l'ensemble atomique, introduisons sa distribution
de {\sc Wigner} :
\begin{equation}
f( {\bf r}, {\bf k} ) =
\int\!{\rm d}{\bf s} \, 
\langle \psi({\bf r} + \demi{\bf s}) \, 
\psi^*({\bf r} - \demi{\bf s} ) \rangle
\exp{ (- {\rm i} {\bf k}\cdot{\bf s} ) }
	\label{eq:def-Wigner}
\end{equation}
Cette distribution a des propri\'et\'es similaires \`a une
densit\'e de probabilit\'e dans l'espace des phases :
\begin{itemize}
\item
elle d\'epend \`a la fois de la position ${\bf r}$ et du
vecteur d'onde ${\bf k}$ ; 
\item
son int\'egrale sur les positions (sur les vecteurs d'onde) donne
la distribution en vecteur d'onde (en position) des atomes ;
\item 
le flux moyen atomique peut s'exprimer comme la valeur moyenne
du vecteur d'onde :
\begin{equation}
{\bf j}( {\bf r} ) = \frac{ \hbar }{ M } 
\mbox{Im } \langle \psi^*( {\bf r} ) 
\nabla \psi( {\bf r} ) \rangle
= \int\! {\rm d}{\bf k} \frac{ \hbar {\bf k} }{ M } 
f( {\bf r}, {\bf k} ) ;
	\label{eq:flux-Wigner}
\end{equation}
\item
notons cependant que la distribution de {\sc Wigner} n'est
pas n\'ecessairement positive, elle ne peut donc pas en toutes
circonstances \^etre interpr\'et\'ee comme une densit\'e de
probabilit\'e.
\end{itemize}

Nous allons souvent consid\'erer des atomes avec une \'energie
$E$ fix\'ee. L'\'equation de {\sc Schr\"odinger} stationnaire
peut alors s'\'ecrire de la fa\c con suivante :
\begin{eqnarray}
\setlength{\arraycolsep}{0.0em}
\nabla^2 \psi( {\bf r} ) + k_0^2 \left[ 1 + \mu( {\bf r} ) \right] 
\psi( {\bf r} ) & = & 0 ;
	\label{eq:Schroedinger-stationnaire}\\
E = \hbar^2 k_0^2 / 2 M , \qquad
\mu( {\bf r} ) & = & - V( {\bf r} ) / E .
	\label{eq:def-mu}
\end{eqnarray}
Cette forme de l'\'equation met en \'evidence l'analogie \`a la
lumi\`ere (dans l'approximation scalaire), o\`u $\mu( {\bf r})$
est la d\'eviation de la constante di\'electrique de l'unit\'e.
L'\'equation de Schr\"odinger stationnaire sera notre point de
d\'epart pour la th\'eorie de la \flqq diffusion multiple des ondes
de mati\`ere\frqq. Une quantit\'e importante dans ce contexte sera
la fonction de corr\'elation de la constante di\'electrique,
donc du potentiel al\'eatoire renormalis\'e :
\begin{eqnarray}
\setlength{\arraycolsep}{0.0em}
\langle \mu( {\bf r}_1 ) \,  \mu( {\bf r}_2 ) \rangle =
E^{-2} \langle V( {\bf r}_1 ) \,  V( {\bf r}_2 ) \rangle & = &
\epsilon^2 g( {\bf r}_2 - {\bf r}_1 ),
	\\
\epsilon & = & \langle V^2 \rangle^{1/2} / E
	\label{eq:def-epsilon-ici}
\end{eqnarray}
Le nombre $\epsilon$ donne donc l'ordre de grandeur du potentiel
al\'eatoire par rapport \`a l'\'energie des atomes. Au chapitre
pr\'ec\'edent, nous avons seulement consid\'er\'e le cas
$\epsilon \ll 1$. Nous ferons de m\^eme ici, notre th\'eorie
ne s'applique donc pas au cas d'un potentiel al\'eatoire
qui \flqq pi\`egerait\frqq\ les atomes dans ses vall\'ees.

\section{Point de vue de {\sc Ryzhik} {\em et al.\/}}
\label{s:Keller-Wigner}

Dans l'article de {\sc Ryzhik}, {\sc Papanicolaou} et {\sc Keller}
\cite{Keller96}, la diffusion des ondes scalaires par un potentiel
al\'eatoire est \'etudi\'ee en guise d'introduction. Nous nous
contenterons ici d'esquisser leur d\'emarche.

\subsection{Equation de transport \`a grande \'echelle}

Ils cherchent une \'equation d'\'evolution pour la distribution
de {\sc Wigner} \`a des \'echelles spatiales grandes par rapport \`a
la port\'ee des corr\'elations du potentiel al\'eatoire,
d'une part, et \`a la longueur d'onde atomique, d'autre part.
En utilisant un d\'eveloppement \flqq multi-\'echelles\frqq, ils
parviennent \`a d\'emontrer que la distribution de {\sc Wigner}
v\'erifie une \'equation du type {\em transfert radiatif\/}
qu'ils \'ecrivent de la fa\c con suivante :
\begin{eqnarray}
\left( \partial_t + \frac{ \hbar{\bf k} }{ M } \cdot
\nabla_r \right) f( {\bf r}, {\bf k} ) & = &
\frac{ 4 \pi M }{ \hbar^3 }
\int\!\frac{ {\rm d}{\bf k}' }{ (2\pi)^3 } 
S( {\bf k}' - {\bf k} )  
\,\delta( {\bf k}'^2 - {\bf k}^2 )
\times
	\nonumber\\
&& \quad \times
\left[ f({\bf r}, {\bf k}' ) - f({\bf r}, {\bf k} ) \right]
	\label{eq:transfert-Keller}
\end{eqnarray}
Le dernier terme dans cette \'equation d\'ecrit la diffusion
d'une onde atomique du vecteur d'onde initial ${\bf k}$ 
vers le vecteur d'onde final ${\bf k}'$.
L'efficacit\'e de ce processus est
proportionnelle \`a la densit\'e spectrale du potentiel 
$S({\bf k}' - {\bf k})$ au transfert de vecteur d'onde,
comme c'est le cas dans l'approximation de {\sc Born}. 
L'\'equation du transfert radiatif~(\ref{eq:transfert-Keller})
implique donc les valeurs suivantes pour la section
efficace de diffusion%
\footnote{%
Il s'agit en fait du nombre de diffusions par unit\'e de temps.
Elle est reli\'ee au libre parcours moyen $\ell_{\rm ex}$ par 
$\sigma_{tot} = \hbar k_0 / (M \ell_{\rm ex})$.%
} 
$\sigma_{\rm{tot}}$ et la fonction de phase%
\footnote{%
La fonction de phase~(\ref{eq:fn-phase-Keller}) d\'epend
des deux vecteurs unitaires ${\bf n}'$, ${\bf n}$ ; elle
est normalis\'ee telle que $\int\!\mbox{d}^2{\bf n}\, 
p( {\bf n}', {\bf n} ; k_0 ) = 1$.%
} 
$p( {\bf n}', {\bf n} ; k_0 )$ 
\begin{eqnarray}
\sigma_{\rm{tot}}( k_0 ) & = & \frac{ k M }{ \pi \hbar^3 }
\int\!\frac{ {\rm d}^2{\bf n}' }{ 4\pi } 
S( k_0 ({\bf n}' - {\bf n}) )
	\label{eq:section-efficace-Keller}\\
p( {\bf n}', {\bf n} ; k_0 ) & = &
\frac{ k_0 M }{ 4 \pi^2 \hbar^3 \sigma_{\rm{tot}} }
S( k_0 ({\bf n}' - {\bf n}) )
	\label{eq:fn-phase-Keller}
\end{eqnarray}

\subsection{Discussion}

Tout d'abord, l'\'equation de transport~(\ref{eq:transfert-Keller})
est un r\'esultat utile
pour fixer le comportement \`a grande distance de la distribution
de {\sc Wigner}. D'une fa\c con plus conceptuelle, le travail de
{\sc Ryzhik} {\em et al.\/} fournit donc une d\'emonstration alternative
de l'\'equation de transfert radiatif pour les ondes scalaires :
celle-ci est valable \`a des \'echelles spatiales grandes 
par rapport \`a la longueur d'onde et la longueur de corr\'elation.
On en a vu d'autres d\'emonstrations par {\sc Rytov} 
(\cite{Rytov},~\S~4.3) et {\sc Barabanenkov} et {\sc Finkel'berg}
\cite{Barabanenkov67}.

Ensuite, il ne faut pas oublier que~(\ref{eq:transfert-Keller})
contient implicitement
l'hypoth\`ese que le potentiel al\'eatoire est \flqq faible\frqq\ ;
en effet, nous avons constat\'e que la diffusion par le potentiel
est d\'ecrite dans l'approximation de {\sc Born} (la section efficace
est proportionnelle au carr\'e de la transform\'ee de {\sc Fourier} 
du potentiel). Nous retrouvons l\`a la proposition de {\sc Luck}
\cite{Luck93,Luck96} qui dit que l'\'equation de {\sc Bethe--%
Salpeter} est \'equivalente \`a l'\'equation du transfert
radiatif lorsque l'on tient compte des diagrammes \flqq en \'echelle\frqq\
({\em ladder approximation\/}) ce qui revient \`a une approximation
de diffusion simple et donc de {\sc Born}.

Finalement, nous remarquons que l'\'equation de 
transfert~(\ref{eq:transfert-Keller}) se simplifie
lorsque l'on se place dans le r\'egime semi-classique.
En utilisant la fonction de phase~(\ref{eq:fn-phase-Keller}),
nous constatons que la valeur maximale de l'angle de diffusion 
$\theta$ est donn\'ee par
\begin{equation}
k_0 | {\bf n}' - {\bf n} |_{\max}
= 2 k_0 \sin(\theta_{\max}/2) 
\simeq \frac{ 1 }{ \ell_c } 
\quad \Longrightarrow \quad
2\sin(\theta_{\max}/2) \simeq \frac{ \lambdabar_{dB} }{ \ell_c }  
\ll 1
	\label{eq:limite-sc-ici}
\end{equation}
Dans le r\'egime semi-classique, la diffusion des atomes se produit
donc de pr\'ef\'erence vers l'avant, et nous pouvons 
utiliser la formulation de {\sc Fokker--Planck} de l'\'equation
de transfert radiatif introduite au chapitre~\ref{s:radiatif}.
Nous avons v\'erifi\'e que l'on trouve une \'equation de {\sc Fokker--Planck}
identique \`a celle du chapitre~\ref{s:classique} o\`u nous nous
sommes plac\'e d'embl\'ee dans une description classique.%
\footnote{%
Il faut \`a cet effet exprimer le coefficient de diffusion
par la fonction de corr\'elation du potentiel al\'eatoire,
et utiliser un d\'eveloppement \`a l'ordre le plus bas en
$\lambda_{dB} / \ell_c$ pour retrouver la fonction de corr\'elation
de la force al\'eatoire.%
}

\subsection{Conclusion}

A des \'echelles spatiales grandes par rapport \`a la longueur
de corr\'elation du potentiel al\'eatoire et pour un potentiel
al\'eatoire faible, la distribution
de Wig\-ner des atomes v\'erifie une \'equation de transport du
type transfert radiatif. La fonction de phase dans cette
th\'eorie s'exprime en fonction de la densit\'e spectrale
du potentiel al\'eatoire, comme c'est le cas dans l'approximation
de {\sc Born} (de diffusion simple). Nous retrouvons 
exactement l'image classique du transport 
lorsque la longueur d'onde atomique
est petite par rapport \`a la longueur de corr\'elation du
potentiel ; la distribution de {\sc Wigner} v\'erifie alors
une \'equation de {\sc Fokker--Planck}.

\section{Diffusion multiple des ondes}
\label{s:diff-mult}

Introduisons maintenant un formalisme ondulatoire plus
g\'en\'eral qui est \'egalement en mesure de d\'ecrire
la diffusion multiple des ondes atomiques. Nous allons
pr\'esenter les objets et les \'equations de base d'une telle
th\'eorie, \`a savoir
\begin{itemize}
\item
la fonction de {\sc Green} moyenne,
qui permet de calculer la fonction d'onde moyenne et qui
v\'erifie l'\'equation de {\sc Dyson},
\item
la fonction de coh\'erence atomique et son \'equation d'\'evolution,
l'\'equation de {\sc Bethe--Salpeter} (B.--S.).
\end{itemize}
Au paragraphe~\ref{s:vers-TR} suivant, 
nous \'etudierons de fa\c con g\'en\'erale
le lien entre l'\'equation de B.--S. et la th\'eorie du transfert
radiatif. Nous nous servons \`a cet effet de la repr\'esentation
de {\sc Wigner}.

Dans ce paragraphe, nous nous pla\c cons dans une situation
stationnaire, l'\'energie des atomes est donc fix\'ee par le
vecteur d'onde atomique $k_0$ (valeur dans le potentiel moyen).

\subsection{Fonction de {\sc Green} moyenne}

La quantit\'e centrale pour d\'ecrire la fonction d'onde
atomique est la fonction de {\sc Green} $G( {\bf r}, {\bf r}')$
pour l'\'equation de {\sc Schr\"odinger} 
stationnaire~(\ref{eq:Schroedinger-stationnaire}) :
\begin{equation}
\nabla_r^2 G( {\bf r}, {\bf r}') +
k_0^2 \left[ 1 + \mu( {\bf r} ) \right]  
G( {\bf r}, {\bf r}')
= \delta( {\bf r} - {\bf r}')
	\label{eq:def-Green}
\end{equation}
Cet objet d\'ecrit la fonction d'onde rayonn\'ee par une
source ponctuelle (de fr\'equence $E/\hbar$) situ\'ee \`a la
position ${\bf r}'$ dans le milieu al\'eatoire. 
Pour une source spatialement \'etendue,
nous trouvons la fonction d'onde en sommant les champs
rayonn\'es par tous les points source.

\subsubsection{Equation de {\sc Dyson}}

La fonction de {\sc Green}~(\ref{eq:def-Green})
d\'epend du potentiel al\'eatoire en chaque point, et il est impossible
dans la pratique de la calculer explicitement. L'objet
int\'eressant est donc la {\em fonction de {\sc Green} moyenne\/},
la moyenne \'etant prise sur les configurations du potentiel
al\'eatoire. Nous utiliserons la notation
\begin{equation}
\overline{G}( {\bf r} - {\bf r}' )
= \langle G( {\bf r}, {\bf r}') \rangle
	\label{eq:Green-moyenne}
\end{equation}
pour cette fonction de {\sc Green}. Elle d\'ecrit donc la valeur
moyenne de la fonction d'onde rayonn\'ee par une source ponctuelle.
En \'ecrivant~(\ref{eq:Green-moyenne}), nous avons suppos\'e que
le milieu al\'eatoire est statistiquement homog\`ene : la fonction
de {\sc Green} moyenne ne d\'epend alors que la distance ${\bf r} -
{\bf r}'$ entre le point d'observation et le point source.

Il s'agit maintenant de trouver une \'equation ferm\'ee pour
la fonction de {\sc Green} moyenne. 
Il est utile \`a cet effet de transformer l'\'equation de 
{\sc Schr\"odinger}~(\ref{eq:def-Green}) en une \'equation int\'egrale :
\begin{equation}
G( {\bf r}, {\bf r}' ) = G_0( {\bf r} - {\bf r}' )
+ k_0^2 \int\!{\rm d}{\bf r}_1
G_0( {\bf r} - {\bf r}_1 ) \mu( {\bf r}_1 )
G( {\bf r}_1, {\bf r}' ) , 
	\label{eq:Green-integrale}
\end{equation}
o\`u nous avons utilis\'e la fonction de {\sc Green}
$G_0( {\bf r} - {\bf r}' )$ pour l'espace libre :
\begin{equation}
G_0( {\bf r} - {\bf r}' ) = -
\frac{ \exp{ {\rm i} k_0 | {\bf r} - {\bf r}' |} }{
4 \pi | {\bf r} - {\bf r}' | }
	\label{eq:Green-libre}
\end{equation}
L'approximation de {\sc Born} consiste \`a 
r\'esoudre~(\ref{eq:Green-integrale}) en rempla\c cant au membre
droit, $G$ par la solution en espace libre $G_0$. On constate
que cette proc\'edure est le premier terme d'une solution par
it\'eration. En allant au-del\`a de l'approximation de {\sc Born},
une telle solution g\'en\`ere des produits avec de plus en plus de
facteurs $\mu( {\bf r} )$. L'on peut alors prendre la valeur
moyenne de cette s\'erie, pour faire appara\^{\i}tre les fonctions
de corr\'elation du potentiel al\'eatoire. En toute g\'en\'eralit\'e,
l'on aura affaire \`a des corr\'elations d'ordre arbitrairement
\'elev\'e, comme par exemple
\[
\langle \mu( {\bf r}_1 ) \mu( {\bf r}_2 ) \mu( {\bf r}_3 ) \rangle
\]
Pour simplifier la th\'eorie, nous allons supposer que le potentiel
al\'eatoire \`a une statistique gaussienne ; c'est-\`a-dire que les
fonctions de corr\'elations d'ordre sup\'erieur \`a deux peuvent
s'exprimer au moyen de la fonction de corr\'elation \`a deux points.
Toutes les corr\'elations \`a un nombre de points impair s'annulent
alors, et pour les corr\'elation \`a quatre points, par exemple,
nous avons
\begin{eqnarray}
\lefteqn{
\langle \mu( {\bf r}_1 ) \mu( {\bf r}_2 ) \mu( {\bf r}_3 ) \mu( {\bf r}_4 
\rangle = \langle \mu( {\bf r}_1 ) \mu( {\bf r}_2 ) \rangle \langle
\mu( {\bf r}_3 ) \mu( {\bf r}_4 \rangle \, + }
	\nonumber\\
&& + \,
\langle \mu( {\bf r}_1 ) \mu( {\bf r}_3 ) \rangle \langle
\mu( {\bf r}_2 ) \mu( {\bf r}_4 \rangle +
\langle \mu( {\bf r}_1 ) \mu( {\bf r}_4 ) \rangle \langle
\mu( {\bf r}_2 ) \mu( {\bf r}_3 \rangle .
	\nonumber
\end{eqnarray}

La solution it\'erative reste quand m\^eme complexe et contient
une infinit\'e de termes. Pour les organiser, l'on peut se servir
d'une m\'ethode de diagrammes. Cette approche est expos\'ee dans
les articles de {\sc Frisch} \cite{Frisch66} et dans les livres
de Rytov {\em et al.\/} (\cite{Rytov},~\S~4.1) et de Ping {\sc Sheng}
(\cite{Sheng95}, \S\S~4.3,~4.4). En r\'e-organisant la s\'erie it\'erative,
l'on trouve l'\'equation suivante pour la fonction de {\sc Green}
moyenne, l'\'equation de {\sc Dyson} :
\begin{equation}
\overline{G}( {\bf r} - {\bf r}' ) = G_0( {\bf r} - {\bf r}' )
+  \int\!{\rm d}{\bf r}_1 {\rm d}{\bf r}_2
G_0( {\bf r} - {\bf r}_1 ) m( {\bf r}_1 - {\bf r}_2)
\overline{G}( {\bf r}_2 - {\bf r}' ) , 
	\label{eq:Dyson}
\end{equation}
o\`u $m( {\bf r}_1 - {\bf r}_2)$ est appel\'e l'\flqq op\'erateur
de masse\frqq. Son d\'eveloppement it\'eratif ne contient que des
\flqq diagrammes irr\'eductibles\frqq%
\footnote{%
``{\em strongly connected\/}''%
} et s'\'ecrit :
\begin{eqnarray}
\lefteqn{m( {\bf r}_1 - {\bf r}_2) = 
\epsilon^2 k_0^4 g( {\bf r}_1 - {\bf r}_2) 
G_0( {\bf r}_1 - {\bf r}_2 ) \, + }
	\label{eq:Bourret}\\
&& + \, \epsilon^4 k_0^8
\int\!{\rm d}{\bf r}_3 {\rm d}{\bf r}_4
G_0( {\bf r}_1 - {\bf r}_3 ) 
G_0( {\bf r}_3 - {\bf r}_4 ) G_0( {\bf r}_4 - {\bf r}_2 ) 
\times
	\label{eq:operateur-masse}\\
&& \quad \times
\left[
 g( {\bf r}_1 - {\bf r}_4) g( {\bf r}_2 - {\bf r}_3) 
+ g( {\bf r}_1 - {\bf r}_2) g( {\bf r}_4 - {\bf r}_3) 
\right] \, + \ldots
	\nonumber
\end{eqnarray}
Nous notons que l'op\'erateur de masse ne d\'epend que de
la diff\'erence des positions ${\bf r}_1 - {\bf r}_2$, ceci \'etant
d\^u \`a l'homog\'en\'eit\'e statistique.

Afin d'interpr\'eter l'op\'erateur de masse, revenons
\`a une formulation diff\'erentielle de
l'\'equation de {\sc Dyson}, en appliquant l'op\'erateur
$\nabla_r^2 + k_0^2$ \`a~(\ref{eq:Dyson}) :
\begin{equation}
\left( \nabla_r^2 + k_0^2 \right) 
\overline{G}( {\bf r} - {\bf r}' ) = 
\delta( {\bf r} - {\bf r}' )
+ \int\!{\rm d}{\bf r}_1 m( {\bf r} - {\bf r}_1)
\overline{G}( {\bf r}_1 - {\bf r}' ) , 
	\label{eq:integro-diff}
\end{equation}
Ceci est en fait une \'equation int\'egro-diff\'erentielle
qui exprime le champ moyen rayonn\'ee dans le milieu al\'eatoire
comme la somme d'un champ en espace libre (la fonction $\delta$),
plus une correction qui d\'epend de fa\c con non-locale de la
valeur du champ autour de la source (le deuxi\`eme terme).
L'op\'erateur de masse traduit donc pr\'ecis\'ement la
r\'etro-action du milieu diffusant sur la propagation du
champ moyen.

Nous notons finalement que l'on peut 
r\'esoudre~(\ref{eq:integro-diff}) par une transform\'ee
de {\sc Fourier} puisque le milieu est homog\`ene. L'int\'egrale
du deuxi\`eme terme est en effet un produit de convolution,
et nous avons donc pour les transform\'ees de {\sc Fourier}
\begin{eqnarray}
( k_0^2 - {\bf k}^2 ) \overline{G}({\bf k})
& = & 1 + m({\bf k}) \overline{G}({\bf k})
	\label{eq:integro-diff-Fourier}\\
\mbox{d'o\`u} : \quad
\overline{G}({\bf k}) & = & \frac{ 1 }{
k_0^2 - {\bf k}^2 - m({\bf k}) }
	\label{eq:TF-<G>}
\end{eqnarray}
Nous en d\'eduisons la relation de dispersion des ondes de
mati\`ere dans le milieu d\'esordonne\'e :
\begin{equation}
{\bf k}^2 + m({\bf k}) = k_0^2 
	\label{eq:dispersion}
\end{equation}
Cette \'equation d\'efinit le vecteur d'onde \flqq effectif\frqq\ 
$k_{\rm{eff}}$ de la fonction
d'onde moyenne dans le milieu. Il diff\`ere du vecteur d'onde
dans le vide $k_0$ pour deux raisons : d'une part, la diffusion
par le milieu introduit de l'att\'enuation pour le champ coh\'erent,
ce qui se traduit par une partie imaginaire du vecteur d'onde,
et d'autre part, la diffusion modifie la phase
de propagation des ondes.%
\footnote{%
{\sc Rytov} dit que la longueur des rayons augmente par l'effet
de la diffusion et par cons\'equent aussi la partie r\'eelle de
$k_{eff}$ (\cite{Rytov},~\S~4.2,~p.~137).%
}

\subsection{Approximation de {\sc Bourret}}

Tout ce que nous venons d'exposer ne serait en fait que des
manipulations formelles d'\'equations s'il n'y avait pas
un moyen de calculer l'op\'erateur de masse $m({\bf r}_1 -
{\bf r}_2)$. Or, le d\'eveloppement~(\ref{eq:operateur-masse})
permet de le faire en principe (il contient n\'eanmoins
une infinit\'e de termes), et en particulier dans une limite 
perturbative,
lorsque l'on suppose $\epsilon \ll 1$. A l'ordre le plus
bas, l'on trouve donc l'expression assez 
maniable de la premi\`ere ligne~(\ref{eq:Bourret})
\begin{equation}
m( {\bf r}_1 - {\bf r}_2) \approx \epsilon^2 k_0^4 \,
g( {\bf r}_1 - {\bf r}_2) 
G_0( {\bf r}_1 - {\bf r}_2 ) ,
	\label{eq:Bourret-ici}
\end{equation}
que l'on appelle, suivant les auteurs, l'\flqq approximation
de {\sc Bourret}\frqq\ \cite{Frisch66}, l'\flqq approximation de
r\'egularisation du premier ordre\frqq\ \cite{Frisch66},
la ``{\em ladder approximation\/}'' \cite{Rytov}. Elle doit
\'egalement avoir un lien avec la ``{\em random phase 
approximation\/}'' (RPA) utilis\'ee dans la physique des
solides. 

Nous remarquons que l'approximation de {\sc Bourret}~%
(\ref{eq:Bourret-ici}) ressemble, au premier
abord, \`a l'approximation de {\sc Born} pour l'op\'erateur
de masse. Qu'a-t-on alors gagn\'e par rapport au choix
de faire d'embl\'ee l'approximation de {\sc Born} pour
la fonction de {\sc Green} ? La r\'eponse se trouve
par exemple dans la formule~(\ref{eq:dispersion})
qui permet de calculer les \flqq nouveaux modes\frqq\ dans
le milieu al\'eatoire, alors qu'avec l'approximation de {\sc Born},
la fonction de {\sc Green} donne seulement 
une amplitude de diffusion. Da fa\c con plus profonde, le
formalisme de l'\'equation de {\sc Dyson} permet, dans
l'approximation de {\sc Bourret}, de trouver une fonction
de {\sc Green} moyenne qui, elle, est d\'ej\`a une resommation
partielle d'une infinit\'e de diagrammes.

Pour illustrer la puissance de l'approximation de {\sc Bourret}
pour la diffusion multiple des ondes, citons le r\'esultat
pour le vecteur d'onde effectif dans un milieu faiblement 
diffuseur, donn\'e par {\sc Rytov} (\cite{Rytov},~\'eq.(4.61)) :
\begin{equation}
k_{\rm{eff}}  =  k_0 + \frac{ \pi }{ 4 } \epsilon^2 k_0^2
\int\limits_0^\infty \! \frac{ k' {\rm d}k' }{ (2\pi)^3 }
g(k') \log\!\left( \frac{ 2 k_0 + k' }{ 2 k_0 - k' } \right)^2
+ \, {\rm i} \frac{ \pi^2 }{ 2 } \epsilon^2 k_0^2
\int\limits_0^{2k_0} \! \frac{ k' {\rm d}k' }{ (2\pi)^3 }
g(k') 
	\label{eq:k-eff-Bourret}
\end{equation}
o\`u la fonction de corr\'elation est suppos\'ee isotrope
(sa transform\'ee de {\sc Fourier} $g(k')$ 
ne d\'epend alors que du module $k'$ du vecteur d'onde).
On constate que le vecteur d'onde dans le milieu contient
une partie imaginaire positive (att\'enuation
de l'onde coh\'erente par la diffusion), et une partie r\'eelle
un peu plus grande que dans le vide.

\begin{figure}
\centering
\resizebox{!}{7cm}{%
\includegraphics[1.2in,3.0in][7.5in,8.5in]{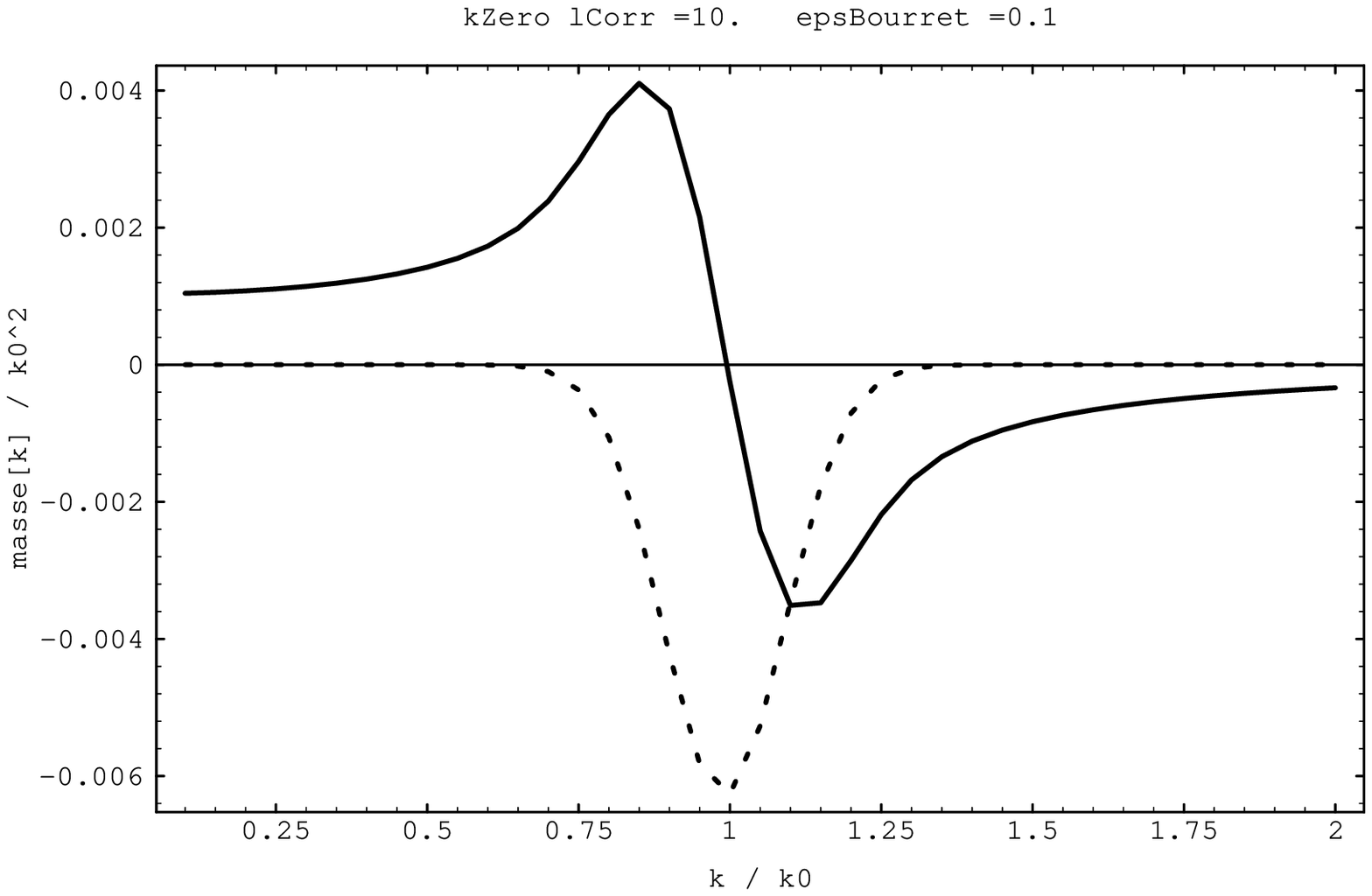}}
\caption[fig:masse]{Op\'erateur de masse pour un potentiel
al\'eatoire \`a corr\'elations gaussienne isotropes.
Trait plein : partie r\'eelle ; tirets : partie imaginaire.
\\
$m( {\bf k} )$ est donn\'e en unit\'es de $k_0^2$,
la longueur de corr\'elation vaut $\ell_c = 10 / k_0$,
et le petit param\`etre de l'approximation de {\sc Bourret}
vaut $(\epsilon k_0 \ell_c)^2 = 0.1$.%
}
\label{fig:masse}
\end{figure}

Pour donner un ordre de grandeur de ces effets, 
prenons un potentiel al\'eatoire
avec une corr\'elation gaussienne. La densit\'e spectrale est
alors gaussienne \'egalement. La transform\'ee de {\sc Fourier}~%
$m({\bf k})$ de l'op\'erateur de masse est alors donn\'ee par
\begin{eqnarray}
m({\bf k}) & = & \frac{ {\rm i} \sqrt{\pi} }{ 2 \sqrt{2} }
\frac{  (\epsilon k_0^2 \ell_c)^2 }{ k \ell_c }
\begin{array}[t]{l}
\left[
{\rm e}^{- (k_0 + k)^2 \ell_c^2 / 2 }
{\rm{erfc}}[-{\rm i} (k_0 + k) \ell_c / \sqrt{2} ] \right. \\
\left. - \, 
{\rm e}^{- (k_0 - k)^2 \ell_c^2 / 2 }
{\rm{erfc}}[-{\rm i} (k_0 - k) \ell_c / \sqrt{2} ]
\right] 
\end{array}
	\label{eq:m(k)-gaussien}\\
&& {\rm{erfc}}(x) = \frac{ 2 }{ \sqrt{\pi} }
\int_x^\infty \! {\rm e}^{- t^2 } \, {\rm d}t 
	\label{eq:def-erfc}
\end{eqnarray}
Cette fonction est repr\'esent\'ee sur la figure~\ref{fig:masse}.
Quant \`a la relation de dispersion~(\ref{eq:dispersion}),
l'on trouve le vecteur d'onde effectif suivant (dans la limite 
semi-classique $k_0 \ell_c \gg 1$)
\begin{equation}
k_{\rm{eff}} \approx k_0 \left( 1 + 
 \frac{ \epsilon^2 }{ 8 } \right)
+ {\rm i} k_0  \frac{ \pi^2 }{ 2 (2\pi)^{3/2} }   
\epsilon^2 k_0 \ell_c .
	\label{eq:k-eff-ordre}
\end{equation}
{\sc Rytov} n'oublie pas de nous rappeler que ce r\'esultat
n'est valable que dans l'approximation de {\sc Bourret}.
En comparant aux corrections d'ordre suivant, 
{\sc Rytov} trouve la condition de validit\'e suivante 
\begin{equation}
\mbox{\sc Bourret}: \quad
\left( \epsilon k_0 \ell_c \right)^2 \ll 1 .
	\label{eq:validite-Bourret}
\end{equation}
Rappelons que dans les m\'ethodes classiques du 
chapitre~\ref{s:classique}, la condition de validit\'e
d'une approche de {\sc Fokker--Planck} \'etait que le potentiel
soit faible par rapport \`a l'\'energie cin\'etique des atomes, 
ce qui revient dans la notation de ce chapitre \`a la condition
$\epsilon^2 \ll 1$ [voir~(\ref{eq:def-epsilon-Keller})
et~(\ref{eq:def-epsilon})]. L'approximation de {\sc Bourret}~%
(\ref{eq:validite-Bourret}) impose donc une limite plus stricte 
au potentiel moyen qui l'est d'autant plus que l'on se place 
dans le r\'egime semi-classique o\`u $k_0 \ell_c \gg 1$.

\paragraph{Remarque.}
L'approximation de {\sc Bourret} peut \^etre g\'en\'eralis\'ee
\`a un milieu al\'eatoire dont la statistique n'est pas
gaussienne. Les fonctions de corr\'elations peuvent alors
\^etre \'ecrites comme une somme sur des corr\'elations
\flqq \'el\'ementaires\frqq\ $g_k( {\bf r}_1, \ldots,
{\bf r}_k )$ pour un groupe de $k$ points 
(d\'eveloppement en essaims%
\footnote{%
``{\em cluster expansion\/}''}%
). {\sc Finkel'berg} montre alors que pour calculer l'op\'erateur
de masse \`a l'ordre le plus bas, il suffit de retenir les
diagrammes irr\'eductibles qui ne contiennent qu'un seul
groupe de points \cite{Finkelberg67}. Il appelle cette 
approximation la ``{\em single-group approximation\/}''.
Son avantage est qu'elle permet \'egalement de tenir compte
de la diffusion par des agglom\'erats de plus que deux
particules. {\sc Finkel'berg} donne la condition de
validit\'e suivante pour cette approximation :
\begin{equation}
\mbox{{\em single group\/} :} \quad
\left| \frac{ {\rm d} m }{ {\rm d} k^2 } \right| \ll 1.
	\label{eq:Finkelberg}
\end{equation}
Il montre en particulier que cette condition est suffisante pour 
que la diffusion soit faible (libre parcours moyen beaucoup
plus grand que la longueur d'onde).

\subsection{Equation de {\sc Bethe--Salpeter}}

Pour finir ce paragraphe, il ne reste qu'\`a pr\'esenter
l'\'equation d'\'evolution pour la fonction de corr\'elation
de la fonction d'onde atomique. Nous allons \`a cet effet
consid\'erer non plus la fonction de {\sc Green}, mais
l'onde rayonn\'ee par une source $\varrho({\bf r})$ quelconque
\begin{equation}
\psi({\bf r}) = \int\!{\rm d}{\bf r}' \, G( {\bf r}, {\bf r}')
\, \varrho({\bf r}')
	\label{eq:champ-rayonne}
\end{equation}
L'onde moyenne $\langle\psi({\bf r})\rangle$ s'exprime
alors par une \'equation analogue o\`u intervient la
fonction de {\sc Green} moyenne 
$\overline{G}( {\bf r} - {\bf r}')$. En ce qui concerne
la fonction de coh\'erence des atomes, elle s'exprime
par une corr\'elation entre deux fonctions de {\sc Green} :
\begin{equation}
\langle \psi({\bf r}_1) \psi^*({\bf r}_2) \rangle = 
\int\!{\rm d}{\bf r}'_1 \, {\rm d}{\bf r}'_2 \,
\langle G( {\bf r}_1, {\bf r}'_1 )
G^*( {\bf r}_2, {\bf r}'_2 ) \rangle
\,\varrho( {\bf r}'_1 ) \varrho^*( {\bf r}'_2 )
	\label{eq:def-corr-psi}
\end{equation}
Cette fonction de corr\'elation entre deux fonctions de 
{\sc Green} est l'objet de l'\'equation de 
{\sc Bethe--Salpeter}. Elle se d\'eduit de l'\'equation de
{\sc Schr\"odinger}~(\ref{eq:Schroedinger-stationnaire})
de fa\c con analogue \`a l'\'equation de {\sc Dyson}~%
(\ref{eq:Dyson}) (voir \cite{Frisch66} et \S~4.3 de \cite{Sheng95}
pour plus de d\'etails). L'\'equation de {\sc Bethe--Salpeter}
permet d'\'ecrire l'\'equation ferm\'ee
suivante pour la fonction de coh\'erence du champ atomique :
\begin{eqnarray}
\langle \psi({\bf r}_1) \psi^*({\bf r}_2) \rangle & = &
\langle \psi({\bf r}_1) \rangle
\langle \psi^*({\bf r}_2) \rangle +
\int\!{\rm d}{\bf r}'_1 \, {\rm d}{\bf r}'_2 \,
{\rm d}{\mbf \rho}_1 \, {\rm d}{\mbf \rho}_2 \,
	\label{eq:Bethe-Salpeter}\\
&& \quad
\overline{G}( {\bf r}_1 - {\bf r}'_1)
\overline{G}^*( {\bf r}_2 - {\bf r}'_2) \,
K( {\bf r}'_1, {\mbf \rho}_1 | {\bf r}'_2 , {\mbf \rho}_2 ) \,
\langle \psi({\mbf \rho}_1) \psi^*({\mbf \rho}_2) \rangle
	\nonumber
\end{eqnarray}
Dans cette \'equation appara\^{\i}t la fonction
$K( {\bf r}_1, {\mbf \rho}_1 | {\bf r}_2 , {\mbf \rho}_2 )$
que {\sc Frisch} appelle l'\flqq op\'erateur d'intensit\'e\frqq.
Chez d'autres auteurs, elle porte le nom de \flqq vertex 
irr\'eductible\frqq\ \cite{Tsang85,Sheng95}.
D'un point de vue physique, cet objet d\'ecrit la corr\'elation 
aux positions ${\bf r}_{1,2}$
entre les champs rayonn\'es par deux sources situ\'ees
\`a ${\mbf \rho}_{1,2}$.  
Son d\'eveloppement perturbatif (en diagrammes) contient
les premiers termes suivants :
\begin{eqnarray}
&& K( {\bf r}_1, {\mbf \rho}_1 | {\bf r}_2, {\mbf \rho}_2 )
= \epsilon^2 k_0^4\, g( {\bf r}_1 - {\bf r}_2 ) \,
\delta( {\bf r}_1 - {\mbf \rho}_1 ) \, 
\delta( {\bf r}_2 - {\mbf \rho}_2 ) 
+
	\label{eq:K-Bourret}\\
&& \quad +\,
\epsilon^4 k_0^6 \, g( {\bf r}_1 - {\bf r}_2 ) \, 
g( {\mbf \rho}_1 - {\mbf \rho}_2 ) 
\left[
\delta( {\bf r}_1 - {\mbf \rho}_1 ) \,
G_0( {\bf r}_2 - {\mbf \rho}_2 ) + \{ \mbox{$1 \leftrightarrow 2$} \} 
\right]
+
	\nonumber\\
&& \quad +\,
\epsilon^4 k_0^8 \, g( {\bf r}_1 - {\mbf \rho}_2 ) \, 
g( {\mbf \rho}_1 - {\bf r}_2 ) 
G_0( {\bf r}_1 - {\mbf \rho}_1 ) \,
G_0( {\bf r}_2 - {\mbf \rho}_2 ) 
+ 
	\nonumber\\
&& \quad +\,
\epsilon^4 k_0^8 \, g( {\bf r}_1 - {\mbf \rho}_1 ) \,
\delta( {\bf r}_2 - {\mbf \rho}_2 )
\int\!{\rm d}{\bf r}_1'\,
G_0( {\bf r}_1 - {\bf r}_1' ) \,
g( {\bf r}_1' - {\bf r}_2 ) \,
G_0( {\bf r}_1' - {\mbf \rho}_1 )
+
	\nonumber\\
&& \quad \quad + \, \{ \mbox{$1 \leftrightarrow 2$} \} + \ldots
	\label{eq:operateur-K}
\end{eqnarray}
L'op\'erateur d'intensit\'e est donc un objet plut\^ot
encombrant. Son approximation de {\sc Bourret} est donn\'ee
par la premi\`ere ligne, l'\'eq.~(\ref{eq:K-Bourret}),
o\`u l'on constate qu'il revient essentiellement \`a la
fonction de corr\'elation du potentiel si ce dernier est
faible. 

Notons encore la propri\'et\'e g\'en\'erale suivante
de l'op\'erateur~(\ref{eq:operateur-K}), elle aussi due
\`a l'homog\'en\'eit\'e du milieu diffuseur :
il est invariant par une translation globale
de ses quatre arguments
\begin{equation}
K( {\bf r}_1 + \Delta{\bf r}, {\mbf \rho}_1 + \Delta{\bf r} | 
{\bf r}_2 + \Delta{\bf r} , {\mbf \rho}_2 + \Delta{\bf r} )
=
K( {\bf r}_1, {\mbf \rho}_1 | {\bf r}_2 , {\mbf \rho}_2 )
	\label{eq:invariance-K}
\end{equation}
Par cons\'equent, il ne peut d\'ependre que des diff\'erences
des positions, comme on le constate aussi dans le
d\'eveloppement~(\ref{eq:operateur-K}). 
Ceci se traduit par la propri\'et\'e suivante
pour la transform\'ee de {\sc Fourier} de l'op\'erateur d'intensit\'e :
\begin{eqnarray}
&& K( {\bf k}_1, {\mbf \kappa}_1 | {\bf k}_2, {\mbf \kappa}_2 )
= 
\int\!{\rm d}{\bf r}_1\, {\rm d}{\bf r}_2\, 
{\rm d}{\mbf \rho}_1\, {\rm d}{\mbf \rho}_2\,
	\nonumber\\
&&\qquad
K( {\bf r}_1, {\mbf \rho}_1 | {\bf r}_2 , {\mbf \rho}_2 )
\exp{\rm i}\left(
- {\bf k}_1\!\cdot\!{\bf r}_1 + {\mbf \kappa}_1\!\cdot\!{\mbf \rho}_1 
+ {\bf k}_2\!\cdot\!{\bf r}_2 - {\mbf \kappa}_2\!\cdot\!{\mbf \rho}_2 
\right)
	\label{eq:def-TF-K}\\[0.3\jot]
&& = (2\pi)^3 \,
\delta\!\left(
{\bf k}_1 - {\mbf \kappa}_1 - ({\bf k}_2 - {\mbf \kappa}_2) \right)
\tilde{K}( {\bf k}_1, {\mbf \kappa}_1 | {\bf k}_2, {\mbf \kappa}_2 )
	\label{eq:TF-K}
\end{eqnarray}
Les signes des vecteurs d'onde ont \'et\'e choisis tels que 
la T.-F. de l'op\'erateur d'intensit\'e admet l'interpr\'etation 
suivante : il d\'ecrit la corr\'elation entre les amplitudes 
pour les deux processus de diffusion 
\[
{\mbf \kappa}_1 \to {\bf k}_1 \quad \mbox{et} \quad
{\mbf \kappa}_2 \to {\bf k}_2 
\]
La fonction $\delta$ dans~(\ref{eq:TF-K}) exprime alors que
ces deux processus ne sont corr\'el\'es que lorsqu'ils font
intervenir le m\^eme transfert de vecteur d'onde :
\begin{equation}
K \ne 0 \quad \Longleftrightarrow \quad
{\bf k}_1 - {\mbf \kappa}_1 = {\bf k}_2 - {\mbf \kappa}_2 
	\label{eq:diffusions-correlees}
\end{equation}
Cette propri\'et\'e se comprend ais\'ement dans l'approximation
de {\sc Born} : l'amplitude de diffusion pour le
processus ${\mbf \kappa}_1 \to {\bf k}_1$ est proportionnelle
\`a la composante de {\sc Fourier} $\mu({\bf k}_1 - {\mbf \kappa}_1)$ 
du potentiel al\'eatoire, et les composantes
de {\sc Fourier} du potentiel al\'eatoire sont 
\flqq $\delta$-corr\'el\'ees\frqq\ lorsque le potentiel est
statistiquement homog\`ene :
\begin{equation}
\langle \mu({\bf q}_1) \mu^*({\bf q}_2) \rangle = \delta({\bf q}_1 -
{\bf q}_2) (2\pi)^3 \epsilon^2 g({\bf q}_1) .
	\label{eq:Vq1-Vq2}
\end{equation}

Dans la pratique, il n'est pas possible de r\'esoudre
l'\'equation de {\sc Bethe--Salpeter}~(\ref{eq:Bethe-Salpeter}), 
m\^eme dans l'approximation de {\sc Bourret}. 
Au paragraphe suivant, nous retra\c cons
une reformulation de l'\'eq. de B.--S. qui permet de retrouver
l'approche du transfert radiatif.

\section{Lien \`a l'\'equation du transfert radiatif}
\label{s:vers-TR}

L'int\'er\^et de ce paragraphe est de pr\'eciser le domaine de
validit\'e de l'\'equation du transfert radiatif, en la
d\'eduisant du formalisme microscopique de la diffusion multiple
des ondes. Une pr\'esentation semblable
a \'et\'e donn\'ee par {\sc Barabanenkov} et {\sc Finkel'berg}
\cite{Barabanenkov67} et par {\sc Rytov} {\em et al.\/} 
(\cite{Rytov},~\S~4.3) pour des ondes scalaires. Pour des
ondes \'electromagn\'etiques, une g\'en\'eralisation de 
l'\'equation du transfert radiatif (portant sur le
vecteur de {\sc Stokes}) est d\'ej\`a formul\'ee dans le livre
de {\sc Chandrasekhar}. Une justification microscopique
en a \'et\'e donn\'ee par {\sc K.~M. Watson} \cite{Watson69}
dans une situation de diffusion faible, et
J.-M.~{\sc Tualle} a fourni une pr\'esentation
plus g\'en\'erale \cite{TualleT}. 
Une bonne r\'ef\'erence est \'egalement 
l'article de revue de {\sc Lagendijk} et {\sc van Tiggelen} 
\cite{vanTiggelen96}.

Nous verrons que l'\'equation du transfert radiatif
(ETR) est valable \`a grande \'echelle
spatiale (ce qui n'est gu\`ere \'etonnant, vu
le r\'esultat de {\sc Ryzhik} {\em et al.\/}), mais
aussi quelle que soit la force du potentiel diffuseur.
Ce dernier point semble ne pas avoir \'et\'e soulev\'e
par d'autres auteurs parce que l'on se place dans la plupart
des cas dans le r\'egime de diffusion faible, en utilisant
l'approximation de {\sc Bourret}. {\sc Luck}, par exemple,
dit dans ses papiers \cite{Luck93,Luck96},
au tournant d'une phrase : 
\flqq l'\'equation de {\sc Bethe--Salpeter}, dans 
l'approximation de {\sc Bourret}, est \'equivalente \`a
l'\'equation du transfert radiatif\frqq. 
Nous allons par contre rester assez g\'en\'eral et 
ne pas faire l'approximation de {\sc Bourret}. 
Ainsi le formalisme permet-il de voir \`a quel endroit 
l'\'equation du transfert radiatif est moins g\'en\'erale
que celle de B.--S.

Nous allons seulement faire l'hypoth\`ese suivante
(qui ne semble pas tr\`es restrictive)
\begin{itemize}
\item
le milieu est statistiquement homog\`ene. La fonction de
{\sc Green} moyenne $\overline{G}$, qui intervient dans
l'\'equation de B.--S.~%
(\ref{eq:Bethe-Salpeter}), admet alors le d\'eveloppement
de {\sc Fourier}~(\ref{eq:TF-<G>})%
\footnote{%
Nous nous servons de la notation abr\'eg\'ee 
\[
\mbox{d}\kern-.6em\hbox{\raise1.2ex
\hbox{\kern.25em\vrule width.4em height.3pt depth.1pt}}%
{\bf k} = \frac{ \mbox{d}{\bf k} }{ (2\pi)^3 } .
\]
}
\begin{eqnarray}
\overline{G}( {\bf r} - {\bf r}' )
& = &\int\!\dbar{\bf k} \frac{ \exp{\rm i}{\bf k}\cdot(
{\bf r} - {\bf r}' ) }{ k_0^2 - {\bf k}^2 - m({\bf k}) }
	\label{eq:TF-<G>-ici}
\end{eqnarray}
\end{itemize}
o\`u $m( {\bf k} )$ est l'op\'erateur de masse.

\subsection{{\sc Bethe--Salpeter} dans {\sc Wigner}}

La premi\`ere \'etape consiste \`a exprimer l'\'equation
de B.--S.~(\ref{eq:Bethe-Salpeter}) en repr\'esentation de
{\sc Wigner}. En introduisant les transform\'ees de {\sc Fourier}
des fonctions de {\sc Green} moyenn\'ees et de l'op\'erateur
d'intensit\'e, l'on trouve l'\'equation suivante :
\begin{equation}
f( {\bf r}, {\bf k} ) =  f_{\rm{coh}}( {\bf r}, {\bf k} )
+ f_{\rm{diff}}( {\bf r}, {\bf k} )
	\label{eq:BS-0}
\end{equation}

Calculons d'abord le premier terme qui correspond \`a la transform\'ee de
{\sc Wigner} du produit des champs moyens. Nous exprimons le
champ moyen \`a l'aide de la fonction de {\sc Green} moyenn\'ee
({\em cf.\/}~\'eq.~\ref{eq:champ-rayonne})
et ins\'erons ensuite la d\'ecomposition de {\sc Fourier}~%
(\ref{eq:TF-<G>-ici}) de cette derni\`ere :
\begin{eqnarray}
\lefteqn{ 
f_{\rm{coh}}( {\bf r}, {\bf k} ) =
}
	\nonumber\\
&& =
\int\!{\rm d}{\bf s} \, {\rm e}^{- {\rm i} {\bf k}\!\cdot\!{\bf s} }
\int\!{\rm d}{\mbf \rho}_1 \, {\rm d}{\mbf \rho}_2 \,
\overline{G}( {\bf r} + \demi {\bf s} - {\mbf \rho}_1 ) \,
\overline{G}^*( {\bf r} - \demi {\bf s} - {\mbf \rho}_2 ) \,
\varrho( {\mbf \rho}_1 ) \, \varrho^*( {\mbf \rho}_2 ) 
	\nonumber\\
&& = 
\int\!{\rm d}{\bf s} \, \dbar{\bf k}_1 \, \dbar{\bf k}_2 \,
{\rm d}{\mbf \rho}\, 
\overline{G}( {\bf k}_1 ) \, \overline{G}^*( {\bf k}_2 ) \,
f_\varrho[ {\mbf \rho}, \demi({\bf k}_1 + {\bf k}_2 ) ] \times
	\nonumber\\
&& \quad \times
\exp{\rm i}\left[
- {\bf k}\!\cdot\!{\bf s} 
+ \demi ({\bf k}_1 + {\bf k}_2)\!\cdot\!{\bf s} +
 ({\bf k}_1 - {\bf k}_2)\!\cdot\! ({\bf r} - {\mbf \rho}) 
\right] 
	\label{eq:coh-1}
\end{eqnarray}
Nous avons effectu\'e une des deux int\'egrations sur les positions
de la source $\varrho({\mbf \rho})$ pour faire appara\^{\i}tre
sa transform\'ee de {\sc Wigner} 
$f_\varrho[ {\mbf \rho}, \demi({\bf k}_1 + {\bf k}_2 ) ]$.
L'int\'egration sur ${\bf s}$ est maintenant imm\'ediate
et l'\'equation~(\ref{eq:coh-1}) devient 
(${\bf q} = {\bf k}_1 - {\bf k}_2$, ${\bf k} = \demi
({\bf k}_1 + {\bf k}_2 )$) :
\begin{equation}
f_{\rm{coh}}( {\bf r}, {\bf k} ) =
\int\!
{\rm d}{\mbf \rho}\, 
\dbar{\bf q} \, 
\tilde{A}( {\bf q} ; {\bf k} ) \,
f_\varrho( {\mbf \rho}, {\bf k} ) \,
\exp{\rm i} {\bf q}\!\cdot\!({\bf r} - {\mbf \rho}) 
	\label{eq:coh-2}
\end{equation}
avec l'abbr\'eviation suivante pour le produit des
transform\'ees de {\sc Fourier} des fonctions de {\sc Green}
moyenn\'ees :
\begin{equation}
\tilde{A}( {\bf q} ; {\bf k} ) = 
\overline{G}( {\bf k} + \demi{\bf q} )
\overline{G}^*( {\bf k} - \demi{\bf q} )
	\label{eq:def-A}
\end{equation}
Nous constatons donc que le terme coh\'erent relie la
luminance observ\'ee
\`a celle de la source au m\^eme vecteur d'onde ${\bf k}$.
Ecrivons ce r\'esultat encore sous la forme int\'egrale
d'une \'equation de transfert radiatif. Le terme coh\'erent
s'exprime comme une int\'egrale sur les points sources,
pond\'er\'es avec une \flqq fonction d'att\'enuation\frqq\ (la
transform\'ee de {sc Fourier} de $\tilde{A}({\bf q} ; {\bf k} )$) :
\begin{eqnarray}
f_{\rm{coh}}( {\bf r}, {\bf k} ) & = &
\int\!{\rm d}{\mbf \rho} \,
A( {\bf r} - {\mbf \rho} ; {\bf k} ) f_\varrho( {\mbf \rho}, {\bf k} ) ,
	\label{eq:coh-fin}\\
A( {\bf r} - {\mbf \rho} ; {\bf k} ) & = &
\int\!\dbar{\bf q} \tilde{A}( {\bf q} ; {\bf k} ) 
\exp{ {\rm i} {\bf q}\!\cdot\! ({\bf r} - {\mbf \rho}) }
	\label{eq:def-att}
\end{eqnarray}

Pour le deuxi\`eme terme $f_{\rm{diff}}$ dans~%
(\ref{eq:BS-0}), nous obtenons d'abord, par une proc\'edure
similaire, l'expression suivante qui ressemble \`a~(\ref{eq:coh-1}) :
\begin{eqnarray}
&& f_{\rm{diff}}( {\bf r}, {\bf k}) =
\int\!{\rm d}{\bf s}\, \dbar{\bf k}_1 \, \dbar{\bf k}_2 \,
{\rm d}{\mbf \rho} \, \dbar{\mbf \kappa}_1 \, \dbar{\mbf \kappa}_2 \,
	\nonumber\\
&& \quad 
K( {\bf k}_1, {\mbf \kappa}_1 | {\bf k}_2, {\mbf \kappa}_2 ) \,
f[ {\mbf \rho}, \demi({\mbf \kappa}_1 + {\mbf \kappa}_2)] \,
\overline{G}( {\bf k}_1 ) \, \overline{G}^*( {\bf k}_2 ) \times
	\nonumber\\
&& \quad \times
\exp{\rm i}\left[
- {\bf k}\!\cdot\!{\bf s}
+ \demi ({\bf k}_1 + {\bf k}_2)\!\cdot\!{\bf s}
+ ({\bf k}_1 - {\bf k}_2)\!\cdot\!{\bf r} 
- ({\mbf \kappa}_1 - {\mbf \kappa}_2) \!\cdot\!{\mbf \rho}
\right] 
	\label{eq:BS-1}
\end{eqnarray}
Les int\'egrations sur ${\bf s}$ et sur la demi-somme 
${\bf k} = \demi ( {\bf k}_1 + {\bf k}_2 )$ des vecteurs
d'onde s'effectuent de la m\^eme fa\c con 
qu'en~(\ref{eq:coh-1}) pour faire appara\^{\i}tre la
fonction $\tilde{A}( {\bf q} ; {\bf k})$~(\ref{eq:def-A}).
Pour l'int\'egration sur ${\mbf\kappa}_{1,2}$, passons \`a
la mi-somme et la diff\'erence, et utilisons l'invariance
par translation~(\ref{eq:TF-K}) de l'op\'erateur d'intensit\'e.
Nous trouvons ainsi l'expression assez compacte similaire
\`a~(\ref{eq:coh-fin}) (avec ${\mbf \kappa} = \demi
({\mbf \kappa}_1 + {\mbf \kappa}_2 )$) :
\begin{equation}
f_{\rm{diff}}( {\bf r}, {\bf k} ) =
\int\!{\rm d}{\mbf \rho} \, \dbar{\mbf \kappa} \,
p_A( {\bf r} - {\mbf \rho} ; {\bf k}, {\mbf \kappa} ) 
f( {\mbf \rho}, {\mbf \kappa} ) 
	\label{eq:diff-fin}
\end{equation}
o\`u intervient une fonction d'att\'enuation diff\'erente
qui ressemble plut\^ot \`a une fonction de phase 
(avec ${\bf q} = {\bf k}_1 - {\bf k}_2 = {\mbf \kappa}_1 - 
{\mbf \kappa}_2$) :
\begin{eqnarray}
&& p_A( {\bf r} - {\mbf \rho} ; {\bf k}, {\mbf \kappa} ) =
	\nonumber\\
&& \int\!\dbar{\bf q} \,
\exp[ {\rm i} {\bf q}\!\cdot\! ({\bf r} - {\mbf \rho}) ] \,
\tilde{A}( {\bf q} ; {\bf k} ) \,
\tilde{K}\!\left( 
{\bf k} + \demi{\bf q}, {\mbf \kappa} + \demi{\bf q} |
 {\bf k} - \demi{\bf q}, {\mbf \kappa} - \demi{\bf q} 
\right)
	\label{eq:def-phase}
\end{eqnarray}
Ce qui est remarquable, c'est que nous
n'avons encore fait aucune approximation pour d\'eduire
ce r\'esultat.

\subsection{Comparaison \`a l'\'equation de transfert}

Regardons d'abord ce que l'on a pu faire : l'\'equation
de {\sc Bethe--Salpeter} a \'et\'e transform\'ee en :
\begin{equation}
f( {\bf r}, {\bf k} ) =
\int\!{\rm d}{\mbf \rho} \,
A( {\bf r} - {\mbf \rho} ; {\bf k} ) f_\varrho( {\mbf \rho}, {\bf k} ) 
+ \int\!{\rm d}{\mbf \rho} \, \dbar{\mbf \kappa} \,
p_A( {\bf r} - {\mbf \rho} ; {\bf k}, {\mbf \kappa} ) 
f( {\mbf \rho}, {\mbf \kappa} ) .
	\label{eq:ETR*}
\end{equation}
La distribution de {\sc Wigner} (rappelons qu'elle est
\'equivalente \`a la luminance pour la lumi\`ere) 
contient un premier terme
\flqq coh\'erent\frqq\ (\ref{eq:coh-fin})) qui
correspond aux champ moyen rayonn\'e par les sources
dans le milieu. Le deuxi\`eme
terme, l'int\'egrale~(\ref{eq:diff-fin}), doit alors correspondre aux
ondes diffus\'ees par le milieu. Il a bien la structure d'une
int\'egrale de diffusion, o\`u intervient la fonction de {\sc Wigner}
pour un vecteur d'onde ${\mbf \kappa}$ diff\'erent. 
En effet, le vecteur d'onde ${\mbf \kappa}$ peut tr\`es bien
s'interpr\'eter comme un \flqq vecteur d'onde entrant\frqq\
qui est diffus\'e vers le vecteur d'onde ${\bf k}$,
avec une probabilit\'e donn\'ee par l'op\'erateur d'intensit\'e.

Les int\'egrales sur ${\mbf \rho}$ dans les deux termes
de~(\ref{eq:ETR*}) ne sont pas si
surprenantes qu'elles le semblent au premier abord, parce que nous
sommes en train d'\'etudier une formulation int\'egrale,
qu'il faut comparer \`a la version int\'egrale de l'\'equation du
transfert radiatif. En utilisant la forme donn\'ee par
{\sc Chandrasekhar}~(\cite{Chandrasekhar}, Chap.~1, \'eq.(50))
et en ajoutant la luminance $S( {\bf r}, {\bf n} )$ des
sources, la version int\'egrale de l'ETR s'\'ecrit
($I$ est la luminance) :
\begin{equation}
I({\bf r}, {\bf n}) = \int\limits_0^\infty \! {\rm d}s \,
{\rm e}^{- s / \ell} S( {\bf r} - s {\bf n}, {\bf n} ) 
+ \int\limits_{0}^\infty \!\frac{ {\rm d}s }{ \ell } \,
{\rm e}^{- s / \ell} \int\!{\rm d}{\bf n}' \,
p( {\bf n'}, {\bf n}) I({\bf r} - s{\bf n}, {\bf n}')
	\label{eq:ETR-Chandra}
\end{equation}
o\`u ${\bf r} - s{\bf n}$ est un point sur le rayon de direction
${\bf n}$ \`a distance $s$ du point d'observation.
L'int\'egrale sur $s$ exprime le fait que la luminance
en ${\bf r}$ est le produit de tous les \'ev\`enements de diffusion
survenus aux positions ant\'erieures \`a ${\bf r}$
sur ce rayon et qui ont amen\'e la lumi\`ere dans la direction
${\bf n}$. Par ailleurs, le premier terme dans~(\ref{eq:ETR-Chandra})
a bien la forme de la partie \flqq coh\'erente\frqq\ du champ,
qui est att\'enu\'ee par la diffusion sur une distance
caract\'eristique \'egale au libre parcours moyen $\ell$. 

A la diff\'erence du transfert radiatif~(\ref{eq:ETR-Chandra}), 
l'int\'egrale sur ${\mbf\rho}$ dans
l'expression exacte~(\ref{eq:ETR*}) s'\'etend {\em a priori\/}
sur tout l'espace. La fonction de {\sc Wigner} v\'erifie donc
une \'equation de transport {\em non-locale\/}. Pour retrouver
l'ETR standard~(\ref{eq:ETR-Chandra}) qui, elle, est locale, 
c'est l'int\'egrale sur ${\bf q}$
dans la fonction d'att\'enuation
$A( {\bf r} - {\mbf \rho} ; {\bf k} )$~(\ref{eq:def-att})
qui doit limiter l'int\'egrale sur ${\mbf \rho}$ de sorte
qu'elle ne porte plus que sur le rayon
arrivant au point d'observation ${\bf r}$ de la direction 
${\bf k}$. Nous allons maintenant pr\'eciser les conditions
pour qu'apparaisse une telle simplification.

\subsection{Solution approch\'ee \`a grandes distances}

\subsubsection{Astuce}

Transformons d'abord le produit des deux d\'enominateurs dans~(%
\ref{eq:def-A})
selon la formule suivante (\'evidente, mais astucieuse\ldots)
\begin{eqnarray}
\lefteqn{
\tilde{A}( {\bf q} ; {\bf k} ) = 
\overline{G}( {\bf k} + \demi{\bf q} )
\overline{G}^*( {\bf k} - \demi{\bf q} )
}
	\nonumber\\
&& = \frac{ 1 }{
[ k_0^2 - ({\bf k} + \demi{\bf q})^2 - m({\bf k} + \demi{\bf q}) ]
[ k_0^2 - ({\bf k} - \demi{\bf q})^2 - m^*({\bf k} - \demi{\bf q}) ] } 
	\nonumber\\
&& = \frac{ 
\overline{G}( {\bf k} + \demi{\bf q} ) - 
\overline{G}^*( {\bf k} - \demi{\bf q} )
 }{ 2 {\bf k}\!\cdot\!{\bf q} + m({\bf k} + \demi{\bf q})  
- m^*({\bf k} - \demi{\bf q}) } 
	\label{eq:astuce}
\end{eqnarray}
Cette expression est encore exacte. 
Si nous cherchons maintenant
le comportement de la distribution de {\sc Wigner}
\`a des \'echelles $|{\bf r} - {\mbf \rho}|$ grandes par rapport
\`a la longueur d'onde $\lambda_{dB}$,
l'int\'egrale~(\ref{eq:def-att}) montre que l'on peut se contenter
des vecteurs d'onde ${\bf q}$ tr\`es petits par rapport aux
vecteurs d'onde atomiques ${\bf k}, {\mbf \kappa}$. 
L'id\'ee vient alors \`a l'esprit de faire un d\'eveloppement
limit\'e en fonction de ${\bf q}$. Nous allons le faire ici
\`a l'ordre le plus bas ; l'on verra que l'on retrouve alors 
la forme~%
(\ref{eq:ETR-Chandra}) de l'\'equation du transfert radiatif.
Au paragraphe~\ref{s:echelle-min} suivant, nous irons un ordre plus
loin pour obtenir une premi\`ere correction par rapport \`a
l'\'equation de transfert.

A l'ordre le plus bas en ${\bf q}$, la formule~(\ref{eq:astuce}) 
devient 
\begin{equation}
\tilde{A}( {\bf q} ; {\bf k} ) \approx
\frac{ - 2 \pi {\rm i} 
\delta( k^2 - {\rm{Re}}\, k_{\rm{eff}}^2 ) }{
2 {\bf k}\!\cdot\!{\bf q} - 2 {\rm i} \,\mbox{Im}\, k_{\rm{eff}}^2 } ,
	\label{eq:simplification-q-petit}
\end{equation}
o\`u nous avons utilis\'e le vecteur d'onde effectif (complexe)
pour \'ecrire le p\^ole du d\'enominateur :
\begin{equation}
m( {\bf k} \pm\demi{\bf q} ) \approx m( {\bf k} )
\approx k_0^2 - k_{\rm{eff}}^2
	\label{eq:approx-masse}
\end{equation}
Nous avons \'egalement approxim\'ee la partie imaginaire
de la fonction de {\sc Green} moyenn\'ee (le terme au num\'erateur
dans~(\ref{eq:astuce})) par une fonction $\delta$.
Celle-ci exprime le fait que le module des vecteurs d'onde atomiques
finaux ${\bf k}$ (apr\`es le processus de diffusion) est fix\'e par
la relation de dispersion dans le milieu (la partie r\'eelle
du vecteur d'onde effectif $k_{\rm{eff}}$). Cette approximation
est justifi\'ee dans le r\'egime de faible diffusion o\`u le
libre parcours moyen est beaucoup plus grand que la longueur
d'onde. Une telle situation correspond \`a l'image intuitive que
l'on se fait de l'ETR. 
La limite oppos\'ee correspond \`a des ondes localis\'ees
dans le milieu al\'eatoire, un r\'egime qui demande une
description au-del\`a de l'ETR standard.

\subsubsection{Retrouver l'ETR diff\'erentielle}

En reportant~(\ref{eq:simplification-q-petit}) dans
l'ETR et en n\'egligeant le petit vecteur d'onde ${\bf q}$ dans
l'op\'erateur d'intensit\'e, nous observons que 
l'ETR~(\ref{eq:ETR*}) se r\'esoud facilement par une
transform\'ee de {\sc Fourier} par rapport \`a la position
${\bf r}$. Nous notons $\tilde{f}( {\bf q}, {\bf k} )$ la
fonction de {\sc Wigner} transform\'ee, et elle v\'erifie
l'\'equation suivante
\begin{eqnarray}
&&
\left(
{\rm i}{\bf k}\!\cdot\!{\bf q} + \mbox{Im}\, k_{\rm{eff}}^2  
\right) \tilde{f}( {\bf q}, {\bf k} ) =
	\nonumber\\
&& \quad
\pi \delta( k^2 - \mbox{Re}\, k_{\rm{eff}}^2 )
\left( 
\tilde{f}_{\varrho}( {\bf q}, {\bf k} ) +
\int\dbar{\mbf \kappa} \,
\tilde{K}( {\bf k}, {\mbf \kappa} | {\bf k}, {\mbf \kappa} )
\tilde{f}( {\bf q}, {\mbf \kappa} )
\right)
	\label{eq:ETR*-diff}
\end{eqnarray}
En revenant dans l'espace directe, nous trouvons donc la forme
diff\'erentielle de l'ETR, dont le libre parcours moyen et
la fonction de phase sont donn\'es par
\begin{eqnarray}
\frac{ 1 }{ \ell } &=& \frac{ \mbox{Im}\,k_{\rm eff}^2 }{ k_p }
= - \frac{ \mbox{Im}\, m(k_p) }{ k_p }
	\label{eq:ell-ETR*}\\
p( {\bf u}, {\bf u}' ) & = &
\frac{ 1 }{ ( 4\pi k_p )^2 } 
\tilde{K}( k_p {\bf u}, k_p {\bf u}' | k_p {\bf u}, k_p {\bf u}' )
	\label{eq:phase-ETR*}\\
k_p & = & ( \mbox{Re}\, k_{\rm eff}^2 )^{1/2}
	\label{eq:def-k-p}
\end{eqnarray}
Nous avons donc trouv\'e ici des expressions microscopiques 
pour les param\`etres qui interviennent dans l'ETR. Des formules
identiques ont \'et\'es \'ecrites par {\sc Barabanenkov} et 
{\sc Finkel'berg} (eq.~(22) de \cite{Barabanenkov67}), mais
en utilisant une forme approch\'ee pour les op\'erateurs
de masse et d'intensit\'e (approximation de {\sc Bourret}). 
La d\'emarche pr\'esent\'ee ici montre
que ces expressions restent valables dans un contexte plus
g\'en\'eral, \`a condition que les ondes ne soient pas
localis\'ees dans le milieu. La seule diff\'erence par
rapport aux r\'esultats
de {\sc Barabanenkov} et {\sc Finkel'berg} est l'apparition
du vecteur d'onde $k_p$ qui d\'ecrit l'indice de r\'efraction
du milieu al\'eatoire. A la limite perturbative, il faut
en effet prendre $k_p = k_0$ pour \^etre consistent avec
la conservation de l'\'energie (l'identit\'e de {\sc Ward},
voir \cite{vanTiggelen96}).

\subsubsection{Retrouver l'ETR int\'egrale}

Nous pouvons \'egalement retrouver la formulation int\'egrale
de l'ETR donn\'ee en~(\ref{eq:ETR-Chandra}). A cet effet,
nous calculons l'int\'egrale sur ${\bf q}$ dans~(\ref{eq:ETR*})
dans l'approximation~(\ref{eq:simplification-q-petit}) pour
la fonction d'att\'enuation $\tilde{A}( {\bf q}; {\bf k} )$.
L'int\'egrale a d'ailleurs \'et\'e calcul\'ee au chapitre pr\'ec\'edent
(voir la note~\ref{fn:delta-transverse} en bas de
la page~\pageref{fn:delta-transverse}):
\begin{eqnarray}
\lefteqn{
\int\!\dbar{\bf q} \frac{
\exp{\rm i}{\bf q}\!\cdot\!( {\bf r} - {\mbf \rho}) }{
2 {\bf k}\!\cdot\!{\bf q} - 2 {\rm i} \,\mbox{Im}\, k_{\rm{eff}}^2 }
= }
	\label{eq:retrouver-Chandra}\\
&& = \left\{
\begin{array}{l}
( {\rm i} / 2 k ) 
\delta\!\left( {\bf r}_\perp - {\mbf \rho}_\perp \right)
\exp[ - ( \mbox{Im}\, k_{\rm{eff}}^2 / k^2 ) 
{\bf k}\!\cdot\!({\bf r} - {\mbf \rho}) ] ,
	\\
\begin{array}[b]{ll}
&\mbox{lorsque } 
{\bf k} \!\cdot\! ( {\bf r} - {\mbf \rho} ) > 0 ;
\\
0 , & \mbox{lorsque } 
{\bf k} \!\cdot\! ( {\bf r} - {\mbf \rho} ) < 0 .
\end{array}
\end{array}
\right.
	\nonumber
\end{eqnarray}
(L'indice \flqq $\perp$\frqq\ fait r\'ef\'erence au
vecteur ${\bf k}$.) Nous en concluons que l'int\'egrale sur ${\mbf \rho}$
ne porte en fait que sur un demi-rayon qui aboutit au point
d'observation ${\bf r}$ en se propageant le long de la direction 
${\bf k}$. Le long de ce rayon, l'intensit\'e du champ est
att\'enu\'ee avec une longueur caract\'eristique 
\begin{equation}
\ell = \frac{ k }{ \mbox{Im } k_{\rm{eff}}^2 }
	\label{eq:def-ell}
\end{equation}
ce qui correspond bien \`a l'att\'enuation du champ coh\'erent
par diffusion. Le terme coh\'erent s'\'ecrit donc comme une int\'egrale
sur les sources \flqq intercept\'ees\frqq\ le long de ce rayon,
qui rayonnent dans la direction $\hat{\bf k}$ :
\begin{equation}
f_{\rm{coh}}( {\bf r}, {\bf k} ) =
\frac{ \pi }{ k } \,
\delta( k^2 - \mbox{Re}\, k_{\rm{eff}}^2 ) 
\int\limits_0^\infty\!{\rm d}s' \, 
{\rm e}^{ - s' / \ell } \,
f_\varrho( {\bf r} - s' \hat{\bf k}, {\bf k} )
	\label{eq:coh-retrouve}
\end{equation}
Nous retrouvons ici le premier terme de l'\'equation du transfert
radiatif de {\sc Chandrasekhar}~(\ref{eq:ETR-Chandra}).
La fonction $\delta$ assure que les seuls vecteurs d'onde de la
source qui rayonnent sont ceux qui correspondent \`a la relation
de dispersion dans le milieu al\'eatoire.
(Rappelons encore que la source $\varrho( {\bf r} )$ 
dans~(\ref{eq:champ-rayonne})
et (\ref{eq:coh-retrouve}) n'a pas la m\^eme dimension que le champ,
ceci est \`a l'origine du pr\'efacteur.)

Quant au deuxi\`eme terme de~(\ref{eq:ETR-Chandra}), 
qui d\'ecrit la diffusion, nous 
remarquons qu'\`a l'ordre le plus bas, 
l'op\'erateur d'intensit\'e dans~(\ref{eq:def-phase})
ne d\'epend pas du vecteur d'onde ${\bf q}$ :
\begin{equation}
\tilde{K}\!\left( 
{\bf k} + \demi{\bf q}, {\mbf \kappa} + \demi{\bf q} |
 {\bf k} - \demi{\bf q}, {\mbf \kappa} - \demi{\bf q} 
\right) \approx
\tilde{K}\!\left( 
{\bf k}, {\mbf \kappa} | {\bf k}, {\mbf \kappa} \right) 
	\label{eq:approx-intensite}
\end{equation}
On peut alors le sortir des int\'egrales sur ${\bf q}$ et ${\mbf \rho}$
et d\'efinir une fonction de phase locale comme suit
(attention,
sa normalisation est diff\'erente de~(\ref{eq:phase-ETR*}))
\begin{eqnarray}
p_A( {\bf r} - {\mbf \rho} ; {\bf k}, {\mbf \kappa} )
& = &
A( {\bf r} - {\mbf \rho} ; {\bf k} ) \, p( {\bf k}, {\mbf \kappa} )
	\nonumber\\
p( {\bf k}, {\mbf \kappa} ) & = & \frac{ \pi }{ k }
\delta( k^2 - {\rm{Re}}\, k_{\rm{eff}}^2 ) \,
\tilde{K}\!\left( {\bf k}, {\mbf \kappa} | {\bf k}, {\mbf \kappa} \right) .
	\label{eq:fn-phase-generale}
\end{eqnarray}
En utilisant le r\'esultat~(\ref{eq:retrouver-Chandra})
pour la fonction d'att\'enuation $A( {\bf r} - {\mbf \rho} ;
{\bf k})$, le terme de diffusion~(\ref{eq:diff-fin}) prend donc
la forme
\begin{equation}
f_{\rm{diff}}( {\bf r}, {\bf k} ) = 
\int\limits_0^\infty\! {\rm d}s' \,
{\rm e}^{- s' / \ell} \,
\int\!\dbar{\mbf \kappa} \,
p( {\bf k}, {\mbf \kappa} ) \, 
f( {\bf r} - s' \hat{\bf k}, {\mbf \kappa} )
	\label{eq:diff-retrouve}
\end{equation}
Puisque la relations de dispersion fixe le module du vecteur
d'onde ${\mbf \kappa}$, le terme de diffusion 
est exactement celui donn\'e par {\sc Chandrasekhar} en~%
(\ref{eq:ETR-Chandra}). (La fonction de 
phase~(\ref{eq:fn-phase-generale}) n'est pas normalis\'ee 
de la m\^eme fa\c con que celle de {\sc Chandrasekhar}, ceci
explique la diff\'erence entre les pr\'efacteurs.)

Nous constatons donc qu'aux grandes \'echelles spatiales, le
transport de la distribution de {\sc Wigner} des atomes (ou
du champ \'electro-magn\'etique) est r\'egi par une \'equation
de transport radiatif. Ceci est vrai quelle que soit la force
du potentiel al\'eatoire. Pour un potentiel faible, 
l'approximation de {\sc Bourret}~(\ref{eq:K-Bourret})
de l'op\'erateur d'intensit\'e montre que la fonction de 
phase est proportionnelle \`a la densit\'e spectrale
du potentiel al\'eatoire
\begin{eqnarray}
\mbox{{\sc Bourret}}: &&
\tilde{K}\!\left( 
{\bf k} + \demi{\bf q}, {\mbf \kappa} + \demi{\bf q} |
 {\bf k} - \demi{\bf q}, {\mbf \kappa} - \demi{\bf q} 
\right) =
	\nonumber\\
&& \quad = \epsilon^2 k_0^4 \, g( {\bf k} - {\mbf \kappa} ) 
= \frac{ 4 M^2 }{ \hbar^4 } S( {\bf k} - {\mbf \kappa} ) 
	\label{eq:K(k)-Bourret}
\end{eqnarray}
Nous retrouvons alors l'approche de {\sc Ryzhik} {\em et al.\/},
ainsi que l'id\'ee que s'est faite {\sc Luck} de l'\'equation
de transfert. La d\'emonstration que nous venons de donner
de cette \'equation montre cependant qu'elle est \'egalement
valable pour des potentiels plus forts o\`u l'approximation
de {\sc Bourret} n'est plus possible. 

\subsubsection{Conclusion}

Nous avons constat\'e qu'aux grandes \'echelles spatiales,
la th\'eorie de la diffusion multiple des ondes prend la
forme d'une \'equation du transfert radiatif. Ce r\'esultat
repose seulement sur l'hypoth\`ese d'un milieu invariant
par translation et de la diffusion faible (libre parcours
moyen beaucoup plus grand que la longueur d'onde). Dans ce
r\'egime, la transform\'ee de {\sc Wigner} de la fonction
de corr\'elation d\'ecrit la luminance du champ, et elle
v\'erifie une ETR~(\ref{eq:ETR*-diff}) dont les param\`etres 
macroscopique s'expriment \`a l'aide des op\'erateurs de masse 
et d'intensit\'e.

Nous avons obtenu l'ETR ainsi g\'en\'eralis\'ee par 
un d\'eveloppement \`a l'ordre le plus bas par rapport
au vecteur d'onde ${\bf q}$, la variable conjugu\'ee
au profil spatial en ${\bf r}$ de la luminance. Cette proc\'edure
(la \flqq limite de {\sc Kubo}\frqq\ \cite{vanTiggelen96})
traduit formellement la limite d'une \flqq grande \'echelle
spatiale\frqq. Au paragraphe suivant, nous allons 
pousser le d\'eveloppement jusqu'au second ordre en ${\bf q}$,
afin de pr\'eciser l'\'echelle spatiale minimale au-del\`a
de laquelle l'\'equation du transfert radiatif est 
justifi\'ee.

\subsection[Echelle spatiale minimale pour l'ETR]{%
Echelle spatiale minimale{\protect \\}
pour l'\'equation du transfer radiatif}
\label{s:echelle-min}

Dans le calcul de la fonction de {\sc Wigner} du champ aux grandes
distances, nous avons pour l'instant seulement retenu les
termes \`a l'ordre le plus bas en ${\bf q}$. Nous allons ici
aller plus loin et analyser les termes de l'ordre suivant
pour trouver une limite sup\'erieure $q_{\max}$ \`a ce
d\'eveloppement. Dans l'espace r\'eel,
la valeur $q_{\max}$ se traduit par une \'echelle spatiale
minimale $L_{\min} = 1 / q_{\max}$, avec la signification
physique la suivante : il faut que la fonction de {\sc Wigner}
du champ, en fonction de la position ${\bf r}$,
varie peu \`a l'\'echelle $L_{\min}$ pour que 
l'\'equation du transfert radiatif soit valable. 

Nous allons nous contenter d'\'etudier la d\'ependance de 
${\bf q}$ de la fonction d'att\'enuation 
$\tilde{A}( {\bf q} ; {\bf k} )$~(\ref{eq:def-att}). 
Quant \`a la variation de l'op\'erateur d'intensit\'e 
$\tilde{K}$~(\ref{eq:approx-intensite})
avec ${\bf q}$, l'on en trouvera une discussion dans
l'article de revue de {\sc Kravtsov} et {\sc Apresyan}
(\S~6.1 de \cite{Kravtsov96}) : 
elle peut \^etre reli\'ee \`a l'effet
de m\'emoire de la figure de tavelures lorsque l'on fait
varier la direction d'incidence du faisceau.
Nous notons \'egalement
que pour un potentiel al\'eatoire faible (approximation de
{\sc Bourret}), la formule~(\ref{eq:K(k)-Bourret}) montre
que $\tilde{K}$ est ind\'ependant de ${\bf q}$. Un moyen
d'appr\'ecier sa d\'ependance de ${\bf q}$ est alors de calculer des termes
sup\'erieurs~(\ref{eq:operateur-K}) qui vont au-del\`a de 
l'approximation de {\sc Bourret}. 

Poussons donc le d\'eveloppement de la fonction d'att\'enuation~%
$\tilde{A}( {\bf q} ; {\bf k} )$ jusqu'\`a
l'ordre quadratique. Nous allons n\'egliger la variation
avec ${\bf q}$ de la diff\'erence des fonctions de {\sc Green}
\[
\overline{G}( {\bf k} + \demi{\bf q}) 
- \overline{G}^*( {\bf k} - \demi{\bf q}) 
\approx - 2 \pi {\rm i} 
\delta( k^2 - {\rm{Re}}\,k_{\rm{eff}}^2 )
\]
et seulement analyser le d\'enominateur dans~(\ref{eq:astuce})
qui pr\'esente un p\^ole en ${\bf q}$. A l'ordre quadratique,
celui-ci devient :
\begin{eqnarray}
\lefteqn{
2 {\bf k}\!\cdot\!{\bf q} + m({\bf k} + \demi{\bf q})  
- m^*({\bf k} - \demi{\bf q}) \approx
}
	\nonumber\\
&& 
2 {\bf k}\!\cdot\!{\bf q} - 2 {\rm i} \,{\rm{Im}}\, k_{\rm{eff}}^2
+ {\bf q}\!\cdot\!{\rm{Re}}\,\nabla_k m
+ \frac{ {\rm i} }{ 4 } \sum_{ij} q_i q_j \,{\rm{Im}}\,
\frac{ \partial^2 m }{ \partial k_i \partial k_j }
	\label{eq:pole-q2}
\end{eqnarray}
Les d\'eriv\'ees de l'op\'erateur de masse $m( {\bf k})$
sont prises ici pour ${\bf k}$ v\'erifiant la relation de 
dispersion. 
Pour continuer le calcul, nous allons faire l'hypoth\`ese
suppl\'ementaire suivante :
\begin{itemize}
\item
le milieu al\'eatoire est statistiquement isotrope.
La relation de dispersion s'\'ecrit alors $k = k_{\rm{eff}}$
et l'op\'erateur de masse $m( {\bf k})$ d\'epend 
seulement du module $k$ du vecteur d'onde ${\bf k}$.
\end{itemize}
Par exemple, le milieu \`a fonction de corr\'elation gaussienne
dont nous avons donn\'e l'op\'erateur de masse en~%
(\ref{eq:m(k)-gaussien}), est isotrope.
Par cons\'equent, les d\'eriv\'ees de l'op\'erateur
de masse de simplifient :
\begin{eqnarray}
\nabla_k m  & = &  \hat{\bf k} \frac{ {\rm d} m }{ {\rm d}k }
	\nonumber\\
\frac{ \partial^2 m }{ \partial k_i \partial k_j }
& = &
\delta_{ij} \left( \frac{ 1 }{ k }
\frac{ {\rm d} m }{ {\rm d}k } \right)
+
\hat{k}_i \hat{k}_j \left(
\frac{ {\rm d}^2 m }{ {\rm d}k^2 } -
\frac{ 1 }{ k } \frac{ {\rm d} m }{ {\rm d}k } \right)
	\label{eq:dm-dk}
\end{eqnarray}
Le d\'enominateur~(\ref{eq:pole-q2}) peut alors \^etre \'ecrit
sous la forme suivante :
\begin{eqnarray}
&&
2 {\bf k}\!\cdot\!{\bf q} + m({\bf k} + \demi{\bf q})  
- m^*({\bf k} - \demi{\bf q}) \approx
2 {\bf k}'\!\cdot\!{\bf q} - 
{\rm i} \gamma 
+ {\rm i} \alpha {\bf q}^2
- {\rm i} \beta ( {\bf k}\!\cdot\!{\bf q} )^2 
	\label{eq:pole-q2-2}\\
&&\qquad {\bf k}' = {\bf k} \left( 1 + \frac{ 1 }{ 2 k }
{\rm{Re}}\, \frac{ {\rm d} m }{ {\rm d}k } \right)
	\label{eq:k-prime}\\
&&\qquad  \gamma = - 2 \, {\rm Im}\, m( k_{\rm eff} ) = 
2 \,{\rm{Im}}\, k_{\rm{eff}}^2
	\label{eq:gamma}\\
&&\qquad  \alpha = 
{\rm{Im}}\left( \frac{ 1 }{ 4 k } \frac{ {\rm d} m }{ {\rm d}k } \right)
	\label{eq:alpha}\\
&&\qquad  \beta = 
\frac{ 1 }{ 4 k^2 } {\rm{Im}}
\left(
\frac{ 1 }{ k } \frac{ {\rm d} m }{ {\rm d}k } 
- \frac{ {\rm d}^2 m }{ {\rm d}k^2 } 
\right)
	\label{eq:beta}
\end{eqnarray}
La d\'eriv\'ee premi\`ere de l'op\'erateur de masse
change donc simplement le module du vecteur d'onde ${\bf k} 
\mapsto {\bf k}'$. Le d\'eriv\'ee seconde ajoute des termes
quadratiques en ${\bf q}$. 

Avant de continuer le calcul, donnons une estimation jusqu'\`a
quel vecteur d'onde maximal ce d\'eveloppement est justifi\'e.
La figure~\ref{fig:masse} et l'expression~(\ref{eq:m(k)-gaussien})
montrent que l'op\'erateur de masse $m( {\bf k})$
varie sur une \'echelle caract\'eristique de $1 / \ell_c$
en fonction de $k$. Nous trouvons donc la condition de 
validit\'e
\begin{equation}
| {\bf q} | \lesssim \frac{ 1 }{ \ell_c }
	\label{eq:condition-q-petit}
\end{equation}
Dans l'espace r\'eel, cette condition traduit le fait que dans
le formalisme du transfert radiatif, la luminance ne r\'esout
pas la structure microscopique (longueur de corr\'elation)
du milieu diffusant.

\subsubsection{Calcul de la fonction d'att\'enuation}

Nous avons donc l'int\'egrale suivante \`a calculer :
\begin{eqnarray}
\lefteqn{
\int\!\dbar{\bf q} \,
\frac{ \exp{\rm i} {\bf q}\!\cdot\!{\bf r} }{
2 {\bf k}'\!\cdot\!{\bf q} - {\rm i} \gamma 
+ {\rm i} \alpha {\bf q}^2
- {\rm i} \beta ( {\bf k}\!\cdot\!{\bf q} )^2 }
}        \nonumber\\
&& = \int\!\dbar{\bf Q} \, \dbar q \,
\frac{ \exp{\rm i} \left( {\bf Q}\!\cdot\!{\bf R} + q z \right) }{
2 k' q - {\rm i} \gamma 
+ {\rm i} \alpha Q^2 
+ {\rm i} ( \alpha - \beta k^2 ) q^2 } .
	 \label{eq:integrale-q}
\end{eqnarray}
Nous avons d\'ecompos\'e les vecteurs ${\bf q}$ et
${\bf r}$ en composantes $\left( {\bf Q}, q \right), \,
\left( {\bf R}, z \right)$ perpendiculaires
et parall\`eles au vecteur ${\bf k}'$. La coordonn\'ee $z$
mesure donc la distance du point d'observation d'un point source 
situ\'e sur le rayon qui arrive de la direction $\hat{\bf k}$
au point d'observation.

Il se trouve que l'on peut discuter de fa\c con analytique
le comportement de l'int\'egrale~(\ref{eq:integrale-q}).
Nous prenons d'abord l'int\'egration sur la composante
longitudinale $q$, en utilisant le th\'eor\`eme des
r\'esidus (nous avons abr\'eg\'e $\bar{\gamma} = 
\gamma - \alpha Q^2, \,
\bar{\alpha} = \alpha - \beta k^2$) :
\begin{equation}
\int\!\dbar q \, 
\frac{ \exp{\rm i} q z }{
2 k' q - {\rm i} \bar{\gamma} + {\rm i} \bar{\alpha} q^2 }
= \frac{ {\rm{sgn}} \, z }{ \bar{\alpha} ( q_1 - q_2 ) } 
\left[
\Theta(z \mbox{Im}\, q_1 )
\, {\rm e}^{ {\rm i} q_1 z }
- 
\Theta(z \mbox{Im}\, q_2 )
\, {\rm e}^{ {\rm i} q_2 z }
\right]
	\label{eq:solution-residus}
\end{equation}
o\`u $\Theta(z)$ est la fonction de marche d'escalier et
les deux quantit\'es imaginaires $q_{1,2}$ sont les
solutions d'une \'equation quadratique
\begin{eqnarray}
&& 2 k' q - {\rm i} \bar{\gamma} + {\rm i} \bar{\alpha} q^2 = 0
	\nonumber\\
&& q_{1,2} = \frac{ {\rm i} k' }{ \bar{\alpha} } 
\left( 1 \mp \sqrt{ 1 - \frac{ \bar{\gamma} \bar{\alpha} }{
k'^2 } } \right)
	\label{eq:q-12}
\end{eqnarray}
(Les signes sont choisis tels que $|q_1| < |q_2|$.)

Pour interpr\'eter et simplifier ce r\'esultat, 
supposons que le potentiel
al\'eatoire est faible (approximation de {\sc Bourret}). 
L'op\'erateur de masse $m( {\bf k} )$ est alors petit 
par rapport \`a $k_0^2$, et le
vecteur d'onde effectif $k_{\rm{eff}} = k_{\rm p} + 
{\rm i} / 2 \ell$ diff\`ere peu du vecteur 
d'onde dans le vide. Par cons\'equent, nous avons
\begin{equation}
|\gamma| = | - 2\,{\rm{Im}}\,m( k_{\rm{eff}} ) | 
\ll k^2 \simeq k'^2 ,
\qquad
| \alpha |,\, | \bar{\alpha} | \ll 1 , 
	\label{eq:gamma-petit}
\end{equation}
et nous pouvons d\'evelopper la racine carr\'e dans~(\ref{eq:q-12})
pour trouver :
\begin{equation}
q_1 \approx \frac{ {\rm i} \bar{\gamma} }{ 2 k' } 
\qquad
q_2 \approx \frac{ 2 {\rm i} k' }{ \bar{\alpha} }
	\label{eq:deux-poles}
\end{equation}
Le premier terme dans~(\ref{eq:solution-residus})
qui correspond \`a la racine $q_1$, repr\'esente donc
une att\'enuation exponentielle le long de la direction
du vecteur ${\bf k}$ 
\begin{equation}
{\rm{sgn}}\,q_1\,
\Theta( z \mbox{Im}\,q_1 )\, {\rm e}^{ {\rm i} q_1 z } \approx
{\rm{sgn}}\,\bar{\gamma}\,
\Theta( z \bar{\gamma} )\, {\rm e}^{ - z / \ell' } ,
\quad \frac{ 1 }{ \ell' } = \frac{ \gamma }{ 2 k' }
= \frac{ {\rm{Im}}\, k_{\rm{eff}}^2 }{ k' }
\: \mbox{pour ${\bf Q} = {\bf 0}$.}
	\label{eq:racine-normale}
\end{equation}
Le libre parcours moyen qui appara\^{\i}t ici est modifi\'e
par rapport au r\'esultat~(\ref{eq:def-ell})
parce que  $k' \ne k$, \`a cause du terme proportionnel \`a 
$\nabla_k m$ dans~(\ref{eq:k-prime}).

La deuxi\`eme racine $q_2$ donne dans cette approximation
\begin{equation}
\Theta(\bar{\alpha} z)\, {\rm e}^{ {\rm i} q_2 z } \approx
\Theta(\bar{\alpha} z)\, {\rm e}^{ - |z| / \ell_2 } ,
\quad \ell_2 = \frac{ | \bar{\alpha} | }{ 2 k' }
	\label{eq:racine-bizarre}
\end{equation}
elle correspond donc \`a une att\'enuation tr\`es rapide,
sur une \'echelle en dessous de la longueur d'onde
dans le milieu. En outre, suivant le signe de la quantit\'e
$\bar{\alpha}$, l'expression~(\ref{eq:racine-bizarre})
est non nulle pour des distances $z$ positives ou negatives,
donc pour des sources en amont ou en aval du point d'observation.
Dans l'approximation de {\sc Bourret}, l'on s'attend \`a
\[
\bar{\alpha} =  
\frac14 {\rm{Im}}\, \frac{ {\rm d}^2 m }{ {\rm d}k^2 } > 0
\]
parce que la partie imaginaire de l'op\'erateur de masse
(reli\'ee au libre parcours moyen) pr\'esente un minimum
autour de $k \approx k_0$ (voir la figure~\ref{fig:masse}).
L'expression~(\ref{eq:racine-bizarre}) porte alors sur les
positions $z$ positives (des sources en amont du point
d'observation).

Il est justifi\'e de n\'egliger
cette deuxi\`eme contribution \`a la fonction d'att\'enuation.
Nous cherchons en effet
la port\'ee la plus grande en fonction de la distance du
point d'observation, et la premi\`ere racine donne 
une port\'ee beaucoup plus grande que la deuxi\`eme.%
\footnote{%
Nous notons aussi que {\sc Rytov} a calcul\'e la fonction
de {\sc Green} moyenne par une technique analogue
(\'eq.~(4.54) de~\cite{Rytov}). 
Il trouve \'egalement deux contributions,
dont l'une s'att\'enue \`a l'\'echelle du libre parcours
moyen et l'autre a une port\'ee de l'ordre de la
longueur de corr\'elation $\ell_c$. Il n\'eglige ensuite cette
derni\`ere contribution, en se pla\c cant \`a une
\'echelle spatiale plus grande que $\ell_c$.}
Si nous nous pla\c cons \`a une \'echelle spatiale
plus grande que la longeur d'onde, nous pouvons donc
n\'egliger la contribution de la racine $q_2$. 
Le r\'esultat de l'int\'egrale~(\ref{eq:solution-residus})
sur le vecteur d'onde longitudinal $q$ est alors :
\begin{equation}
\int\!\dbar q \,
\frac{ \exp{\rm i} q z }{
2 k' q - {\rm i} \gamma 
+ {\rm i} \alpha Q^2 
+ {\rm i} \bar{\alpha} q^2 }
=
\frac{ 2 {\rm i} k' \, {\rm{sgn}}\,z \,
\Theta[ z ( \gamma - \alpha Q^2 ) ] }{
4 k'^2 - \bar{\alpha} ( \gamma - \alpha Q^2 ) }
\exp[ - ( \gamma - \alpha Q^2 ) z / 2 k' ] 
	 \label{eq:resultat-integrale-q}
\end{equation}
o\`u nous avons r\'e-exprim\'e $\bar{\gamma}$ en fonction de
${\bf Q}$. Nous notons que la fonction $\Theta$ assure que
l'exponentielle est toujours d\'ecroissante en fonction de
$|z|$.

Il reste maintenant \`a effectuer l'int\'egration sur le
vecteur d'onde transverse ${\bf Q}$. L'int\'egrale prend la
forme suivante :
\begin{equation}
2 {\rm i} k' \, {\rm sgn}\, z
\int\!\dbar{\bf Q}
\frac{ 
\Theta[ z ( \gamma - \alpha Q^2 ) ] \, 
{\rm e}^{{\rm i} {\bf Q}\cdot{\bf R} }
}{ 4 k'^2 - \bar{\alpha} ( \gamma - \alpha Q^2 ) }
\exp[ - ( \gamma - \alpha Q^2 ) z / 2 k' ]
	\label{eq:integrale-Q}
\end{equation}
Il convient de distinguer
entre les cas $z > 0$ et $z < 0$ (positions sur le rayon
lumineux ant\'erieures et post\'erieures au point d'observation).

\paragraph[Le cas causal.]{Le cas \flqq causal\frqq\ $z > 0$.}
Le module du vecteur d'onde ${\bf Q}$ est limit\'e \`a l'intervalle
$ 0 \le Q \le ( \gamma / \alpha )^{1/2}$. Nous effectuons 
l'int\'egration sur l'angle azimuthal du vecteur ${\bf Q}$.
En utilisant une nouvelle variable d'int\'egration, l'int\'egrale
s'\'ecrit
\begin{equation}
\frac{ {\rm i} }{ 8 \pi k' }
\int\limits_0^{\gamma / \alpha } \!
{\rm d}s \,
J_0( \sqrt{ \gamma / \alpha - s } R ) \,
{\rm e}^{ - \alpha s z / 2 k' } ,
\quad s = \gamma / \alpha - Q^2 
	\label{eq:integrale-causale}
\end{equation}
Nous avons \'egalement n\'eglig\'e le terme proportionnel \`a
$\bar{\alpha} \alpha Q^2$ au d\'enominateur 
de~(\ref{eq:integrale-Q}). Comme $Q^2$ est limit\'e par
$\gamma / \alpha$, ce terme est petit par rapport \`a
$k'^2$ en vertu de l'hypoth\`ese~(\ref{eq:gamma-petit})
que nous avons d\'ej\`a utilis\'ee ci-dessus (pour calculer
$q_{1,2}$). 

Il est facile maintenant de d\'eterminer l'\'echelle
caract\'eristique pour la distance transverse $R$. A partir
de l'argument de la fonction de {\sc Bessel} $J_0$, nous
obtenons :
\begin{equation}
\delta R \simeq \sqrt{ \frac{ \alpha }{ \gamma } }
\sim
\sqrt{ \frac{ \ell_c / \ell }{ 2 k' / \ell } }
\simeq \sqrt{ \demi \ell_c \lambdabar }
	\label{eq:echelle-transverse}
\end{equation}
(Nous avons utilis\'e l'\'echelle de variation $1 / \ell_c$
de l'op\'erateur de masse pour estimer $\alpha \sim
\ell_c (k_0 / \ell) / k_0 = \ell_c / \ell$.)
Avant de donner une interpr\'etation physique de ce r\'esultat,
\'etudions le deuxi\`eme cas.

\paragraph[Le cas acausal.]{Le cas \flqq acausal\frqq\ $z < 0$.}
Le nom \flqq acausal\frqq\ provient du fait, rappelons-le,
que les positions $z < 0$ se trouvent en amont du point 
d'observation : elles d\'ecrivent donc l'influence sur
la luminance du milieu que le rayon n'a pas encore travers\'e
d'un point de vue g\'eom\'etrique.

L'int\'egrale~(\ref{eq:integrale-Q}) porte maintenant sur
les vecteurs d'onde $ ( \gamma / \alpha )^{1/2} \le
Q \le \infty$. Avec une changement de variables similaire,
elle prend la forme
\begin{equation}
- \frac{ {\rm i} }{ 8 \pi k' }
\int\limits_0^{\infty} \!
{\rm d}s \,
J_0( \sqrt{ \gamma / \alpha + s } R ) \,
{\rm e}^{ - \alpha s | z | / 2 k' },
\quad s = Q^2 - \gamma / \alpha 
	\label{eq:integrale-acausale}
\end{equation}
Nous avons encore n\'eglig\'e la contribution en $Q^2$ dans
le d\'enominateur. Sur les distances qui contribuent \`a
l'int\'egrale ($\alpha s |z| / 2 k' \sim 1$), elle est
n\'egligeable si $z \gg \bar{\alpha} / 2 k'$. Puisque
cette distance est bien en-dessous de la longueur d'onde,
cette approximation est justifi\'ee aux grandes \'echelles
qui nous int\'eressent ici.

Pour l'\'echelle transverse $\delta R$, nous trouvons alors
deux limites suivant la distance $|z|$ du point d'observation :
\begin{equation}
\delta R \simeq
\left\{
\begin{array}{cl}
\sqrt{ \alpha / \gamma } \simeq \sqrt{ \demi \ell_c \lambdabar }
& \mbox{pour } z \gg \ell ,
\\
\sqrt{ \alpha |z| / 2 k' } \simeq \sqrt{ \demi \ell_c \lambdabar 
 ( |z| / \ell ) }
& \mbox{pour } z \ll \ell .
\end{array}
\right.
	\label{eq:largeur-acausale}
\end{equation}

\subsubsection{Discussion}

\begin{figure}
\centering
\resizebox{!}{5cm}{%
\includegraphics{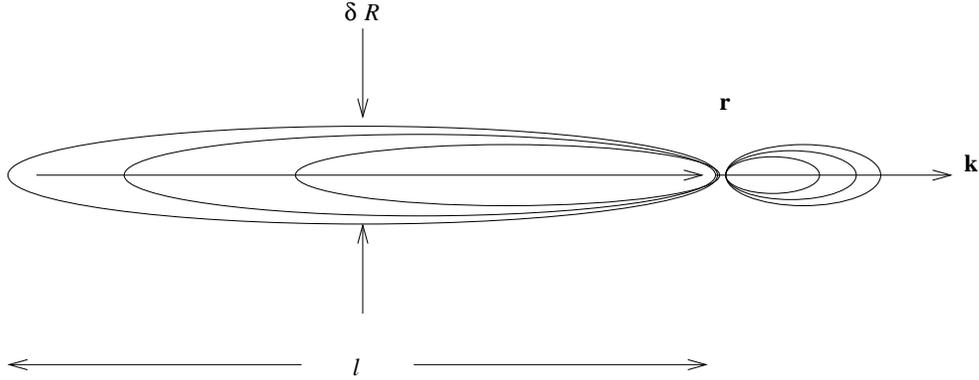}}
\caption[fig:lobe]{Repr\'esentation de la fonction de phase
nonlocale. Les processus de diffusion \`a l'int\'erieur
des ellipses \`a la luminance au point d'observation ${\bf r}$.
Les ellipses repr\'esentent leur poids r\'espectif par des
courbes de niveaux. Elles sont \'etir\'ees autour du rayon
arrivant au point ${\bf r}$ dans la direction d'observation
${\bf k}$. La partie \flqq en amont\frqq donne \'egalement
une contribution.}
\label{fig:lobe}
\end{figure}

Nous constatons qu'en un point d'observation ${\bf r}$ donn\'e, 
la luminance est la somme
des processus de diffusion qui ont amen\'e de la lumi\`ere dans
la direction d'observation ${\bf k}$ et {\em qui ont eu lieu
dans un \flqq lobe\frqq\ autour du rayon g\'eom\'etrique
de direction ${\bf k}$ qui aboutit \`a ${\bf r}$\/}. 
Ce ne sont plus des processus
de diffusion qui ont eu lieu exactement sur le rayon
qui contribuent \`a la luminance observ\'ee : le terme
de diffusion de l'\'equation du transfert radiatif est devenu 
non-local.  La largeur transverse des \flqq lobes\frqq est
de l'ordre de $\delta R \simeq ( \ell_c \lambda )^{1/2}$.
En outre, une contribution non nulle de la luminance provient
de positions \flqq acausales\frqq\ sur le rayon et en aval du
point d'observation.  La situation est illustr\'ee sur
la figure~\ref{fig:lobe} par des courbes de niveaux.

Nous constatons que la longueur de non-localit\'e transverse
est \'egale \`a la
taille de la zone de {\sc Fresnel} lorsque l'on observe
une source ponctuelle de longueur d'onde $\lambda$ \`a la
distance $\ell_c$ de la longueur de corr\'elation.
La non-localit\'e de l'ETR que nous mettons ici en
\'evidence est donc reli\'ee \`a la diffraction dans le
champ interm\'ediaire (diffraction de {\sc Fresnel}
par rapport \`a celle de {\sc Fraunhofer} dans le champ
lointain). En outre, l'\'echelle $\delta R$ sugg\`ere
que la d\'eviation par rapport \`a l'ETR locale est due
\`a la diffusion recurrente entre deux diffuseurs qui sont
plac\'es \`a l'int\'erieur d'un rayon de corr\'elation et
qui s'\'eclairent mutuellement par leurs champs proches
et interm\'ediaires. C'est en effet une des pr\'edictions
de {\sc Lagendijk} et {\sc van Tiggelen} pour les
ph\'enom\`enes physiques au-del\`a de l'ETR standard
\cite{vanTiggelen96}.

Notre calcul permet d'obtenir encore un autre r\'esultat
curieux qui concerne le comportement \`a grande distance
$|z| \gg \ell$ de la fonction d'att\'enuation. Dans
cette limite, nous pouvons calculer les int\'egrales
sur $s$~(\ref{eq:integrale-causale}, \ref{eq:integrale-acausale})
de fa\c con approch\'ee.  Leurs contributions dominantes 
proviennent en effet de la r\'egion $s \to 0$, de sorte
que nous obtenons, pour le cas causal par exemple,
\begin{equation}
\frac{ {\rm i} }{ 8 \pi k' }
\int\limits_0^{\gamma / \alpha } \!
{\rm d}s \,
J_0( \sqrt{ \gamma / \alpha  } R ) \,
{\rm e}^{ - \alpha s z / 2 k' } 
= 
\frac{ {\rm i} J_0( R / \delta R ) 
}{ 4 \pi \alpha }
\frac{ 1 }{ z }
	\label{eq:decroissance-en-1/z}
\end{equation}
Si nous pouvons r\'esoudre l'\'echelle transverse $\delta R$,
la fonction d'att\'enuation est donc de longue port\'ee,
avec une d\'ecroissance en $1/z$. Nous n'avons pas encore
trouv\'e d'interpr\'etation physique simple de ce r\'esultat.
Il est d'autant plus surprenant que l'on trouve un
comportement analogue pour le cas \flqq acausal\frqq.

Il est cependant \'evident que ce r\'esultat est d\^u aux
vecteurs d'onde transverses $Q^2 \simeq \gamma / \alpha$,
qui s'accompagnent d'une d\'ecroissance tr\`es lente de la fonction
d'att\'enuation en fonction de $|z|$ 
(voir~(\ref{eq:integrale-Q})).
L'on peut alors revenir \`a la condition~(\ref{eq:condition-q-petit})
sur les vecteurs d'onde ($Q \le |q| \lesssim 1 / \ell_c$)
et se demander si elle est encore
v\'erifi\'ee pour la situation que nous envisageons ici.
En imposant $ ( \gamma / \alpha )^{1/2} \lesssim 1 / \ell_c$ et
en utilisant l'estimation pour $\alpha$ donn\'ee ci-dessus,
nous trouvons la condition suivante :
\begin{equation}
\ell_c \lesssim \lambdabar 
	\label{eq:lambda-grand}
\end{equation}
Nos conclusions sont donc seulement valables lorsque la longueur
de corr\'elation est plus petite que la longueur d'onde.
Ceci n'est pas tr\`es \'etonnant parce que c'est seulement dans
ce r\'egime que la longueur de non-localit\'e $\delta R$ est plus
grande que la longueur de corr\'elation (sur laquelle il faut
toujours moyenner dans une th\'eorie statistique de transport).
Ce r\'egime appara\^{\i}t fr\'equemment dans le domaine
optique lors de la diffusion par des tr\`es petites particules.
Il est par contre plus difficile \`a r\'ealiser avec les atomes.

Les atomes se trouvent g\'en\'eralement dans le r\'egime oppos\'e
o\`u la longueur d'onde est petite par rapport \`a la longueur
de corr\'elation du milieu al\'eatoire (r\'egime semi-classique). 
Dans ce r\'egime,
l'on n'a pas le droit d'\'etendre le domaine d'int\'egration
\`a des vecteurs d'onde plus grand que $1 / \ell_c$. En
limitant les int\'egrales sur ${\bf Q}$ \`a cet intervalle,
c'est alors la longueur de corr\'elation qui donne
l'\'echelle spatiale minimale pour l'ETR :
\begin{equation}
\mbox{r\'egime semi-classique :}\quad
\delta R \simeq \ell_c
	\label{eq:deltaR-semicl}
\end{equation}
En outre,
la fonction d'att\'enuation ne contient qu'une partie causale,
avec une port\'ee longitudinale donn\'ee par le libre parcours
moyen $\ell$.

\subsubsection{Conclusion}

L'\'equation du transfert radiatif est valable lorsque
la luminance varie lentement \`a l'\'echelle d'une
longueur de {\sc Fresnel}, d\'efinie par la longueur
d'onde et la longueur de corr\'elation. Les d\'eviations
par rapport \`a l'ETR proviennent alors de la diffusion
recurrente de la lumi\`ere par des diffuseurs qui
s'\'eclairement mutuellement par leurs champs proche
et interm\'ediaire. Cette description suppose des
diffuseurs petits par rapport \'a la longueur d'onde.
Dans la limite oppos\'ee (r\'egime semi-classique),
la longueur minimale pour l'ETR est donn\'ee par la
longueur de corr\'elation.

Nous avons obtenu ce r\'esultat en calculant une correction
\`a l'\'equation du transfert lorsque l'on se rapproche 
des \'echelles spatiales plus courtes. Ce calcul met en
outre en \'evidence une correction au libre parcours moyen 
qui fait intervenir la d\'eriv\'ee de l'op\'erateur de masse
(le passage de $k$ \`a $k'$ dans~(\ref{eq:k-prime})).
Cette correction est probablement li\'ee \`a la vitesse de
transport du champ dans le milieu qui n'est donn\'ee ni par
la vitesse de groupe ni celle de phase \cite{vanTiggelen96}.
Elle appara\^{\i}t encore plus nettement dans une approche
au transport d\'ependante du temps, que nous n'avons pas suivie ici.

\paragraph{Relation de dispersion non-triviale.}
Pour finir, une question d'ordre exp\'erimental :
comment est-ce possible que l'op\'erateur de masse pr\'esente
une variation importante avec le vecteur d'onde ${\bf k}$ ?
L'on aurait alors une vitesse de transport $k'$ tr\`es diff\'erente
de celle de propagation libre, ainsi qu'un libre parcours
moyen $\ell'$ fortement corrig\'e. On peut esp\'erer que tel
est le cas dans les solides
o\`u la relation de dispersion pr\'esente des fortes d\'eviations
par rapport \`a l'espace libre, notamment aux bords de la
zone de {\sc Brilluoin} (aux bords des bandes permises). 
Une autre voie peut \^etre ouverte par des particules diffuseurs
avec une \flqq structure interne\frqq, des billes di\'electriques
avec un noyau creux, par exemple. En effet, la fonction de
corr\'elation de la constante di\'electrique $g( r )$
va alors retracer, \`a courte distance,
le profil de l'indice des billes, et ceci se traduira par une
variation importante de l'op\'erateur de masse avec le vecteur
d'onde. Ce dernier est en effet donn\'e, 
dans l'approximation de {\sc Bourret} et pour
des corr\'elation isotropes, par
\[
m({\bf k}) = \frac{ {\rm i} k_0^4 }{ 2 k } 
\int\limits_0^\infty \!{\rm d}r \, g( r ) 
\left( {\rm e}^{ {\rm i} ( k_0 + k ) r } - 
{\rm e}^{ {\rm i} ( k_0 - k ) r } \right) 
\]
ce qui revient essentiellement \`a la transform\'ee de 
{\sc Fourier} du profil d'indice radial.
D'un point de vue exp\'erimental, l'on pourrait donc faire
varier la structure interne des billes pour observer un
comportement diff\'erent du libre parcours moyen. Malheureusement,
ceci est une variable qui a \'egalement une influence sur
d'autres quantit\'es (le contraste di\'electrique total,
moyenn\'e sur la bille, par exemple). Une astuce peut 
\'eviter cet inconv\'enient : se placer dans une g\'eom\'etrie
en deux dimensions o\`u le vecteur d'onde est fix\'e par
la projection sur le plan de sym\'etrie. En choisisant
l'angle d'incidence, l'on peut alors faire varier le
vecteur d'onde sans changer la fr\'equence.

Il reste \`a \'evaluer quel contraste di\'electrique et quelle
concentration de diffuseurs sont n\'ecessaires pour que
l'effet soit observable. En premi\`ere approche, l'on peut
envisager d'utiliser l'approximation de {\sc Bourret} \`a
cet effet.

\section{R\'eciprocit\'e et r\'etro-diffusion exalt\'ee}

Pour finir, nous voudrions \'etudier la cons\'equence
de la r\'eciprocit\'e de la diffusion pour l'\'equation de
{\sc Bethe--Salpeter} dans la pr\'esente formulation. Sans
en donner une justification pr\'ecise, la r\'eciprocit\'e
exprime que les deux processus de diffusion suivants
ont la m\^eme amplitude de diffusion :
\[
{\mbf \kappa} \to {\bf k} \quad \mbox{et} \quad
\mbox{$- {\bf k}$} \to - {\mbf \kappa}
\]
parce qu'ils s'obtiennent l'un de l'autre par un simple
renversement du temps. Cette invariance existe pour notre
th\'eorie scalaire, elle n'existe plus pour la
diffusion \'electro-magn\'etique en pr\'esence d'un champ 
magn\'etique, par exemple.
Si les deux processus ont la m\^eme amplitude, ceci
reste vrai aussi apr\`es la moyenne sur le potentiel
al\'eatoire. Par cons\'equent, l'on s'attend \`a une
intensit\'e augment\'ee (par interf\'erence constructive)
dans ces directions.  

En comparant aux arguments de l'op\'erateur d'intensit\'e
dans~(\ref{eq:def-phase}), nous constatons qu'il correspond
\`a la corr\'elation entre deux processus r\'eciproques
(donc de m\^eme amplitude) lorsque nous avons
\begin{equation}
\left.
\begin{array}{rcl}
{\bf k} + \demi{\bf q} & = & - ( {\mbf \kappa} - \demi{\bf q} )
	\\[0.5\jot]
{\mbf \kappa} + \demi{\bf q} & = & - ( {\bf k} - \demi{\bf q} )
\end{array}
\right\} \quad \Longrightarrow \quad
{\bf k} = - {\mbf \kappa} .
	\label{eq:def-retrodiffusion}
\end{equation}
C'est donc le cas pour la diffusion vers l'arri\`ere.
Dans l'approximation de {\sc Bourret}, 
l'op\'erateur d'intensit\'e prend la forme~(\ref{eq:K(k)-Bourret})
et la probabilit\'e relative de diffusion vers l'arri\`ere
et vers l'avant est donn\'ee par
\[
\frac{ \mbox{ arri\`ere } }{ \mbox{ avant } } =
\frac{ S( 2 k ) }{ S( 0 ) }
\]
($S(k)$ est la densit\'e spectrale du potentiel al\'eatoire,
suppos\'ee isotrope ici.)
Cette probabilit\'e est faible lorsque la longueur d'onde
est petite par rapport \`a la longueur de corr\'elation.
En outre, elle ne pr\'esente pas de trace particuli\`ere
d'un effet d'interf\'erence dans la direction arri\`ere.
Ceci est d\^u au fait que l'approximation de {\sc Bourret}
ne prend en compte que les processus de simple diffusion.
Par contre, si l'on utilise une expression pour l'op\'erateur
d'intensit\'e
qui aille au-del\`a de l'approximation de {\sc Bourret}, 
l'on peut tr\`es bien d\'ecrire la r\'etro-diffusion exalt\'ee
dans le formalisme de l'\'equation du transfert radiatif.
La bonne nouvelle est qu'il suffit d'inclure une s\'erie
certes infinie de diagrammes 
(les {\em maximally crossed diagrams\/}), mais que leur calcul
n\'ecessite pas, dans la pratique, d'aller au-del\`a de
l'approximation de {\sc Bourret} : l'on s'en sort donc
en \flqq am\'eliorant le formalisme\frqq. Nous renvoyons aux
travaux de {\sc Tsang} \cite{Tsang85} ainsi qu'au~\S~6 du livre
de Ping {\sc Sheng} \cite{Sheng95} pour plus de details.%

\section{Conclusion et perspectives}

L'\'equation du transfert radiatif ne s'en sort pas mal de
cette discussion : sa forme traduit bien le transport de
la luminance (la transform\'ee de {\sc Wigner} du champ)
aux grandes \'echelles spatiales, m\^eme si le potentiel des
diffuseurs est fort et que l'approximation de {\sc Born} \'echoue.
Il convient cependant de rappeler que l'ETR g\'en\'eralis\'ee
a \'et\'e obtenue pour un milieu statistiquement homog\`ene,
donc sans prendre en compte les effets de bords ni les
interfaces. Il faut donc mod\'eliser ceux-ci au cas par
cas avec des coefficients de r\'eflexion et de transmission,
par exemple.
En outre, l'ETR est seulement valable lorsque l'on
peut parler de la propagation d'ondes dans le milieu :
formellement, ceci se traduit par un libre parcours moyen
beaucoup plus grand que la longueur d'onde (r\'egime de
faible diffusion). L'ETR ne donne donc aucun acc\`es \`a
la localisation des ondes.

Dans la litt\'erature, l'on rencontre souvent des affirmations
moins ambitieuses quant au statut de l'ETR : celle-ci ne
serait valable que dans l'approximation de {\sc Bourret}
\cite{Kravtsov96,Luck93} ou de ``{\em single group\/}''.
En effet, l'analyse de {\sc Barabanenkov} et {\sc Finkel'berg}
\cite{Barabanenkov67} montre que la validit\'e de l'approximation
de {\sc Bourret} entra\^{\i}ne le r\'egime de faible diffusion :
le champ se propage dans le milieu,
les diffuseurs se voient dans le champ lointain et
la dispersion spatiale (la d\'ependance de l'op\'erateur
de masse avec le vecteur d'onde) est n\'egligeable.
Sous ce point de vue, l'approximation de {\sc Bourret}
appara\^{\i}t donc comme une {\em condition suffisante\/} 
pour l'application de l'ETR. Le libre parcours moyen et
la fonction de phase sont alors donn\'es par les op\'erateurs
de masse et d'intensit\'e que l'on calcule en tenant compte
d'un groupe de diffuseurs avec $k = 1,2,...$ particules
(caract\'eris\'e par une fonction de corr\'elation \`a $k$ points).
Cette ``{\em single group approximation\/}''
revient \`a l'approximation de {\sc Born} (les formules
de {\sc Ryzhik, Papanicolaou} et {\sc Keller}) pour la diffusion
par un potentiel al\'eatoire continue. Pour des particules
diffuseurs, des corr\'elations entre particules peuvent donc en
principe \^etre d\'ecrites par le formalisme, mais il semble
difficile dans la pratique de calculer les sections efficaces
de diffusion (voir \cite{Tsang85,MacKintosh89} pour des 
exemples).

L'\'equation du transfert radiatif g\'en\'eralis\'ee 
(\flqq ETR$^*$\frqq, \'eq.~\ref{eq:ETR*-diff}) que nous
avons trouv\'ee ici ne d\'epend pas, {\em a priori\/}, de
l'approximation de {\sc Bourret}. Elle va donc au-del\'a
de l'approximation de la diffusion simple, tout en supposant
que la diffusion est \flqq faible\frqq\ (ondes non-localis\'ees).
Les travaux des Russes indiquent les effets physiques nouveaux
qui sont d\'ecrits par l'ETR$^*$ :
\begin{itemize}
\item
la diffusion devient d\'ependante (diffuseurs s'\'eclairant dans le
champ proche et interm\'ediaire, diffusion recurrente)
\item
la relation de dispersion s'\'elargit (dispersion spatiale :
la partie imaginaire de la fonction de {\sc Green} moyenn\'ee
n'est plus une fonction $\delta$)
\item
il appara\^{\i}t une vitesse de transport pour l'intensit\'e 
des ondes.
\end{itemize}
Nous avons identifi\'e les traces de ces effets en d\'eveloppant
la partie \flqq propagation\frqq\ de l'\'equation de {\sc Bethe--%
Salpeter} au-del\`a de l'approximation habituelle des grandes \'echelles.
Il faut cependant rester prudent sur ces r\'esultats
parce que les autres termes dans l'\'equation de {\sc B.--S.}
donnent des contributions analogues, la fonction de {\sc Green}
moyenn\'ee et l'op\'erateur d'intensit\'e. En outre, ces
contributions ne sont pas ind\'ependantes parce qu'elles sont
reli\'ees entre elles par l'identit\'e de {\sc Ward} (une
g\'en\'eralisation du th\'eor\`eme optique), qui est 
valable \`a toute \'echelle. La liste des effets physiques
ci-dessus semble quand m\^eme qualitativement correcte ;
{\sc Lagendijk} et {\sc van Tiggelen} dressent en effet un
r\'epertoire analogue \cite{vanTiggelen96}.

Finalement, tout semble indiquer que l'\'equation du transfert
radiatif ne peut pas d\'ecrire la localisation (forte) des ondes
dans un milieu al\'eatoire. Notre \'etude peut tout au plus
sugg\`erer l'image physique suivante : nous avons constat\'e
que l'\'equation de {\sc Bethe--Salpeter} est non-locale, 
avec une fonction de phase non-positive qui pr\'esente des 
changements de signe. 
Il est alors concevable que dans l'int\'egrale sur les
processus de diffusion, ceux-ci \flqq interf\`erent de fa\c con
destructive\frqq\ :
aux grandes distances, le champ ne se propagerait pas et
serait {\em localis\'e\/} dans le milieu diffusant.
Rappelons dans ce contexte que la fonction de {\sc Wigner}
n'est pas forc\'ement positive, et que ses \flqq n\'egativit\'es\frqq\ 
sont pr\'ecis\'ement le r\'esultat d'interf\'erences.




\chapter*{Conclusion}
\addcontentsline{toc}{chapter}{Conclusion}
\setcounter{footnote}{0}
\setcounter{figure}{0}

\section*{Le transport des ondes}
\addcontentsline{toc}{section}{Le transport des ondes}

Nous avons \'etudi\'e dans ce rapport le transport d'une onde
(scalaire, \'electro-magn\'etique ou de mati\`ere) \`a travers
un milieu al\'eatoire diffusant. L'on peut distinguer trois
niveaux de description dans la th\'eorie :
\begin{enumerate}
\item
le niveau \flqq microscopique\frqq, o\`u l'on tient compte
de l'aspect ondulatoire, en \'etudiant le champ moyen et
la fonction de coh\'erence. A ce niveau,
les effets d'interf\'erence sont inclus dans la th\'eorie,
dont les \'equations de base sont celles de {\sc Dyson}
et de {\sc Bethe--Salpeter}.
\item
un niveau \flqq m\'esoscopique\frqq, o\`u le champ est
d\'ecrit par une luminance : elle en donne
l'intensit\'e, distribu\'ee en position
et selon les directions de propagation. La th\'eorie du
transfert radiatif, mais aussi la m\'ecanique classique
statistique si situent \`a ce niveau th\'eorique. 
Les \'equations de base sont celle du transfert radiatif,
de {\sc Boltzmann} ou encore l'\'equation de {\sc Fokker--Planck}%
\footnote{%
Il d\'epend du rapport entre la longueur d'onde et
la longueur de corr\'elation si c'est l'ETR ou l'\'equation
de {\sc F.--P} qui est bien adapt\'ee au probl\`eme.
La deuxi\`eme correspond au r\'egime des petites
longueurs d'onde o\`u la diffusion se produit de
pr\'ef\'erence aux petits angles.}%
. Le lien \`a la
th\'eorie microscopique passe par l'identification
de la luminance avec la fonction de corr\'elation
(de coh\'erence) du champ en repr\'esentation de
{\sc Wigner}. L'on obtient l'\'equation du transfert
radiatif \`a partir de celle de {\sc Bethe--Salpeter}
dans la limite des \'echelles spatiales grandes 
par rapport \`a la longueur de corr\'elation du milieu
(les diffuseurs se voient en champ lointain).
\item
Finalement, la diffusion dans le milieu redistribue
l'intensit\'e de l'onde du faisceau collimat\'e
incident vers la partie diffuse de la distribution
angulaire. Aux grandes distances, la distribution
angulaire devient isotrope, et le champ est d\'ecrit
seulement par son densit\'e locale d'\'energie.
Sur de telles \'echelles \flqq macroscopiques\frqq,
le transport de l'onde est gouvern\'e par une
\'equation de la diffusion (spatiale). Dans une
exp\'erience avec les atomes o\`u le temps d'interaction
est souvent limit\'e, ce sont seulement les atomes
les plus lents qui entrent dans le r\'egime du transport
diffusif. 
\end{enumerate}
Ces diff\'erents niveaux th\'eoriques sont repr\'esent\'es
sur la figure suivante.
\begin{center}
\resizebox{!}{12cm}{%
\setlength{\unitlength}{3947sp}%
\begingroup\makeatletter\ifx\SetFigFont\undefined%
\gdef\SetFigFont#1#2#3#4#5{%
  \reset@font\fontsize{#1}{#2pt}%
  \fontfamily{#3}\fontseries{#4}\fontshape{#5}%
  \selectfont}%
\fi\endgroup%
\begin{picture}(5121,5424)(811,-5023)
\thicklines
\put(2390,-308){\framebox(1336,697){}}
\put(881,-1934){\framebox(1451,755){}}
\put(3668,-1992){\framebox(1974,813){}}
\put(823,-3327){\framebox(1858,697){}}
\put(881,-5011){\framebox(1683,813){}}
\put(1461,-4198){\vector( 0, 1){871}}
\put(1403,-2630){\vector( 0, 1){696}}
\put(1519,-1179){\vector( 3, 2){1326.923}}
\put(4597,-1179){\vector(-3, 2){1326.923}}
\put(2332,-1643){\vector( 1, 0){1336}}
\put(2622,-110){\makebox(0,0)[lb]{\smash{\SetFigFont{9}{10.8}{rm}{}{}la diffusion}}}
\put(1113,-1736){\makebox(0,0)[lb]{\smash{\SetFigFont{9}{10.8}{rm}{}{}transfert radiatif}}}
\put(1635,-2398){\makebox(0,0)[lb]{\smash{\SetFigFont{8}{9.6}{rm}{}{}diffusion simple}}}
\put(5874, 41){\makebox(0,0)[lb]{\smash{\SetFigFont{9}{10.8}{it}{}{}macroscopique}}}
\put(5932,-4430){\makebox(0,0)[lb]{\smash{\SetFigFont{9}{10.8}{it}{}{}microscopique}}}
\put(2506,-830){\makebox(0,0)[lb]{\smash{\SetFigFont{8}{9.6}{rm}{}{}luminance isotrope}}}
\put(2564, 99){\makebox(0,0)[lb]{\smash{\SetFigFont{9}{10.8}{rm}{}{}équation de }}}
\put(4132,-1469){\makebox(0,0)[lb]{\smash{\SetFigFont{9}{10.8}{rm}{}{}équation}}}
\put(3784,-1736){\makebox(0,0)[lb]{\smash{\SetFigFont{9}{10.8}{rm}{}{}de {\sc Fokker--Planck}}}}
\put(5874,-1527){\makebox(0,0)[lb]{\smash{\SetFigFont{9}{10.8}{it}{}{}mésoscopique}}}
\put(1171,-1469){\makebox(0,0)[lb]{\smash{\SetFigFont{9}{10.8}{rm}{}{}équation du}}}
\put(2506,-1527){\makebox(0,0)[lb]{\smash{\SetFigFont{8}{9.6}{rm}{}{}$\Delta\theta \ll 1$}}}
\put(997,-3037){\makebox(0,0)[lb]{\smash{\SetFigFont{9}{10.8}{rm}{}{}ETR généralisée}}}
\put(1810,-3676){\makebox(0,0)[lb]{\smash{\SetFigFont{8}{9.6}{rm}{}{}grande échelle spatiale}}}
\put(1229,-4488){\makebox(0,0)[lb]{\smash{\SetFigFont{9}{10.8}{rm}{}{}équation de}}}
\put(997,-4756){\makebox(0,0)[lb]{\smash{\SetFigFont{9}{10.8}{rm}{}{}{\sc Bethe--Salpeter}}}}
\end{picture}
}\\
{\it Repr\'esentation sch\'ematique
des diff\'erents niveaux d'approximation pour le transport
des ondes dans un milieu al\'eatoire.}
\end{center}

\section*{Perspectives}
\addcontentsline{toc}{section}{Perspectives}

Les atomes froids plac\'es dans les tavelures lumineuses 
r\'ealisent un probl\`eme de transport en milieu al\'eatoire 
dont on peut facilement 
varier les param\`etres caract\'eristiques (amplitude
du potentiel, longueur de corr\'elation par rapport
\`a la longueur d'onde (temp\'erature) des atomes, 
dur\'ee de l'interaction,
poids relatif des effets r\'eactifs et dissipatifs).
Le transport d'atomes repr\'esente donc une sorte de
\flqq laboratoire\frqq\ o\`u l'on peut \'etudier des
r\'egimes diff\'erents du transport. Etant donn\'e la
difficult\'e de refroidir des atomes en dessous de la
limite de recul (longueur d'onde atomique plus grande
que la longueur de corr\'elation des tavelures), 
c'est d'abord le r\'egime mesoscopique
du transport que l'on explorera. Ce n'est pas un domaine
d\'epourvu d'int\'er\^et : l'on peut m\^eme envisager
de mod\'eliser, en pr\'esence du champ de pesanteur,
un mod\`ele pour la conduction des \'electrons dans un solide
avec des d\'efauts (potentiel al\'eatoire plus force
constante).

La structure interne des atomes offre 
une autre possibilit\'e int\'eressante : il est possible
d'\'etudier par exemple le transport d'un ensemble de
spins polaris\'es \`a travers un champ magn\'etique
al\'eatoire (statique). Ce probl\`eme est pertinent pour
l'interf\'erom\'etrie atomique \`a \'etats internes :
un champ magn\'etique r\'esiduel induit en effet une rotation
al\'eatoire du spin et r\'eduit le contraste des franges
d'interf\'erence. D'un point de vue th\'eorique, l'on
pourra exploiter l'analogie au transfert radiatif de
la lumi\`ere polaris\'ee. Le formalisme de {\sc Ryzhik,
Papanicolaou} et {\sc Keller} est \'egalement suffisamment
g\'en\'eral pour trouver rapidement les \'equations de
transport du spin atomique
dans l'approximation d'un faible champ r\'esiduel.

Dans le domaine du transfert radiatif proprement dit,
les consid\'erations de ce rapport indiquent 
que l'\'equation du transfert radiatif 
est suffisamment g\'en\'erale pour d\'ecrire le r\'egime
de la diffusion d\'ependante, \`a condition de g\'en\'eraliser
la section efficace de diffusion \`a la diffusion par des
agglom\'erats de particules. De la m\^eme fa\c con,
l'ETR peut incorporer dans une certaine mesure des effets 
d'interf\'erence comme la r\'etrodiffusion exalt\'ee.
Elle se limite cependant \`a une description \`a grande
\'echelle spatiale. A plus courte \'echelle, le transport
est r\'egi par une \'equation non-locale, et il n'est plus
possible de d\'efinir une relation de dispersion pour
les ondes dans le milieu. L'ETR est \'egalement limit\'ee
au r\'egime de faible diffusion o\`u le libre parcours
moyen est beaucoup plus grand que la longueur d'onde.
Le r\'egime oppos\'e correspond \`a la localisation des
ondes pour laquelle il faut revenir \`a une
description microscopique \`a l'aide de l'\'equation
de {\sc Bethe--Salpeter}. Il est \'etonnant qu'il faille en 
\'elaborer une th\'eorie \`a un niveau si fondamental bien que la
localisation se manifeste par l'absence de transport
\`a tr\`es grande \'echelle spatiale
(voir le titre de l'article d'{\sc Anderson}). Il 
semble que se trouve l\`a une explication pour la
difficult\'e notoire du probl\`eme de la localisation
qui reste ouvert m\^eme apr\`es quarante ans de
recherches ardues.


\bigskip

\begin{flushright}
Carsten Henkel\\
Ch\^atenay--Malabry, juillet 1997
\end{flushright}


\end{document}